\documentclass[11pt]{article}
\usepackage{fullpage}
\usepackage{media9}
\addmediapath{./Movies/}
\usepackage{multimedia}
\usepackage{subcaption}
\usepackage[english]{babel}
\usepackage[utf8]{inputenc}
\usepackage{amsmath}
\usepackage{amssymb}
\usepackage{float}
\usepackage{multicol}
\usepackage{dirtytalk}
\usepackage{graphicx}
\usepackage{caption}
\usepackage{adjustbox}
\usepackage{ulem}
\usepackage[margin=1in]{geometry}
\usepackage{graphicx}
\usepackage[colorinlistoftodos]{todonotes}
\usepackage{algorithm}
\usepackage{algpseudocode}
\usepackage{siunitx}
\usepackage{bm}
\usepackage{natbib}
\def\Var{\text{Var}}
\def\Cov{\text{Cov}}
\usepackage{setspace}
\usepackage{color}

\newcommand{\mathi}{\mathrm{i}}
\newcommand{\expe}{\mathrm{e}}
\newcommand{\grad}{\boldsymbol{\nabla}}
\DeclareMathOperator{\ord}{ord}
\newcommand{\jp}{j^{\prime}}
\newcommand{\jpp}{j^{\prime\prime}}
\newcommand{\celllen}{l}
\newcommand{\colrad}{a}
\newcommand{\flagang}{\alpha}
\newcommand{\flagdisp}{S}
\newcommand{\flagdispi}{\flagdisp_i}

\newcommand{\flagposj}{\mathbf{l}_j}

\newcommand{\Xcol}{\mathbf{X}^{(c)}}
\newcommand{\hXcol}{\hat{\mathbf{X}}^{(c)}}
\newcommand{\Xcolnd}{\tilde{\mathbf{X}}^{(c)}}
\newcommand{\Xcolavg}{\bar{\mathbf{X}}^{(c)}}
\newcommand{\Thetacol}{\Theta^{(c)}}
\newcommand{\Thetacolavg}{\bar{\Theta}^{(c)}}
\newcommand{\Thetacolfluc}{\tilde{\Theta}^{(c)}}
\newcommand{\hThetacol}{\hat{\Theta}^{(c)}}
\newcommand{\thetacol}{\theta^{(c)}}
\newcommand{\Thetacelli}{\Theta_{i}}
\newcommand{\Thetacellnd}{\tilde{\Theta}}
\newcommand{\Thetacellndi}{\Thetacellnd_{i}}
\newcommand{\bThetacellnd}{\tilde{\bm{\Theta}}}
\newcommand{\difd}{\mathrm{d}}
\newcommand{\bX}{\mathbf{X}}
\newcommand{\bx}{\mathbf{x}}
\newcommand{\bW}{\mathbf{W}}

\newcommand{\gradnorm}{g}
\newcommand{\gradang}{\theta_g}
\newcommand{\graddir}{\hat{\mathbf{e}}_g}
\newcommand{\gradlen}{\ell_g}
\newcommand{\ktaxis}{k_T}
\newcommand{\ktaxiseff}{k_T^{(c)}}
\newcommand{\ktaxisnd}{m_T}
\newcommand{\kkin}{k_K}
\newcommand{\kkinnd}{m_K}
\newcommand{\kkincol}{k_K^{(c)}}
\newcommand{\sigmaThe}{\sigma_{\Theta}}
\newcommand{\derat}{\beta}

\newcommand{\trho}{\rho_{\bThetacellnd|\Thetacol}}
\newcommand{\trhoi}{\rho_{\Thetacellndi|\Thetacol}}
\newcommand{\hthvar}[1]{y_{{#1}}}
\newcommand{\thvarnd}[1]{y_{{#1}}}
\newcommand{\hbthvar}{\mathbf{y}}
\newcommand{\tbthvar}{\mathbf{y}}
\newcommand{\hthcvar}{z}
\newcommand{\thcvar}{z}
\newcommand{\pstatthctax}{\rho_{\Thetacol}}
\newcommand{\pstatthc}{\rho_{\hThetacol}}

\newcommand{\geotaxamp}{\chi}
\newcommand{\geotaxphase}{\phi}
\newcommand{\twoflagavg}{M_2}
\newcommand{\twoflagvar}{V_2}
\newcommand{\twoflagmod}{\twoflagvar^{\ast}}
\newcommand{\threeflagcov}{M_3}
\newcommand{\threeflagcovconj}{\threeflagcov^{\ast}}
\newcommand{\flagangv}{\omega}
\newcommand{\drifteff}{\mathbf{V}_{\infty}}
\newcommand{\drifteffnd}{\tilde{\mathbf{V}}_{\infty}}
\newcommand{\drifteffmean}{\langle \drifteff \rangle}
\DeclareMathOperator{\Real}{\mathrm{Re}}
\DeclareMathOperator{\Imag}{\mathrm{Im}}
\newcommand{\Diffrotcol}{D_{\mathrm{r}}}
\newcommand{\CI}{\mathrm{CI}}
\newcommand{\Cordrift}{C}
\title{Stochastic Analysis of Taxis and Kinesis Properties of Colonial Protozoa} 
\author{Yonatan L. Ashenafi and Peter R. Kramer}

\newcommand\PRK[1]{\textcolor{blue}{\textbf{PRK:  {#1}}}}

\begin{document}
\maketitle 
\begin{abstract}
   Protozoan colonies  undergo  stimulus driven motion for purposes such as 
   nutrient acquisition. Colonial response to a stimulus is mediated through a mechanical aggregation of the response properties of members of the colony. We develop and apply asymptotic analysis to a stochastic model for the integration of two classes of stimulus responses of the constituent cells -- taxis and kinesis.  We investigate in particular the maintenance of effectiveness of taxis and kinesis in the transition from unicellular to multicellular organisms, using experimental observations of chemotaxis and aerotaxis of protozoa as a reference.   Our taxis model based on a steering response of individual cells actually leads to a counterproductive drift of the colony down the stimulus gradient, together with a constructive drift up the gradient which is proportional to a measure of asymmetry of the flagellar placement.  The strength of taxis drift up the stimulus gradient decreases with colony size while the counterproductive term does not, indicating a failure for colonial taxis based on a steering response of individual cells.  Under a kinesis response of the cellular flagellar motion, enhancing the noise as the cell is facing away from the stimulus gradient, the colony does drift up the gradient with a speed independent of colony size, even under a completely symmetric placement of flagella.
\end{abstract}


\section{Introduction}  \label{sec:Intro}
The microhydrodynamics of microorganisms
in response to stimuli of their fluid environments is an essential aspect of life.  One broad class of stimulus response is taxis, in which microorganisms endeavor to realign their motion up a favorable environmental gradient (or down an unfavorable one).  
Another form of stimulus response is kinesis, in which microorganisms vary the vigor of some disorganized or diffusive motion in response to the local environment.  The general goal in kinesis is to upregulate noise when the organisms is in a currently unfavorable configuration and downregulate noise when the current configuration is sensed to be favorable.
Processes of taxis and kinesis are involved in various functions of life such as avoidance of toxins for bacteria \citep{Adler}, reproduction for sea urchins and other animals where sperm cells navigate to egg cells~\citep{Hildebrand,VOGEL1982189,Armon}, and nutrient acquisition for ameboid cells \citep{FIRTEL2000421}.

Examples of environmental gradients driving taxis and kinesis are chemical (chemotaxis) and oxygen (aerotaxis) gradients. For chemotaxis, prokaryotes like bacteria and eukaryotes have different mechanisms. Bacteria use temporal sampling of concentration, where they tumble less frequently than normal when they sense a positive growth of concentration in time and tumble more frequently than normal for a negative temporal gradient~\citep{Block,Macnab2509}. On the other hand, eukaryotic cells have been shown to sense chemo-attractant gradients spatially where they use their relatively larger diameter to compute differences of chemical concentration across their cell body \citep{Levine}. This has also been documented by microfluidic experiments for aerotaxis in animal-like protozoa called choanoflagellates, which sense and migrate up the relative spatial gradient of oxygen~\citep{aerotax, Sherratt1994}.  In this paper, we focus on a spatial gradient sensing mode which would be appropriate for chemotaxis and aerotaxis in protozoa.    


The question of how the cells in the colony aggregate their stimulus response properties to cause effective stimulus response of the colony remains largely analytically open.
The collective swimming effect of ciliary arrays have of course been well analyzed in terms of their hydrodynamic and internal coupling which can induce metachronal synchronization~\citep{JoannyFlow}.  The colonial alga \emph{Volvox carteri} was also found to exhibit a substantial degree of metachronal coordination in its flagella~\citep{BrumleyVolvox}.
The flagella in many other protozoan colonies are by contrast far less numerous and geometrically organized, being roughly uniformly but certainly not periodically nor isotropically arranged on the surface~\citep{DayelChoano2011}. \emph{S. rosetta} for example exhibits morphological variations between cells in a colony \citep{Naumann-morphology}, and the flagella from different \emph{S. rosetta} cells were found experimentally to beat in an uncoordinated and approximately independent manner~\citep{colonial-motility,Roper}.

So rather than considering how the flagella from different cells on the colony cooperate to boost stimulus driven motion, the question in such organisms seems to be how well the colony can still manage a coherent stimulus response from independent  responses of the constituent cells which are moreover geometrically constrained to frustrate favorable alignment of most of the cells with the environmental gradient.  \citep{Roper} point out the relatively inefficient swimming behavior of \emph{S. rosetta} colonies relative to individual cells due to the disorganization of the flagella, but provide theoretical arguments for how they can nonetheless improve the fluid supply to enhance the feeding of the colony on suspended prey. Subsequent finer computational studies of the cellular geometry, particularly the feeding collar, brought into question the conclusion of actual filter feeding enhancement of colonies~\citep{Kirkegaard2016Filter,KoehlSelective,Fauci_morphology}.

We endeavor here to apply some general mathematical modeling, similar in spirit to~\citep{Roper}, to quantify analogously the effectiveness of taxis or kinesis of a colony of spatially gradient sensing cells in the absence of coordination between them.   Our goal is similarly a baseline theory which would likely require modification in detail for quantitative accuracy for particular protozoans.  Stochastic effects are central, because the simplest approximation of uniformly distributed, identical flagella normal to a spherical surface would lead to no motion due to cancellation of the flagellar forces and torques!  Our model framework comprises two sources of stochasticity:  1) dynamical stochasticity, meaning the flagellar beating exhibits some irregular noisy behavior as a function of time~\citep{colonial-motility}, and 2) demographic stochasticity, meaning the flagellum of a cell is not perfectly placed in the center of the portion exposed to the fluid due to variations in how the cells are oriented within the colony.  We endeavor to quantify how each of these sources of randomness contribute, to leading order, to the  taxis and kinesis properties of the colony.  Our focus will be entirely on the effects of mechanical coupling between the cells in a colony in response the the environmental stimulus, and do not concern ourselves here with considerations of the means or quality of the sensing of the environmental gradient.  In particular, following~\citet{aerotax}, we distinguish kinesis and taxis purely by whether the redirection response is by direct steering or indirect modulation and not on whether the sensing is temporal or spatial~\citep{MachemerOrientation2001}.
The question of how cells may communicate, either mechanically or chemically, their sensory information with other cells in the colony to improve their measurement of the environmental gradient has been addressed, for example, in~\citep{Colizzi-Pottsmodel}.  \citet{CamleyLeader,MuglerChemotaxis} in particular conduct complementary studies to ours in considering how the geometric arrangement of gradient-sensing cells in a cluster affects the quality of the collective motion up the gradient as determined by a simple mechanical weighting of the signal from each sensing cell.  These studies of taxis in cell clusters are typically not concerned with geometric constraints of motion apart from the often rigid model for the cluster arrangement, as cells which creep rather than swim can move flexibly in arbitrary directions without a body reorientation. The ``emergent chemotaxis'' model considered in~\citet{MuglerChemotaxis}, however, does constrain cell polarization to be always outwardly normal to the colony due to contact inhibition with the neighboring cells, and we will briefly compare and contrast it with our models in Section~\ref{sec:conclusion}.  \cite{CamleyReview} provides an excellent overview of the questions and analyses of various models for how gradient-sensing creeping cells interact within a cluster to determine the collective behavior.

Taxis and kinesis have been extensively studied, particularly the analysis and derivation of Keller-Segel partial differential equation models to describe the dynamics of suspensions of microorganisms \citep{kellersegel,KSmodelSummary, KSmodelkinesis,Othmer}.  \citet{CelaniStrategies} in particular analyze partial differential equation models to consider which temporal structures for the taxis response optimize nutrient uptake in smooth or fluctuating environments.  We focus here on the statistical dynamics of an individual colony, similarly to recent work simulating the chemotaxis response of aggregates of cells based on the cellular Potts model \citep{Colizzi-Pottsmodel}.  
We will represent taxis and kinesis mathematically building on the simple phenomenological models considered in~\citep{aerotax}.  The swimming cell or microorganism  is taken to move at speed $ \nu $ in a plane 
along a dynamically varying orientation $ \Theta(t)$, so that the dynamics of its position are governed simply by:
\begin{equation}  \label{colony_position}
\begin{aligned}
\difd\bX (t)=\nu \begin{pmatrix}
\cos(\Theta(t)) \\
\sin(\Theta(t))
\end{pmatrix}\, \difd t + \sqrt{2 D_t} \, \difd \bW^{X} (t),
\end{aligned}
\end{equation}
where the stochastic Brownian motion term describes translational diffusion with diffusivity $ D_t$.  Next we assume a favorable environmental gradient along the direction $ \theta=\pi/2$.  A model for taxis would have the orientation $ \Theta (t) $ tend toward the favorable direction:  
\begin{equation} \label{directed_flagellum_orientation_taxis}
\begin{aligned}
d\Theta(t)=r_T\cos(\Theta)dt+\sqrt{2D_r}dW^{\Theta}(t).
\end{aligned}
\end{equation}
Here $ r_T $ would describe the strength of the taxis response and the stochastic term describes rotational diffusion with diffusivity $ D_r$.  On the other hand, a model for kinesis would rather modulate the strength of rotational diffusion depending on the current orientation:  
\begin{equation} \label{directed_flagellum_orientation_kinesis}
\begin{aligned}
d\Theta(t)=\sqrt{2D_r(1-\kkinnd \sin(\Theta))}dW^{\Theta}(t).
\end{aligned}
\end{equation}
This noise modulation can be viewed as a continuous version of the bacterial run and tumble process which has been extensively studied theoretically~\citep{Othmer}. The cell reduces the noisiness of their rotational motion when the local chemo-attractant gradient agrees with the orientation of the cell.  The models~\eqref{directed_flagellum_orientation_taxis} and~\eqref{directed_flagellum_orientation_kinesis} were actually applied in~\citep{aerotax} phenomenologically to colonies as a whole, with the kinesis model found to better represent the experimental observations of aerotaxis in \emph{S. rosetta}.  

In Section~\ref{sec:model}, we apply rather these basic taxis and kinesis for individual cells in a colony, and model  the translational and rotational dynamics of a colony by the resultant forces and torques exerted by the cells in response to the spatial gradient of an environmental stimulus.  To focus on the essential ideas, we restrict ourselves to dynamics in two spatial dimensions (flat colonies) to avoid the intricacies of three-dimensional rotational dynamics.  We present first the results and analyses of the idealized case of a circular disc geometry for the colony, with each cell taking up an equal portion of the cell boundary.  We do allow for random irregularity of where the flagellum is placed on each cell's boundary, and these demographic variations between cells (and therefore also between colonies) are important because, as our results show, exact regular spacing of the flagella would lead to complete cancellation of the response to the environmental gradient.  More precisely, the effective strength of the colony's response to an environmental gradient, under both taxis and kinesis models, is proportional to a shape parameter governed by the irregularity of the flagellar placement.  \cite{CamleyReview} also emphasizes the importance of studying demographic variability between cells and summarizes their effects on collective response via a different shape parameter in the context of creeping, rather than swimming, colonies of cells.

In Section \ref{sec:nondim}, we nondimensionalize our models and, based on physical parameter values adopted from the experimental literature, identify small parameters that motivate an asymptotic analysis in Section \ref{sec:analysis} of the effective colonial response to an environmental stimulus gradient.  Our calculation is essentially a homogenization of the contributions of relatively fast dynamics of the individual cell flagella on the relatively slow rotational and translational dynamics of the colony.  
Monte Carlo simulations are employed in Section~\ref{sec:simulations} to validate the asymptotic analysis for our model.  Our main result is a leading order approximation for the mean rate of progress of the colony up the stimulus gradient in terms of the size of the colony, the physical properties of the cells, and the effects of both dynamic and demographic stochasticity.
We conclude in Section \ref{sec:conclusion} with a summary and discussion for what our mathematical results suggest for the stimulus response properties of protozoan colonies, including how our results are expected to generalize in three spatial dimensions.

The results from our analysis show that the kinesis model~\eqref{directed_flagellum_orientation_kinesis} does generally drive the colony 
up the stimulus gradient with approximately linearly increasing effectiveness as the kinesis response factor increases. The speed at which the colony moves up the stimulus gradient decreases with increasing colony size, but does asymptotically approach a positive value which is independent of the colony size and geometry.
Under the taxis model~\eqref{directed_flagellum_orientation_taxis}, the colony typically moves up the stimulus gradient only for small to moderate values of the taxis response factor.  At all values of the taxis response factor, colonies can in fact move at low speeds in the wrong direction depending on the particular flagellar arrangement.  The  motion of the colony down the stimulus gradient becomes more likely and prominent at larger values of the taxis response factor, and is attributable to a sensible steering response for an individual cell  producing a counterproductive steering response in a colony.  As with the kinesis model, the progress of the colony up the stimulus gradient becomes less effective under the taxis model as the colony size increases.

\section{Mathematical Model}
\label{sec:model}
 We present in Subsection \ref{subsec:model_col} a variation of the basic modeling framework for a swimming colony from our companion paper~\citep{Ashenafi-mobility}, which provides more motivational discussion.  We augment this model in Subsections  \ref{subsec:model_taxis} and \ref{subsec:model_kinesis} by a taxis or kinesis response of the cells to an environmental stimulus gradient.

\subsection{Colony Dynamics Model} \label{subsec:model_col}
\begin{figure}[H]
    \begin{subfigure}{0.45\textwidth}
     \centering
     \includegraphics[scale=0.30]{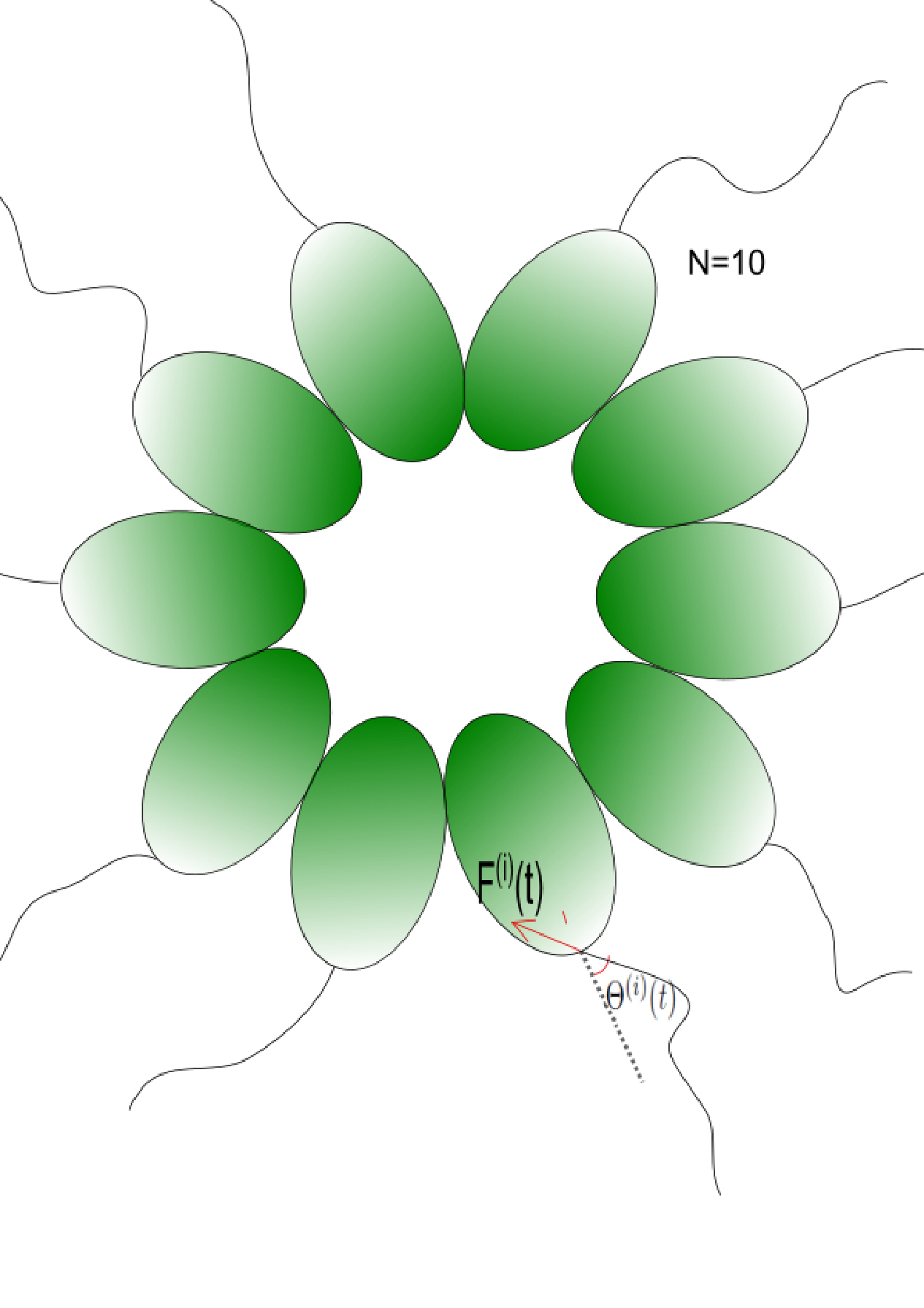}\\
  \end{subfigure} 
      \begin{subfigure}{0.45\textwidth}
     \centering
     \includegraphics[height=80mm, width=105mm]{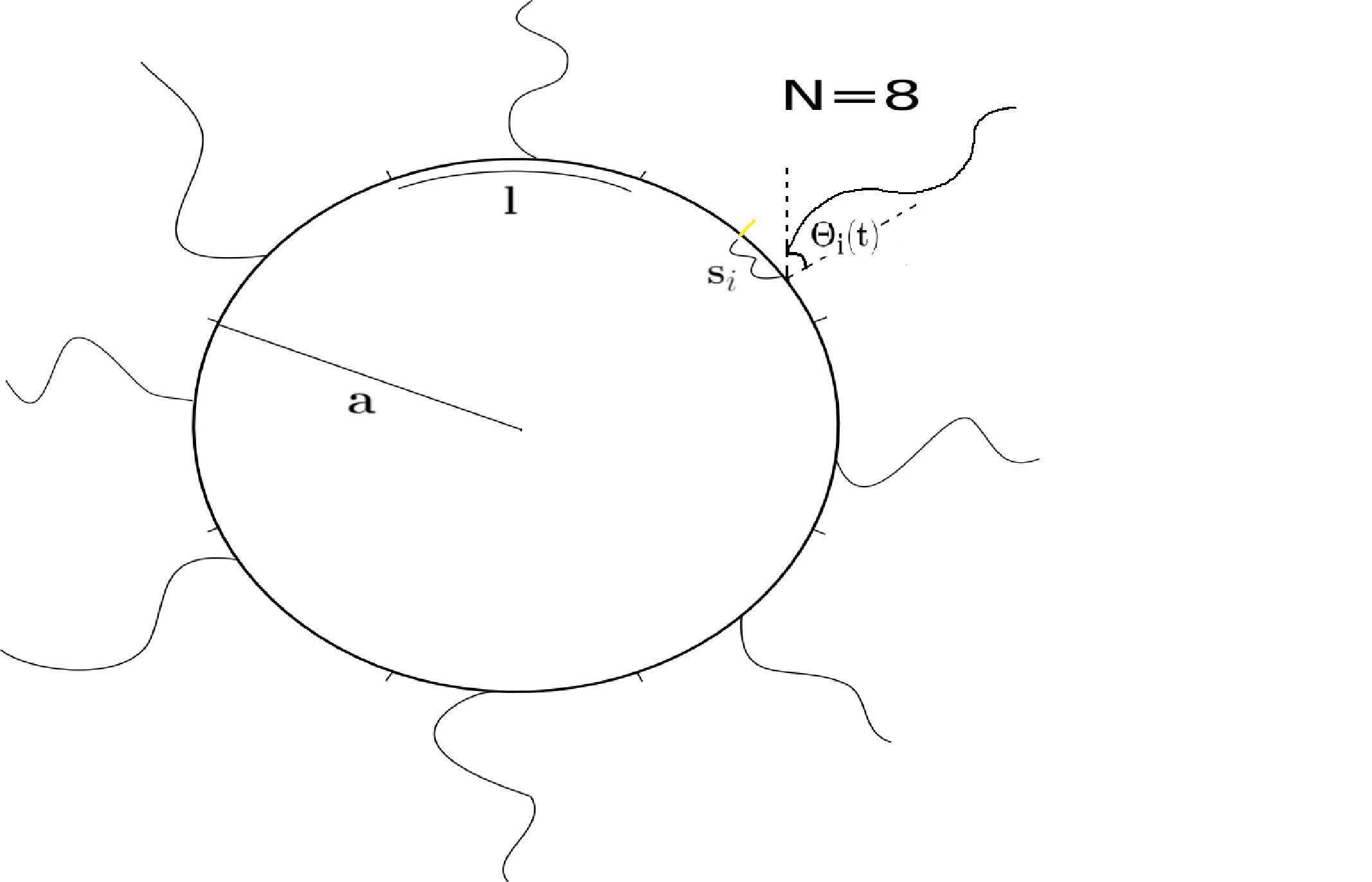}\\
\end{subfigure} 
\caption{Left: A schematic of a Choanoflagellate colony of $N= 10$ cells with $F^{(i)}(t)$ and $\Theta_i(t)$ being the time-dependent magnitude and direction of the propulsive forcing on the $i^{th}$ cell. Right: A detailed representation of the mathematical model with $ N=8 $ cells, indicating the cell radius $ \colrad$, cell segment length $ \celllen$, and displacement $ \flagdispi$ of the flagellar attachment point from the center of the cell arc.}
 \label{fig:model_schematic}
\end{figure}

We assume a maximally simple geometric configuration for the structure of a colony, illustrated in Figure~\ref{fig:model_schematic}.    The physical structure of the colony is represented as a rigid circular disc moving and rotating in a two-dimensional fluid.  The colony is composed of $N $ constituent cells, each of which occupies an arc of length $\celllen $ on the circular surface.  Thus, the colony has radius $ \colrad= \frac{N \celllen}{2\pi}  $.  Each cell is associated with a flagellum which we shall assume to be centered \emph{on average} normally to its point of attachment to the circular cell surface, but we will allow the attachment point to be displaced by an amount $ \flagdispi \in (-\celllen/2,\celllen/2) $ from the center of circular arc representing the surface of the cell exposed to the fluid environment.  This arrangement is inspired by the morphology of animal-like protozoa like choanoflagelatte \emph{S. rosetta} and plant like protozoa like \emph{Eudorina elegans} which are approximately spheroidal with the member cell's flagellum pointing out from the cellular surface \citep{Leadbeater, Gottlieb}.  The rigidity of the geometric structure in our model is a reasonable idealization due to the intercellular bridges~\citep{DayelChoano2011}. Our two-dimensional simplification is motivated entirely by the desire to avoid the complexities of three-dimensional rotational dynamics in our analysis.  Choanoflagellates actually do have some linear chain and planar sheet forms~\citep{BrunetMulticell}, but these would have stronger asymmetries which would otherwise complicate our method of analysis.  The colonial swimming properties of such shapes, without stimulus response, is covered by the analysis in~\citet{Ashenafi-mobility}. The displacements $ \flagdispi $ of the flagellar basal attachment points to the colony will be taken to be random or irregular in general, and therefore represent demographic stochasticity between colonies of a given size in our model.   We remark that the model in~\citep{Ashenafi-mobility} also allows irregularity in the mean angle at which the flagellum is oriented relative to the surface, but we neglect this feature here because the variations are small and would complicate the calculations.   Random placement of flagella and random variation of their orientations was also recently employed in a three-dimensional computational model study of a choanoflagellate colony~\citep{fauci2025multicellular} to examine variability across colonies for their swimming speed and feeding flux.

The dynamical state of the colony will be represented by a two-dimensional center of mass $ \Xcol (t) $ and a rotation angle $ \Thetacol (t)$ of a reference body point with respect to the $ x_1$ coordinate direction in the environment.  Each flagellum, indexed $ i=1,\ldots,N$, is assumed to exert a constant force $ F $ but at an angle $ \Thetacelli (t) $ with respect to the normal which varies in a stochastic dynamical way to be described below.  (Variations in the force magnitude are considered in the colonial swimming analysis in~\citet{Ashenafi-mobility}.)  
Given such a model for the dynamics of the flagellar force, the dynamics of the colony is given by adding up the flagellar forces or torques, and adding thermally induced translational and rotational diffusion:
\begin{subequations}\label{Colony_dynamicsb}
\begin{equation} \label{Colony_orientationb} 
\begin{aligned}
\difd\Theta^{(c)}(t)=-\sum\limits_{j=1}^{N} \frac{Fa}{\gamma_r}\sin(\Theta_{j}(t))dt+\sqrt{\frac{2k_BT}{\gamma_r}}dW^{\Theta,c}(t)
\end{aligned}
\end{equation}
\begin{equation} \label{Colony_positionb} 
\begin{aligned}
d\mathbf{X}^{(c)}(t)=-\sum\limits_{j=1}^{N} \frac{F}{\gamma_t}\begin{pmatrix}\cos(\alpha_{j}+\Theta_{j}(t)+\Theta^{(c)}(t))\\
\sin(\alpha_{j}+\Theta_{j}(t)+\Theta^{(c)}(\tilde{t}))\end{pmatrix}\, dt+\sqrt{\frac{2k_BT}{\gamma_t}}d\mathbf{W}^{X,c}(t)
\end{aligned}
\end{equation}
\end{subequations}
These are stochastic differential equations driven by standard independent Brownian motions $ W^{\Theta,c} (t)$ and $ \mathbf{W}^{X,c} (t)$~\cite[Sec. 4.3]{cwg:hsm}.  The angular positions of the flagella on the colony are given by
\begin{equation}
\alpha_{j}=2\pi[\frac{j-\frac{1}{2}+\frac{S_{j}}{l}}{N}]. \label{eq:flagdisp}\end{equation}
In some analytical and numerical calculations, we use a specific statistical model for the demographic variation of flagellar displacement:
\begin{equation}
    S_{j'} \sim U(-u,u) \textrm{ with } 0\leq u \leq l/2 \label{eq:unifmodel}
\end{equation}
meaning a uniform distribution of the flagellar attachment points over a region of width $ 2u$ about the center of the exposed cellular arc.

The translational drag coefficient $ \gamma_t $ and rotational drag coefficient $ \gamma_r $ of the colony are modeled as if the colony were a spheroid in the oblate limit in a three-dimensional fluid of dynamic viscosity $\eta$, discarding consideration of forces or torques that would move the limiting spheroid (a disc) out of its plane~\citep{kim}:
$$
\gamma_t=\frac{32}{3}\eta a, \hspace{0.6 cm}  \gamma_r=\frac{32}{3}\eta a^3.
$$ 
We remark that $\gamma_r=a^2\gamma_t$.

What remains is to model the dynamics of the angles $ \{\Theta_j\}_{j=1}^N $ of the flagellar forces with respect to their local normal.  We describe separate models for the flagellar force orientations for taxis in Subsection \ref{subsec:model_taxis} and for kinesis in Subsection \ref{subsec:model_kinesis} .  We will in both cases assume the cells are sensing local \emph{relative} spatial gradients, which appears relevant for aerotaxis of choanoflagellates \citep{aerotax} and for taxis of other eukaryotes~\citep{SKUPSKY,Janetopoulos8951} as well as \emph{E. coli} bacteria~\citep{WuLogSensing,MesibovRange}.  This means the response of a flagellum with index $j$ to an environmental concentration field $ c (\bx) $ is proportional to the  quantity 
\begin{equation*}
    \grad \ln c (\Xcol + \flagposj (\Thetacol)).
\end{equation*}
Here
$$\flagposj(\theta)\equiv a\begin{pmatrix}\cos(\alpha_{j}+\theta)\\
\sin(\alpha_{j}+\theta)\end{pmatrix}$$
denotes the location of the flagellar base of cell $j $ relative to the center of mass of the colony when the colony is rotated by an angle $ \theta$ relative to its reference orientation.  To simplify the notation in our mathematical model equations, we decompose the relative spatial gradient
\begin{equation*}
 \grad \ln c (\bx) = \gradnorm (\bx)    \begin{pmatrix}
\cos \gradang  (\bx) \\
\sin \gradang (\bx)
\end{pmatrix}
\end{equation*}
into its magnitude $ \gradnorm (\bx) $ and direction $ \gradang (\bx)$.  We sometimes find it  useful to refer to the unit vector 
\begin{equation}
    \graddir (\bx)= 
    \begin{pmatrix} \cos \gradang (\bx) \\ \sin \gradang (\bx) \end{pmatrix}
\end{equation}

In general the environmental concentration field $ c (\bx) $ would be a time-dependent solution of an advection-diffusion-reaction equation, but here we will only consider a prescribed steady-state profile.  In our analysis, we will consider only the situation where both the concentration gradient and concentration magnitude are sufficiently slowly varying over the scale of colony motion that we may approximate the logarithmic concentration gradient as constant: $ \gradnorm (\bx) = \gradnorm $ and $ \gradang (\bx) = \gradang $. This affords a great simplification by decoupling the colony's internal dynamics from the SDE~\eqref{Colony_positionb} governing the spatial motion of the center of mass.  Assuming a slowly varying concentration gradient could be appropriate when the stimulus is emitted steadily over a large region such as the surface of a relatively large microorganism or a spatially extended collection of microorganisms, but would not be relevant for taxis response to chemoattractants emitted from small phytoplankton in the ocean~\citep{slomka2025slower}.



The taxis and kinesis models~\eqref{directed_flagellum_orientation_taxis} and~\eqref{directed_flagellum_orientation_kinesis} from~\citep{aerotax} were posed in terms of the swimming orientation of the cell or microorganism.  If we consider in particular their application to a swimming cell with a single flagellum, we could view the flagellum as inducing the rotational dynamics by modification of its orientation with respect to the cell surface.  Of course flagella have approximately cyclic beating patterns~\citep{colonial-motility}, but following~\citet{Ashenafi-mobility}, we coarse-grain over the detailed beating dynamics and consider only the induced force $ F $ and its orientation $ \Thetacelli $, which somehow reflects asymmetries or deflections from a beating pattern centered about the normal to the cell surface.  Put another way, we coarse-grain the swimming motion into an effective force dipole which allows for various swimming mechanisms of colonial protozoa such as the single flagellar pushing dynamics of Choanoflagellates and the double flagellar breaststroke pulling mechanism of Volvocene green algae \citep{colonial-motility,GoldsteinVolvocine}. We don't represent the reciprocal force of the flagellum on the fluid because we do not consider hydrodynamic interaction effects here, in contrast to the literature on suspensions of swimming microorganisms~\citep{Gompper2015microreview,GoldsteinVolvocine,Marchetti2013hydrosoft,Underhill2011correlations}  or metachronal waves~\citep{JoannyFlow,BrumleyVolvox}.

We apply, therefore, the spirit of the mathematical models for taxis~\eqref{directed_flagellum_orientation_taxis} and kinesis~\eqref{directed_flagellum_orientation_kinesis} of the swimming cell orientation to the orientation of the force which the flagella exert relative to the normal to the cell surface.  One can verify that our flagellar force orientation models would induce a corresponding taxis or kinesis of the associated cell, were it not attached to other cells in a colony.  Naturally, the parameters appearing in the taxis and kinesis models based on flagellar force orientation will differ from those for the cellular orientation models in~\citep{aerotax}, but the mathematical structure will be similar.  By adopting these drift-diffusion models, we are neglecting the possibility of sudden jumps in swimming direction, as are seen for fast \emph{S. rosetta} swimming cells in~\citet{Sparacino_solitary}.  We are unaware of evidence of such jumpiness when the swimming cells are in colonial form.

The parameters used in our study are summarized with approximate physical values in Table~\ref{tab:params}.  We note that the observations of~\citep{Burkhardt_architecture} show that \emph{S. rosetta} cells have somewhat different morphologies as individual cells than when they are in colonies, but our parameter estimates are taken from observations of colonial cells.  Another relevant observation of~\citep{Burkhardt_architecture} for our model is that the mean cell sizes do not seem to depend strongly on the size of the colony.   \\

\subsection{Taxis Model}  \label{subsec:model_taxis}


In analogy to~\eqref{directed_flagellum_orientation_taxis}, our taxis model for the flagellar force orientation of each cell $ j $ will be taken to be: 
\begin{equation} \label{taxismodel}
\begin{aligned}
\difd \Theta_j(t)=&-\gamma_\Theta(\Theta_j(t) +\ktaxis \gradnorm
\sin[\Thetacol (t)+\alpha_{j}-\gradang]) \, \difd t+\sqrt{2\gamma_\Theta \sigma_\Theta^2} \difd W_j^\Theta(t)
\end{aligned}
\end{equation}
In this equation, $ \gradnorm = \gradnorm (\Xcol (t) + \flagposj (\Thetacol (t)))$ and $ \gradang = \gradang(\Xcol (t) + \flagposj (\Thetacol (t))) $; we often take these to be constant.  The stochastic term in~\eqref{taxismodel} is driven by standard mathematical Brownian motion $ W_j^{\Theta} (t)$, assumed independent for different cells $j$ as well as independent of the thermal rotation and translation of the colony in Eqs.~\eqref{Colony_orientationb} and~\eqref{Colony_positionb}.   

\begin{figure}[h!]  
{\centering{\includegraphics[height=80mm,width=130mm]{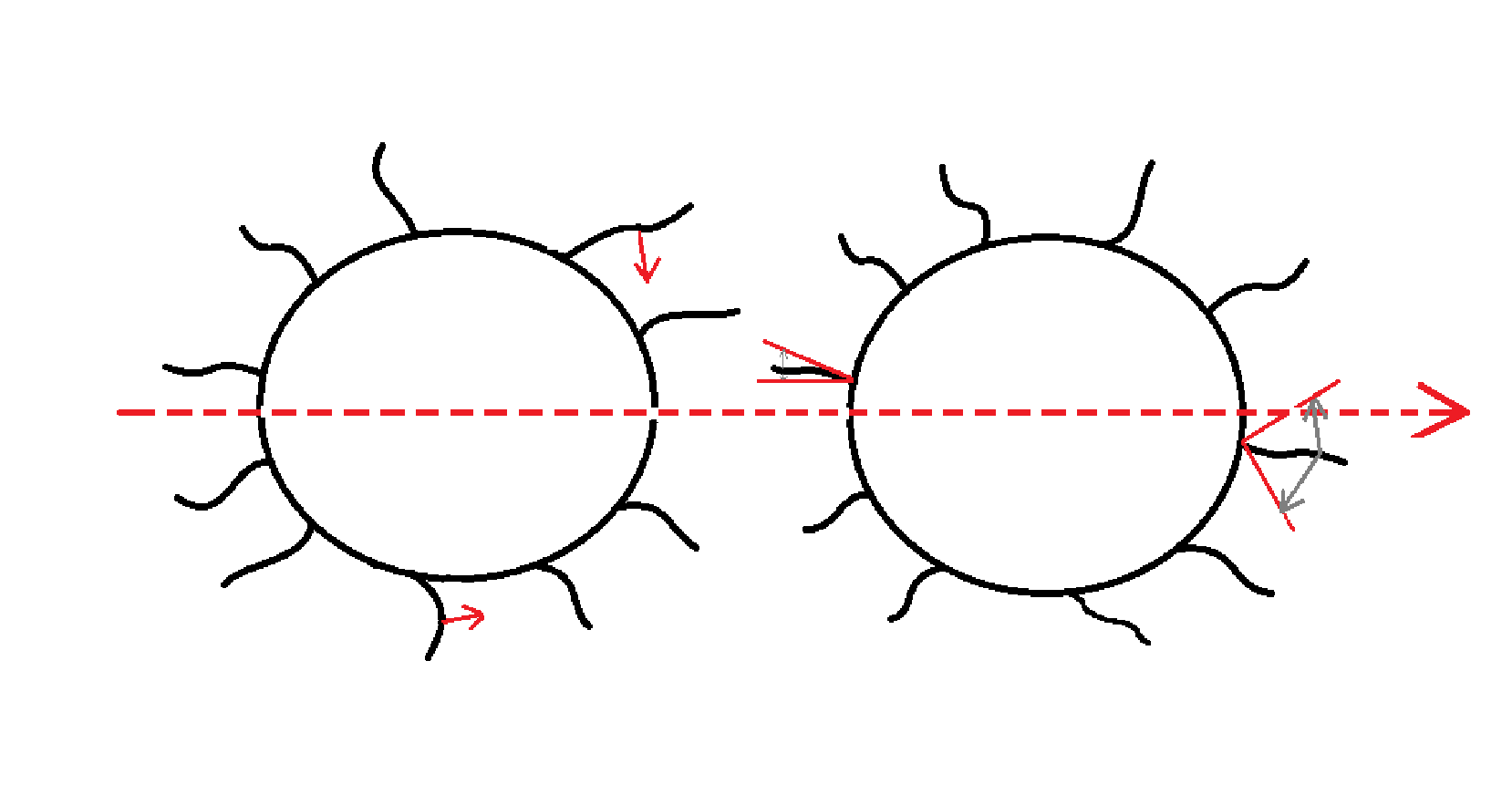}}
\caption{Schematic of taxis and kinesis models of the flagella associated with cells in a colony in the presence of an environmental gradient indicated by the long dashed arrow.
Left: A colony of taxis enabled cells with the red arrow indicating a biasing of the flagellum to apply force in a direction which would cause that cell to rotate toward swimming up the environmental gradient.  
Right: a colony of kinesis enabled cells with the red wedge estimating the range of motion for the beating flagella.
\label{fig:taxis_kinesis_schematic}}}
\end{figure}

Equation~\eqref{taxismodel} is an effectively autoregressive model under which the flagellar orientation reverts at rate $ \gamma_{\Theta}$ toward an angle so that its induced torque would tend to rotate the cell so that the flagellum is facing down the gradient, thereby pushing its cell up the gradient (see the left panel of Figure~\ref{fig:taxis_kinesis_schematic}).  The strength of the flagellar orientation bias is proportional to the strength $ \gradnorm $ of the relative concentration gradient with proportionality factor
$\ktaxis$.  A more realistic model should have the response saturate at large $ \gradnorm $~\citep{aerotax}; we are tacitly restricting attention to the linear response regime.  A simpler similar Langevin model for the taxis of cells in a cluster has been employed by~\citet{CamleyLeader}.  The stochastic noise term induces fluctuations of cycle-averaged flagellar force orientation of variance $ \sigma_{\Theta}^2 $ in absence of the concentration field.  We remark that this noise includes not just thermal noise but also active noise arising from the motors driving the flagella, so we do not express its magnitude in terms of temperature.   

\subsection{Kinesis Model} \label{subsec:model_kinesis}

Our kinesis model similarly adapts~\eqref{directed_flagellum_orientation_kinesis} for the orientation of the cycle-averaged propulsive force on each cell:
\begin{equation} \label{directed_flagellum_orientation_kinesis_m}
\begin{aligned}
d\Theta_j(t)=&-\gamma_\Theta(\Theta_j(t))dt
+\Bigg(2\left[\gamma_\Theta\sigma_\Theta^2+\kkin g \cos \left(\Thetacol (t)+\alpha_{j} - \gradang\right)\right]\Bigg)^{1/2} dW_j^\Theta(t)
\end{aligned}
\end{equation}
where again $ \gradnorm = \gradnorm (\Xcol (t) + \flagposj (\Thetacol (t)))$ and $ \gradang = \gradang(\Xcol (t) + \flagposj (\Thetacol (t))) $.
The Brownian motions $ \{W_j^\Theta (t)\}_{j=1}^N $ are taken to be independent of each other and the Brownian motions driving the thermal rotation and translation of the colony in Eq.~\eqref{Colony_dynamicsb}.  As the coefficient of the noise term in the kinesis model depends on the state, we should specify its interpretation as being in the It\^{o} sense \citep{cwg:hsm}.  But in fact, the results would be the same if we took the Stratonovich interpretation because the noise driving the coefficient is independent of the noise increment.
This autoregressive model has the tendency to return the flagellar orientation at rate $ \gamma_{\Theta} $ to the normal direction to the cell surface, but modulates the noise amplitude to increase when the colony is oriented so that the flagellum is positioned to swim against the favorable gradient.  $ \sigma_{\Theta} $ gives the reference standard deviation of the flagellar orientation fluctuations when swimming neither up nor down the environmental gradient, and 
$\kkin$ is the kinesis response factor which describes the strength of response of the noise amplitude to the environmental gradient as sensed spatially by a cell.  We naturally assume that 
\begin{equation}
    \kkin g < \gamma_{\Theta} \sigma_{\Theta}^2. \label{eq:kkincons}
\end{equation}
Indeed, we are assuming $ \gradnorm$ is sufficiently below the scale of saturation of the kinesis modulation amplitude, so that the stimulus response can be taken to be linearly proportional to $ \gradnorm $.

\begin{table}  
\begin{center}
\begin{tabular}{ c c c c }
 \hline
 Symbol & Definition & Estimated value
 \\
\hline
  N & Number of cells [2D] & 2-10 \\ 
  $F$ & Flagellar Force magnitude & $1-10 \,\si{\pico\newton}$ 
 \\  $ \celllen$ & Arclength of cell exposure & $ 2 \pi \,\si{\micro\metre}$ \\ 
  $\gamma_\Theta$ & Damping rate of force orientation  & $10\, \si{\sec^{-1}}$ \\  
   $\sigma_\Theta^2$ & Variance of force orientation & 0.002   \\
   T &Temperature & 300 \si{K} 
  \\ 
    $\eta$ & Dynamic viscosity & $\SI{0.01}{\gram/\cm \sec}$  \\
   $\ktaxis$ & taxis response factor &   \\
   $\kkin$ & kinesis response factor &  \\
 \hline
\end{tabular}
\end{center}
 \caption{Model parameters and typical values.  The values reported here should be taken as order-of-magnitude estimates.  The estimate for the number of cells is for a two-dimensional cross section of observed three-dimensional colonies~ \citep{FAIRCLOUGH,Yamashita,KoehlCapture,MahadevanChoano}. In particular, the flagellar force magnitude is motivated by observations in \citep{GoldsteinVolvocine, Roper}, the cell arclength on a colony and the variance of force orientation by~\citep{GoldsteinVolvocine, colonial-motility}.
 The damping rate of force orientation was estimated from the correlation times of flagellar orientations of both choanoflagellates and volvox~\citep{colonial-motility, GoldsteinVolvocine}. The above parameter ranges should apply for certain classes of choanoflagellate (animal-like) and volvocine (plant like) protozoa.  
  We do not have meaningful prior dimensional estimates of the kinesis and taxis response factors, other than the constraint~\eqref{eq:kkincons}.}
 \label{tab:params}
 \end{table}


\section{Effective Colony Dynamics}
\label{sec:summary}
As we will describe in Sections~\ref{sec:nondim} and~\ref{sec:analysis}, a plausible separation of time scales between the flagellar orientation fluctuations and the overall dynamics of the colony support a homogenization of the individual flagellar dynamics to obtain an effective dynamical description purely in terms of the orientation and center of mass of the colony.   
We summarize in this section these effective equations, which can be used as the basis of a Fokker-Planck equation model for the taxis or kinesis of a suspension of colonies which is dilute enough for the hydrodynamic interactions to be negligible. We also present a description of the effectiveness of the colony's motion along the environmental gradient as measured by its long term drift, which could be used as the basis of parameterizing the taxis or kinesis sensitivity of colonies in a Keller-Segel type model.   We separately discuss taxis in Subsection~\ref{sec:taxisresults} and kinesis in Subsection~\ref{sec:kinesisresults}. The derivations of the results are presented in Sections~\ref{sec:nondim} and~\ref{sec:analysis}.

\subsection{Measures of Colony Asymmetry}
\label{sec:asymmetry}
The results depend on the geometry of the center of mass of the flagella, which we decompose
in complex polar form: 
\begin{equation*}
    \frac{1}{N}\sum\limits_{j=1}^{N} \expe^{\mathi \alpha_j}= -\geotaxamp N^{-3/2}\expe^{\mathi \geotaxphase}.
\end{equation*}
where the real parameter $ \geotaxamp$ describes the strength of asymmetry and the real parameter $ \geotaxphase$ describes the opposite direction of the center of mass offset.  Note that $ 1-\geotaxamp/N^{3/2} $ is the circular variance of the distribution of flagellar attachment points, though this interpretation isn't particularly useful here.
The scaling by $ N^{-3/2} $ is chosen so the asymmetry strength $ \geotaxamp $ converges to a mean zero random variable with finite, nonzero variance as $ N \rightarrow  \infty $ when the flagella are placed at random perturbations to equally spaced locations ($\flagdispi $ in Eq.~\eqref{eq:flagdisp} with a fixed independent, mean zero distribution, which  induces $ \flagang_j$ to have variance $ \sim N^{-2}$).  The reason we include the minus sign is so that, in absence of fluctuations, the colony would swim at an angle $ \geotaxphase $ with respect to the its reference orientation.  To see this, note that a flagellum applying force normally at angular position $ \alpha_j$ would push the colony in the direction $ \alpha_j \pm \pi $.  More generally, if the flagella were all normally oriented, the colony would swim along the time-varying direction $ \Thetacol (t) + \geotaxphase $ as its orientation dynamically rotates.

The direct expression of the center of mass measures in terms of the angular placement of the flagella on the colony is as follows:
$$
\geotaxamp=N^{1/2}\sqrt{\left[\sum\limits_{j=1}^{N}\cos(\alpha_{j})\right]^2+\left[\sum\limits_{j=1}^{N}\sin(\alpha_{j})\right]^2}, \qquad \qquad 
\geotaxphase=\pi+\tan^{-1} \left(\frac{\sum\limits_{j=1}^{N}\sin(\alpha_{j})}{\sum\limits_{j=1}^{N}\cos(\alpha_{j})}\right),
$$
with the usual $ \pm \pi $ phase ambiguity of the inverse tangent to be resolved by consideration of the signs of the numerator and denominator.  
We similarly will require another asymmetry measure:
\begin{equation*}
    \frac{1}{N}\sum\limits_{j=1}^{N} \expe^{2\mathi (\alpha_j-\geotaxphase)} \equiv -\geotaxamp_2 N^{-3/2}\expe^{\mathi \geotaxphase_2},
\end{equation*}
or equivalently:
$$
\geotaxamp_2=N^{1/2}\sqrt{\left[\sum\limits_{j=1}^{N}\cos(2(\alpha_{j}-\geotaxphase))\right]^2+\left[\sum\limits_{j=1}^{N}\sin(2(\alpha_{j}-\geotaxphase))\right]^2}, \  
\geotaxphase_2=\pi+\tan^{-1}\left(\frac{\sum\limits_{j=1}^{N}\sin(2(\alpha_{j}-\geotaxphase))}{\sum\limits_{j=1}^{N}\cos(2(\alpha_{j}-\geotaxphase))}\right).
$$
These quantities are not well-defined when the center of mass of the flagella is at the origin ($\geotaxamp =0$) but they do not appear in the expressions relevant for this case.

\subsection{Taxis}
\label{sec:taxisresults}
Provided the taxis response strength induces a small change $ \ktaxis \gradnorm \ll 1$ in the preferred flagellum orientation, the effective dynamics of a colony whose individual cellular flagella respond to the environmental gradient via the taxis model in Subsection~\ref{subsec:model_taxis} is given by:
\begin{equation}    \label{eff_taxis_col_orientation}
\begin{aligned}
d\Theta^{(c)}(t)= - \frac{F a}{\gamma_r}  \ktaxiseff g
\sin[\Thetacol -\gradang+\phi]dt+\sqrt{\Diffrotcol}dW^{\Theta,c}(t)
\end{aligned}
\end{equation}
\begin{equation} \label{eff_taxis_col_position}
\begin{aligned}
\difd \Xcol(t)=&-\frac{Fe^{-\frac{1}{2}\sigma_\Theta^2}}{\gamma_t}
\sum\limits_{j=1}^{N}  \begin{pmatrix}\cos(\Thetacol (t)+\alpha_j-k_Tg \sin[\Theta^{(c)} (t)+\alpha_{j}-\gradang])\\ \sin{(\Thetacol (t)+\alpha_j - k_Tg \sin[\Theta^{(c)} (t) +\alpha_{j}-\gradang])}\end{pmatrix}dt
+\sqrt{\frac{2k_BT}{\gamma_t}} \,  d\mathbf{W}^{X,c}(t)
\end{aligned}
\end{equation}
where 
\begin{equation}\ktaxiseff \equiv e^{-\frac{1}{2}\sigma_\theta^2}\geotaxamp \ktaxis N^{-1/2} \label{eq:ktaxiseff}
\end{equation}
is the effective taxis response factor of the colony to the environmental gradient and 
\begin{equation}
    \Diffrotcol = \frac{k_BT}{\gamma_r}+  \frac{N  \sigmaThe^2 \flagangv^2}{\gamma_{\Theta}}
    \label{eq:diffrotcol}
\end{equation}
is the effective rotational diffusivity enhanced by the flagellar activity.  The quantity
\begin{equation}
    \flagangv \equiv \frac{F a}{\gamma_r}
    \label{eq:flagangv}
\end{equation}
is the angular velocity scale induced by a single flagellum on the colony, if it were oriented tangentially.   Of course the flagella are in fact approximately normally oriented relative to the colony surface, so the nominal angular velocity scale~\eqref{eq:flagangv} will always be multiplied by factors  $ \sigmaThe$ to reflect the magnitude of the angular deviations from the normal.  
The  enhancement of the rotational diffusivity in Eq.~\eqref{eq:diffrotcol} from the active but stochastic flagellar activity can be understood via the Kubo formula~\citep{rk:sle} as a square of the rotational frequency magnitude induced by the flagellar torque fluctuations multiplied by the correlation time $ \gamma_{\Theta}^{-1}$.  We estimate $\flagangv$ based on table \ref{tab:params} to be in the range $10^2 - 10^3N^{-2} \si{\sec}^{-1}$, and $\Diffrotcol$ to be dominated by the active second term, ranging from $2N^{-3} \text{ sec}^{-1}$ to $200N^{-3} \text{ sec}^{-1}$. The thermally driven rotational diffusion is on the order of $ 0.4 N^{-3} \si{\sec}^{-1}$. 

The effective equations of motion given above apply to smooth environmental gradients in general, with $g$ and $\theta_g$ now depending on the center of mass $\mathbf{X}^{(c)}(\hat{t})$ and orientation $\Theta^{(c)}(\hat{t})$.
 The asymptotic analysis yielding these effective dynamics are self-consistent provided \begin{equation*}
    N \gg \left(\frac{3 \pi^2 F \geotaxamp \ktaxis \gradnorm}{8\eta l^2\gamma_\Theta \sigmaThe}\right)^{2/5}, \left(\frac{3\pi^3k_B T}{4\eta l^3 \gamma_{\Theta}}\right)^{1/3}, \left(\frac{3 \pi^2F\sigmaThe}{8\eta l
^2 \gamma_{\Theta}}\right)^{2/3}.
\end{equation*}
The first restriction is easily satisfied for a weak taxis response or weak asymmetry in flagellar arrangement.  For the physical parameter values from Table~\ref{tab:params},   the second restriction is generally satisfied for all $ N \geq 2$, and  the third restriction is also generally satisfied for $ F \sim 1 \,\si{\pico\newton}$ but requires $ N \gg 2 $ for the larger force values $ F \sim 10 \, \si{\pico\newton}$.  The analysis could still proceed without the assumption of a small individual taxis response $ \ktaxis \gradnorm \ll 1$, but the resulting expressions would be far more complicated.

We notice from~\eqref{eff_taxis_col_orientation} that the colony's swimming direction $ \Thetacol (t) + \geotaxphase $ is attracted toward the environmental gradient direction $ \gradang$, and thus, with rotational diffusion, will approach a stationary distribution about this direction.  The effective colony taxis coefficient~\eqref{eq:ktaxiseff} governs the strength of the attraction toward the environmental gradient, and the resulting prefactor $ F \colrad \ktaxiseff/ \gamma_r $ may  be interpreted as governing the rate of reorientation of the colony toward the environmental gradient~\citep{CamleyReview}.  We see that the effective colony taxis coefficient $ \ktaxiseff$  increases with asymmetry measure $ \geotaxamp$ but decreases with colony size.  The dependence on colony size is due to the combination of the steering contributions from each cell, but these contributions would cancel from cells in reflection-symmetric position with respect to the environmental gradient direction.  This is why the colony taxis would be completely ineffective under perfect symmetry ($\geotaxamp = 0$).   The inverse square root dependence on colony size arises from the reduction in asymmetry as more flagella produce a more uniform distribution about the cell surface. The translational dynamics sum up contributions from each flagellum, with force directed along its average taxis-induced deviation from the normal based on the current colony orientation, and magnitude depleted by a factor $ \expe^{-1/2 \sigmaThe^2}$ due to fluctuations in the flagellar force orientation.  This same factor mitigates the effective colony taxis strength in Eq.~\eqref{eq:ktaxiseff}.  The deviation of the flagellar force from the normal is largest for those flagella that are positioned to be approximately orthogonal to the environmental gradient.

When the environmental gradient is constant, we can compute the long-time drift by averaging over the rotational dynamics and taking a first order expansion with respect to the taxis response coefficient $ \ktaxis$: 
\begin{equation} \label{eff_taxis_drift}
\begin{aligned}
\drifteff \equiv \lim\limits_{t \rightarrow\infty}\frac{\Xcol (t)}{t} &= \frac{-Fe^{-\frac{1}{2}\sigma_\Theta^2}}{\gamma_t} N J_1 (\ktaxis \gradnorm)
   \begin{pmatrix} \cos \theta_g\\ \sin\theta_g\end{pmatrix} \textrm{ if } \geotaxamp = 0, \\
&=\frac{Fe^{-\frac{1}{2}\sigma_\Theta^2}}{\gamma_t}
  \left[ -\frac{N k_T g}{2}\begin{pmatrix} \cos\theta_g\\ \sin\theta_g \end{pmatrix}- \frac{ \geotaxamp_2 k_T g}{ 2\sqrt{N}}\begin{pmatrix} \cos(\gradang-\geotaxphase_2)\\ \sin(\gradang-\geotaxphase_2) \end{pmatrix} \right.
  \\
&+ \frac{I_1\left(\frac{\ktaxiseff g \flagangv}{\Diffrotcol}\right)}{I_0\left(\frac{\ktaxiseff g \flagangv}{\Diffrotcol}\right)}\left(\frac{\geotaxamp}{\sqrt{N}}\begin{pmatrix} \cos\gradang\\ \sin\gradang \end{pmatrix} + \frac{\Diffrotcol  \expe^{\sigma_\Theta^2/2}\geotaxamp_2}{\flagangv \geotaxamp}\begin{pmatrix} \cos(\gradang-\geotaxphase_2)\\ \sin(\gradang-\geotaxphase_2) \end{pmatrix}\right)
\\&\left.+O(k_T^2 g^2)\right] \textrm{ if } \geotaxamp \neq 0
\end{aligned}
\end{equation}
Here $J_1 $ is an ordinary Bessel function of the first kind, and $I_0$ and $I_1$ are modified Bessel functions of the first kind. First we observe that under symmetric placement of the flagella ($\geotaxamp = 0$), the drift of the colony is in the opposite direction of the attractant gradient!  The reason for this is that our taxis model is based on a response which displaces the flagellum to apply torque to attempt to steer the cell up the concentration gradient.  Under symmetric placement of the flagella, the torque contributions exactly cancel ($\ktaxiseff =0$) and the colony only undergoes rotational diffusion.  The steering effect of taxis is completely lost, but the deflection of the flagella in the futile steering attempts all rotate the flagellar forces to all be more aligned \emph{down} the gradient under our model~\eqref{taxismodel}.  This perverse motion down the gradient would drop out under various model variations, such as having a quick saturation to the steering response as a function of the cell's orientation mismatch, or modulating the flagellar force rather than orientation.  

A productive colonial drift up the environmental gradient is manifested in the term proportional to $ \geotaxamp$, measuring the degree of asymmetry.  We also note off-gradient contributions proportional to the second asymmetry measure $ \geotaxamp_2$.  These can be understood by noting that the colony does tend to orient so the environmental gradient is aligned with the net force induced by the flagella \emph{when positioned on average normally to the cell surface}.  But the flagellar angles are displaced from the normal due to the taxis response, and this creates a higher order drift correction with direction depending on flagellar placement.  The relative magnitude of the various effects is a bit obscured by our hesitation to express the drift in terms of a completely formal first order expansion with respect to the taxis strength, because the argument of the Bessel function involves the parameter combination $ N^{-1/2} \flagangv/\Diffrotcol $
which can be moderately large based on the estimates presented earlier in this subsection.  But with a completely formal first order expansion, we would simply have:
\begin{equation*}
\begin{aligned}
\drifteff = \frac{F}{2\gamma_t}
  \left[\ktaxis \gradnorm\left(\frac{\geotaxamp^2 \flagangv}{N \Diffrotcol}e^{-\sigma_\Theta^2} -N e^{-\frac{1}{2}\sigma_\Theta^2}\right)\graddir
 +O\left(k_T^2 g^2 \left(1+ \frac{\flagangv^2}{N \Diffrotcol^2}\right)\right)\right]
\end{aligned}
\end{equation*}
where $ \graddir $ is a unit vector aligned up the environmental gradient.  The off-gradient terms would appear at second order, mixed together with further second order terms.  At first order we simply see the competition between the productive colonial taxis which relies on asymmetry and the counterforce effect which does not.  The constructive term relies additionally on the relative strength of the flagellar torques to the thermal energy, which is moderately large and reflects the ability of the steering response to keep the colony oriented up the gradient against rotational diffusion.   Noting that the colony radius $ \colrad $ scales in proportion to $N $, the productive colonial taxis is approximately inversely proportional to the number of cells.  The countervailing term is approximately independent of the number of cells and would therefore become dominant for large enough colonies.

Averaging over the demographic variability in the flagellar displacement under the uniform model~\eqref{eq:unifmodel}, we obtain:
\begin{equation*}
    \drifteffmean = \frac{F}{2\gamma_t}
  \left[\ktaxis \gradnorm\left(\frac{N \flagangv }{\Diffrotcol}e^{-\sigma_\Theta^2}
  \left(1-\frac{\sin^2 (u/\colrad)}{(u/\colrad)^2}\right) -N e^{-\frac{1}{2}\sigma_\Theta^2}\right)\graddir
  +O\left(k_T^2 g^2 \left(1+ \frac{\flagangv^2}{N \Diffrotcol^2}\right)\right)\right]
\end{equation*}
We can further simplify this expression using $ 1 - y^{-2} \sin^2 y \sim \frac{1}{3} y^2 + O (y^4)$ for the relevant regime $ u\ll \colrad$ where the random flagellar displacements are small relative to the size of the colony:
\begin{equation}
       \drifteffmean = \frac{F}{2\gamma_t}
  \left[\ktaxis \gradnorm\left(\frac{N \flagangv u^2 }{3\colrad^2 \Diffrotcol}e^{-\sigma_\Theta^2}\left(1+ O ((u/\colrad)^4)\right)
 -N e^{-\frac{1}{2}\sigma_\Theta^2}\right)\graddir
  +O\left(k_T^2 g^2 \left(1+ \frac{\flagangv^2}{N \Diffrotcol^2}\right)\right) \right]. \label{eq:meantaxisdrift}
\end{equation}
The ratio of the productive to counterproductive term  is then seen to be $\sim 50-500 N^{-1}\left(\frac{u}{l}\right)^2 $
for the parameter values in Table~\ref{tab:params}. The counterproductive term is therefore negligible unless the colony size is very large or the random displacements from symmetry very small.  The standard deviation of the colony drift across demographic variations in colony size is approximately half the mean.

We note that we can also compute the chemotactic index using the stationary distribution~\eqref{eq:pstattaxis} for the colony orientation~\citep{MuglerChemotaxis}:
\begin{equation*}
\CI = \int_{-\pi}^\pi \cos (\theta +\geotaxphase- \gradang) \pstatthctax (\theta) \, \difd \theta
= \frac{I_1 (\frac{N^{1/2}\sigma_\theta \ktaxisnd}{\delta^2}e^{-\sigma_\Theta^2/2}\geotaxamp)}{I_0(\frac{N^{1/2}\sigma_\theta \ktaxisnd}{\delta^2}e^{-\sigma_\Theta^2/2}\geotaxamp)}.
\end{equation*}




\subsection{Kinesis}
\label{sec:kinesisresults}
The effective dynamics of a colony whose individual cellular flagella respond to the environmental gradient via kinesis are given by:
 \begin{subequations}
\begin{equation} \label{eff_kinesis_col_orientation}
\begin{aligned}
d\Thetacol (t) = \sqrt{2[\Diffrotcol-\kkincol g 
\cos(\Thetacol-\theta_g+\phi)]}
dW^{\Theta,c}(t)
\end{aligned}
\end{equation}
\begin{equation} \label{eff_kinesis_col_position}
\begin{aligned}
\difd \Xcol(t)=&- \sum\limits_{j=1}^{N} \frac{F}{\gamma_t}e^{-\frac{1}{2}\sigma_\Theta ^2-\frac{k_Kg}{2\gamma_\theta}\cos(\alpha_{j}+{\Theta}^{(c)}-\theta_g)}\begin{pmatrix}\cos(\alpha_{j}+\Thetacol)\\ \sin{(\alpha_{j}+\Thetacol )}\end{pmatrix}dt
+\sqrt{\frac{2k_BT}{\gamma_t}} \,  d\mathbf{W}^{X,c}(t)
\end{aligned}
\end{equation}
\label{eff_kinesis_col}
 \end{subequations}
where $\Diffrotcol $ is the enhanced rotational diffusion of the colony in absence of the environmental gradient, and has the same expression~\eqref{eq:diffrotcol}  as for the taxis response.  The effective kinesis response coefficient of the colony to the environmental gradient is:
\begin{equation*}
    \kkincol = k_K N^{-1/2} \geotaxamp 
    \left(\frac{\flagangv}{\gamma_\Theta}\right)^2,
\end{equation*} 
with $ \flagangv$ defined in the same way~\eqref{eq:flagangv} as for the taxis response.
Note the effective kinesis effect on the colony decreases with colony size $N$.  We note the equations~\eqref{eff_kinesis_col} apply in their current form to general smooth environmental gradients with $ \kkin $ and $ \gradang $ depending on the center of mass $ \Xcol (t)$ and orientation $ \Thetacol (t) $ as described in Subsection~\ref{subsec:model_col}.  The asymptotic derivation of the above results is self-consistent for
\begin{equation*}
N \gg \left(\frac{6\pi^3k_B T}{8\eta l^3 \gamma_{\Theta}}\right)^{1/3}, \left(\frac{3 \pi^2F\sigmaThe}{8\eta l
^2 \gamma_{\Theta}}\right)^{2/3}.
\end{equation*}
For the physical parameter values from Table~\ref{tab:params}, the first restriction is generally satisfied for all $ N \geq 2$ while the second restriction is also generally satisfied for $ F \sim 1 \,\si{\pico\newton}$ but requires $ N \gg 2 $ for the larger force values $ F \sim 10 \, \si{\pico\newton}$.
 
The kinesis response of the colony involves two factors:  1) a measure of the geometric asymmetery $ \geotaxamp$ of the flagellar arrangement, and 2) the squared product of the angular velocity magnitude induced by the flagellar beating and the correlation time  $\gamma_{\Theta}^{-1}$.  The structure  of this enhancement can be understood via the Kubo formula in a similar way to that of the rotational diffusivity~\eqref{eq:pstattaxis}, with the difference in the amplitudes arising from the different amplitudes for the baseline fluctuation strength and the kinesis response strength in the individual cell kinesis model~\eqref{directed_flagellum_orientation_kinesis_m}.

The angular dependence of the kinesis response can be understood by noting from the discussion at the beginning of this section that the flagellar placement asymmetries induce swimming along the direction $ \geotaxphase $ when the colony is in the reference orientation, and thus along $ \Thetacol (t) + \geotaxphase $  in general. Thus, the rotational noise is kinetically reduced the more the colony is swimming along the environmental gradient direction $ \gradang$.  Note the rotational dynamics of the colony shows vanishing response to the environmental gradient as the flagella become more evenly spaced so that the asymmetry measure $ \geotaxamp$ approaches zero.

For the translational dynamics, we see a contribution of force from each flagellum as if it were applied normally to the colony surface, but with the exponential weights reflecting the weakening of the flagellar force due to its angular fluctuations, whose strength is measured by $ \sigmaThe $. Though the translation speed would seem to be simply diminished by increased flagellar fluctuations $ \sigmaThe$, we note that without these fluctuations $ \sigmaThe \downarrow 0$, the effective dynamics are oblivious to the environmental gradient as we must restrict $ \kkin \gradnorm < \gamma_{\Theta} \sigmaThe^2$. 
Like we did for the taxis results, we note that the Fokker Planck equation of (\ref{eff_kinesis_col_orientation}) and (\ref{eff_kinesis_col_position}) can serve as a Keller-Segel type model of kinesis where we have suspensions of colonies instead of single cells and the suspension density is dilute enough for the hydrodynamic interactions to be negligible~\citep{Sherratt1994, Byrne_Chemokinesis}.

The long-time effective drift along a constant environmental gradient
induced by the equations~\eqref{eff_kinesis_col} is given in a first order expansion with respect to the kinesis response coefficient $ \kkin$: 
\begin{equation} \label{eff_kinesis_drift}
\begin{aligned}
\drifteff \equiv \lim\limits_{t \rightarrow\infty}\frac{\Xcol(t)}{t}  &=
\graddir 
\frac{N Fk_K g}{4\gamma_t\gamma_\theta} e^{-\frac{1}{2}\sigma_\Theta^2}
\left[1 +\frac{2\geotaxamp^2\flagangv^2}{N^2\Diffrotcol \gamma_{\Theta}}
+ O\left(\frac{\kkinnd \flagangv^4}{N^4\Diffrotcol^2\gamma_{\Theta}^2}\right)
+ O \left(\frac{\kkinnd\flagangv^2\sigmaThe^2}{N^{1/2} \Diffrotcol \gamma_{\Theta}}
\right)\right]
\\&
\approx \graddir 
\frac{N Fk_K g}{4\gamma_t\gamma_\theta} e^{-\frac{1}{2}\sigma_\Theta^2}
\left[1 +\frac{2\geotaxamp^2}{N^3\sigmaThe^2}\right],
\end{aligned}
\end{equation}
where $ \graddir $ is a unit vector aligned up the environmental gradient ($\graddir = [ \cos \gradang, \sin \gradang ]^{T}$)  and the final approximation is applicable when the rotational diffusion~\eqref{eq:diffrotcol} is dominated by its active component.  This effective drift would also be applicable locally to non-constant environmental gradients over time scales long compared to the rotational diffusion time scale $1/ \Diffrotcol$ but short enough that the colony is moving on a scale small relative to the 
spatial fluctuations of the environmental field.  A more complicated expression would result without assuming $ \kkin g \ll \gamma_{\theta} \sigmaThe^2$.

We see first a natural factor $ NF/\gamma_{t}$ representing the velocity scale which would be induced if all flagella were pushing the colony up the gradient.  This factor is independent of colony size since the drag coefficient grows in proportion to the number of cells.
The other factors will naturally reduce the effective drift from this velocity scale based on the constraints and fluctuations. The second factor $ \kkin \gradnorm/\gamma_{\Theta}$ gives the scale of the fluctuation in flagellar force orientations owing to kinesis effects, in analogy to how $ \sigmaThe^2 = \gamma_{\Theta} \sigmaThe^2/\gamma_{\Theta}$ describes the flagellar force orientation fluctuations in absence of an environmental gradient.  The factor $ \expe^{-\sigmaThe^2/2} $ represents the weakening of the effective force induced by a flagellum due to the fluctuations in the angle in which it is applied to the colony.  Turning next to the factor in the brackets, we may be surprised to see a term which survives in the limit of perfect symmetric placement of the flagella ($\geotaxamp \downarrow 0$) even though the rotational diffusivity of the colony~\eqref{eq:diffrotcol} shows no kinesis effect in this limit!  The reason is that, even in a geometrically regular flagellar arrangement, the flagellar force orientation fluctuates more for flagella pushing in the wrong direction, creating an effective weakening of that flagellum's contribution.  Therefore, the environmental gradient induces a kinesis drift of the geometrically regular colony without a kinesis rotational diffusion effect by simply weakening (via angle rather than force modulation!) the contributions from the flagella pushing in the wrong direction.  The strength of this shape-independent drift due to kinesis is independent of colony size.  The second term in the brackets reflects the contribution from the  modulations of the colony's rotational diffusion induced by kinesis in Eq.~\eqref{eff_kinesis_drift}, and is naturally proportional to (the square of) the asymmetry measure $ \geotaxamp$ which is required for the rotational kinesis of the colony.  The kinesis drift from the rotational modulation is, as with all the activity-induced enhancements, proportional to the angular velocity scale $ \flagangv^2$ and the  correlation time $ \gamma_{\Theta}^{-1}$ of the flagellar force orientation fluctuations.  The factor $ 1/\Diffrotcol$ reflects the disruption to the kinesis drift by random rotation of the colony that is indifferent to the environmental gradient. This shape-dependent contribution to the taxis drift is positive but with a smaller effect for larger colony sizes.

We can make the kinesis drift term dependence on the colony geometry more transparent by considering a model of demographic variations of the flagellar attachmment points as having their displacement $ S_i $ from the center of the cell's exposed arc being uniformly distributed over an interval $ (-u,u) $, with $ 0 < u < l/2 $.  
The kinesis drift~\eqref{eff_kinesis_drift} averaged over these demographic variations is, for $ N \geq 2$:
\begin{equation} \label{avgdrift_kinesis}
\begin{aligned}
\drifteffmean  &\sim
\graddir 
\frac{N Fk_K g}{4\gamma_t\gamma_\theta} e^{-\frac{1}{2}\sigma_\Theta^2} \\
& \qquad  \times \left[1 +\frac{2\flagangv^2}{\Diffrotcol \gamma_{\Theta}}\left(1-\frac{\sin^2( \frac{2\pi u}{Nl})}{(\frac{2\pi u}{Nl})^2}\right)
+O\left(\kkinnd^2\left(\frac{1}{N(\beta^2+N)^2}+\frac{\sigmaThe^2}{\sqrt{N}(\beta^2+N)}\right)\right)
\right] \\ &\sim \graddir 
\frac{N Fk_K g}{4\gamma_t\gamma_\theta} e^{-\frac{1}{2}\sigma_\Theta^2} \\
& \qquad \times \left[1 +\frac{2\flagangv^2  u^2}{3\Diffrotcol \gamma_{\Theta}\colrad^2}\left(1+O((u/\colrad)^2)\right)+O\left(\kkinnd^2\left(\frac{1}{N(\beta^2+N)^2}+\frac{\sigmaThe^2}{\sqrt{N}(\beta^2+N)}\right)\right)\right]
\end{aligned}
\end{equation}

\begin{figure}[H]   
    \includegraphics[height=80mm,width=\linewidth]{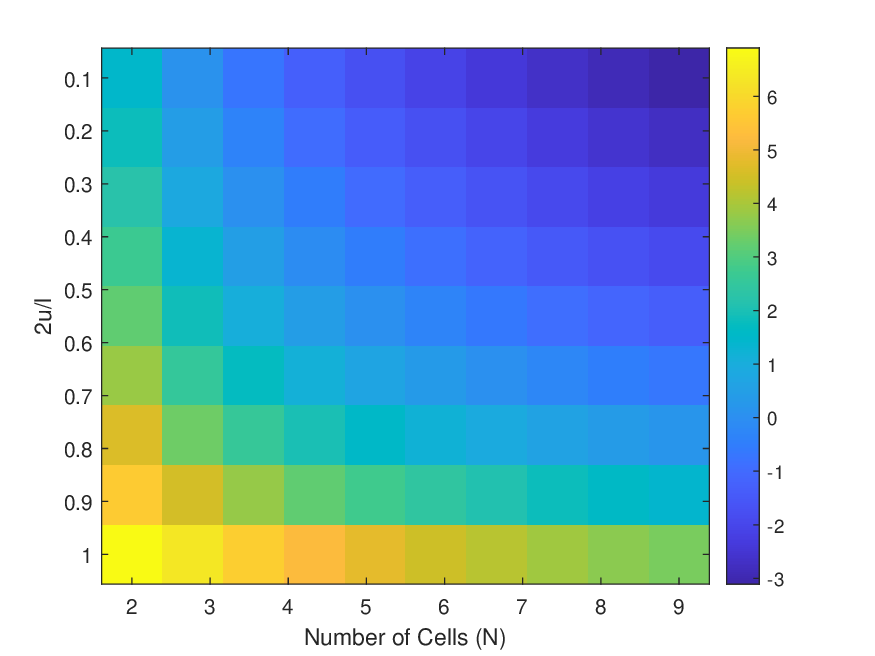}
\caption{The natural logarithm of the magnitude of the geometry-dependent term: $\ln \left[\frac{2 \flagangv^2}{\Diffrotcol \gamma_{\Theta}}\left(1-\frac{\sin^2(\frac{2\pi}{lN}u)}{(\frac{2\pi}{lN})^2u^2}\right)\right]$
in Eq.~\eqref{avgdrift_kinesis}
as a function of colony size $N$ and randomness $ 2 u/l $ of flagellar placement.  The flagellar force is taken as $F=5$ pN and the other parameters are as indicated in Table~\ref{tab:params}.
}
\label{fig:drift_ratio}
\end{figure}

The geometric contribution grows as the flagellar placement variability $u $ increases from $0 $ (the regularly spaced limit).  With parameters chosen with their experimental estimates from Table~\ref{tab:params} and $ F= 5\si{\pico\newton}$, we plot the magnitude of the contribution of the second, geometry sensitive, term in the brackets of Eq.~\eqref{avgdrift_kinesis} in Figure~\ref{fig:drift_ratio}. We note that $\frac{2 \flagangv^2}{\Diffrotcol \gamma_{\Theta}} \approx 2 \sigmaThe^{-2} N^{-1} \approx 1000 N^{-1} $,
 which helps elucidate the relative magnitude of the geometry-dependent term.  In particular, this term diminishes as $ \sim N^{-3}$ for large colony size $N$.  
Thus, for physically relevant parameter values, the geometry dependence significantly enhances the kinesis drift of small colonies but is negligible for larger colonies.  The standard deviation of the colony drift across demographic variations in colony size is approximately half the mean.


We note that we can also compute the chemotactic index using the stationary distribution~\eqref{eq:pstat_kin} for the colony orientation~\citep{MuglerChemotaxis}:
\begin{align*}
\CI &= \int_{-\pi}^\pi \cos (\theta +\geotaxphase - \gradang) \pstatthc (\theta) \, \difd \theta = 
\frac{\Diffrotcol}{\kkincol \gradnorm} \left(1 - \sqrt{1-\left(\frac{\kkincol \gradnorm}{\Diffrotcol}\right)^2}\right) \\
&= \frac{\kkincol\gradnorm}{2\Diffrotcol}\left(1 + O \left(\left(\frac{\kkincol \gradnorm}{\Diffrotcol}\right)^2\right)\right).
\end{align*}
\section{Nondimensionalization} \label{sec:nondim}

We next proceed to nondimensionalize the model equations for the colony dynamics, with the aim of identifying small parameters to exploit in an asymptotic reduction in Section~\ref{sec:analysis}.

We nondimensionalize time with respect to the time scale of flagellar reorientation: $\tilde{t}\equiv \gamma_\Theta t$ and the spatial position by the colony radius: ${\mathbf{\tilde{X}^{(c)}}}\equiv \colrad^{-1} \Xcol $. We furthermore normalize the flagellar angular deviations by their standard deviation scale: $ \Thetacellndi \equiv \sigmaThe^{-1} \Thetacelli $.  The nondimensionalized versions of the equations of motion (\ref{Colony_orientationb}), and (\ref{Colony_positionb}) of the colony become:
\begin{equation}  \label{colony_orientation_nondim}
\begin{aligned}
d\Theta^{(c)}(\tilde{t})=-\sum\limits_{j'=1}^{N} \frac{\zeta}{N^2} \sin{\sigmaThe \Thetacellnd_{j'}(\tilde{t})}d\tilde{t}+\delta\sqrt{\frac{2 \zeta}{N^3}} dW^{\Theta,c}(\tilde{t})
\end{aligned}
\end{equation}
\begin{equation}  \label{colony_position_nondim}
\begin{aligned}
d\mathbf{\tilde{X}^{(c)}}(\tilde{t})=-\sum\limits_{j'=1}^{N} \frac{\zeta}{N^2} \begin{pmatrix}\cos(\alpha_{j'}+\sigmaThe \Thetacellnd_{j'} (\tilde{t}) +\Theta^{(c)} (\tilde{t}))\\
\sin(\alpha_{j'}+\sigmaThe \Thetacellnd_{j'} (\tilde{t})+\Theta^{(c)}(\tilde{t}))\end{pmatrix}\, d\tilde{t}+\delta
\sqrt{\frac{2 \zeta}{N^3}}d\mathbf{W}^{X,c}(\tilde{t})
\end{aligned}
\end{equation}\\
where we have defined two nondimenisonal groups:
\begin{itemize}
\item 
\begin{equation*}
\zeta \equiv \frac{3\pi^2F}{8\eta l^2\gamma_\Theta} \sim 10-100
\end{equation*}
is a measure of the ratio of the time scale $ \gamma_{\Theta}^{-1} $ of flagellar motion to the time scale 
$ \propto \frac{\eta l^2}{F} $ of a cell's motion (across itself) due to propulsion by the flagellar force, and
\item 
\begin{equation}
\delta\equiv\sqrt{\frac{2\pi k_BT}{Fl}} \sim 0.02-0.06. \label{eq:def_delta}
\end{equation}
is, up to a numerical factor, the square root of the ratio of the thermal energy to an active energy scale of the cell.
\end{itemize}
We note the definition of $ \zeta$ is chosen so the nondimensional torque and force scales applied by each cell can be written simply as:
\begin{equation*}
\frac{F}{a\gamma_t\gamma_\Theta}   = \frac{Fa}{\gamma_r\gamma_\Theta}
= \frac{F\frac{Nl}{2\pi}}{(\frac{32}{3})\eta\left(\frac{Nl}{2\pi}\right)^3\gamma_\Theta} =
\frac{\zeta}{N^2}.
\end{equation*}

The nondimensional taxis model~\eqref{taxismodel} reads:
\begin{equation} \label{eg:nondimensional_taxis_model}
\begin{aligned}
\difd \Thetacellndi (\tilde{t})=&-(\Thetacellndi (t)+\ktaxisnd 
\sin[\Thetacol (\tilde{t})+\alpha_{i}-\gradang]) \, \difd \tilde{t}+\sqrt{2} \difd W_i^\Theta(\tilde{t})
\end{aligned}
\end{equation}
where we have defined 
\begin{equation*}
    \ktaxisnd \equiv \ktaxis \gradnorm \sigmaThe^{-1}
\end{equation*}
as a nondimensional measure of strength of the taxis response.  The nondimensional version of the alternative kinesis model~\eqref{directed_flagellum_orientation_kinesis_m} reads:
\begin{equation} \label{directed_flagellum_orientation_kinesis_m_nondim}
\begin{aligned}
\difd \Thetacellndi (\tilde{t}) =&-\Thetacellndi (t) \, \difd t
+\Bigg(2\left[1+\kkinnd \cos \left(\Thetacol (\tilde{t})+\alpha_{i} - \gradang\right)\right]\Bigg)^{1/2} dW_i^\Theta(\tilde{t})
\end{aligned}
\end{equation}
where we have defined 
\begin{equation*}
    \kkinnd \equiv \frac{\kkin \gradnorm}{\gamma_{\Theta} \sigmaThe^{2}} < 1
\end{equation*}
as a nondimensional measure of strength of the kinesis response.  Note that  $ \ktaxisnd$ and $ \kkinnd $ would each depend on the spatial position of the flagellum (and therefore be also labelled by cell index $i$) through the spatial dependence of $ \gradnorm $ but our analysis will be restricted to the case where we treat the environmental logarithmic gradient as effectively constant over the colony motion scale of interest.  We therefore simplify the notation accordingly.

\section{Asymptotic Analysis}  \label{sec:analysis}

\begin{figure}[h!]  
{\centering{\includegraphics[height=80mm,width=130mm]{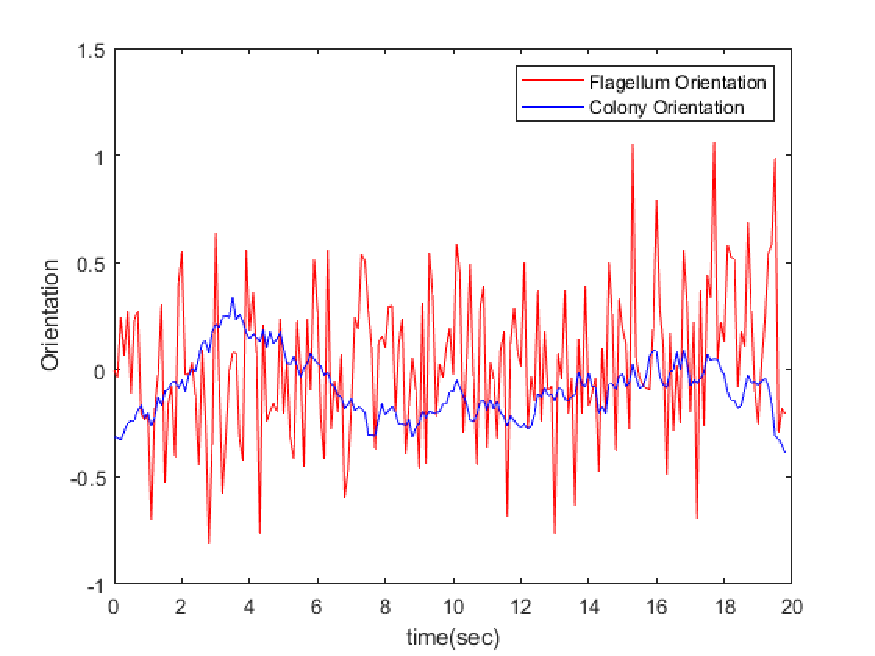}}
\caption{Dynamics of colony orientation $\Theta^{(c)}$ and flagellum orientation $\Theta_i$ for a given cell in a colony of $N=10$ cells exhibiting kinesis with 
rotational diffusivity modulation 
$ \kkin \gradnorm = 0.5 \mathrm{sec}^{-1}$ and flagellar force $ F=\SI{5}{\pico\newton}$.
    The other parameters are as specified in Table~\ref{tab:params}.\label{fig:scale_separation}}}
\end{figure}

Our nondimensionalization makes the time scale for the flagellar dynamics order unity, and we next examine the nondimensional time scale for the orientation $ \Thetacol$ of the colony in Eq.~\eqref{colony_orientation_nondim}. The drift term arising from the torques of each flagellum will be on the order of 
 $ \sim N^{1/2} (\zeta/N^2) \sigmaThe \sim \zeta \sigmaThe/N^{3/2} $ for kinesis, assuming each torque contribution is roughly independent.  For taxis, in principle the bias in the flagellar orientations can induce a drift term in $ \Thetacol$ on the order of $ \sim N \zeta/N^2 \sigmaThe \ktaxisnd \sim \zeta \sigmaThe \ktaxisnd/N$, though in practice there will be cancellations to this order due to the flagella on opposite sides of the colony with respect to the environmental gradient direction biasing themselves oppositely (see Figure~\ref{fig:taxis_kinesis_schematic}).  From the parameter estimates in Table~\ref{tab:params}, we see both the drift and diffusion term for $ \Thetacol $ should typically be somewhat small, provided $ \ktaxisnd \sim O(1)$ especially for larger colony sizes $ N$.  This motivates an asymptotic analysis which treats the flagellar dynamics as fast relative to the rotational (and therefore translational) dynamics of the colony.  More precisely, we will coarse-grain the flagellar fluctuations to obtain approximate closed equations for the slower orientational and translational dynamics of the colony.
The scale separation amounts to assuming that the colonies turn significantly more slowly than flagella force fluctuate. This is depicted graphically with Monte Carlo simulations of $\Theta^{(c)}$ and $\Theta_i$ in Figure~\ref{fig:scale_separation}.  
 To set up the asymptotic analysis, we define
 \begin{equation}
     \epsilon \equiv \sigma_\Theta\frac{\zeta}{N^2}
 \end{equation}
 as the formal small parameter.  As noted above, the drift term in the colony orientation dynamics~\eqref{colony_orientation_nondim} will actually be somewhat larger by a factor of $ N^{1/2} $ or $N $ (or by $ \ktaxisnd$ if it is large in the case of taxis), and so the validity of asymptotic analysis isn't quite determined just by the smallness of $ \epsilon $.  Moreover, we need to specify how the strength of the thermal noise relates to the drift terms, which can be done via a linkage between the parameters $ \delta $ and $ \epsilon $. We choose the distinguished limit
\begin{equation}
    \delta = \derat \sqrt{\epsilon \sigmaThe N}
\end{equation}
with the nondimensional parameter $ \derat$ nominally assumed order unity.  This distinguished limit is the one that would make the thermal rotation of the colony comparable to the one induced by the active flagellar forces.  Actually the results of our calculation also apply even if $ \derat$ is large or small, as the contribution from thermal or active forces is simply additive.  From our physical parameter choices in Table~\ref{tab:params}, we find
\begin{equation*}
    \derat N^{-1/2} = \frac{\delta}{\sigmaThe \zeta^{1/2}} = \sqrt{\frac{16k_B T \eta l\gamma_{\Theta}}{3\pi  F^2 \sigmaThe^2}} \sim 0.1 - 1.
\end{equation*}
so that the thermal contribution to rotation is comparable to the contribution from active forces for colonies of moderate size, and somewhat weaker for small colonies.  We proceed formally with this setup, and will scrutinize the validity of this asymptotic approximation immediately after the calculations. 

The mathematical coarse-graining procedures for taxis and kinesis differ somewhat, so we now separately consider the two models.

\subsection{Taxis}  \label{subsec:analysis_taxis}
Making the above notational changes to Eqs.~\eqref{colony_orientation_nondim} and~\eqref{eg:nondimensional_taxis_model}, we have the fast-slow system for the flagellar angles and the colony orientation:
\begin{equation} \label{eg:nondimensional_rescaled_taxis_model}
\begin{aligned}
d\Thetacellndi=-(\Thetacellndi(\tilde{t})+\ktaxisnd\sin(\Theta^{(c)}(\tilde{t})+\alpha_i-\gradang))d\tilde{t}+\sqrt{2}dW_i^{\Theta}(\tilde{t})
\end{aligned}
\end{equation}
\begin{equation}
\begin{aligned}
\label{eg:nondimensional_rescaled_colony_model}
d\Theta^{(c)}=-\epsilon\sigmaThe^{-1}\sum\limits_{i=1}^{N}  \sin{\sigmaThe \Thetacellndi(\tilde{t})}d\tilde{t}+\sqrt{2} \derat \epsilon \, dW^{\Theta,c}(\tilde{t}).
\end{aligned}
\end{equation}
We first use the assumed scale separation $ \epsilon \ll 1 $ to simply average the drift term for the slow colony orientation $ \Theta^{(c)} $with respect to the joint conditional stationary distribution $ \trho (\tbthvar|\thcvar)$ of the flagellar force orientations $\{\Thetacellndi\}_{i=1}^N$ given the current value of the colony orientation $ \Thetacol (\tilde{t}) = \thcvar$. The flagellar force orientations are evidently conditionally independent, and the conditional stationary distribution of each $ \Thetacellndi$ in the taxis case is simply that of an Orstein-Uhlenbeck process with shifted mean:
\begin{equation}  \label{flagella_distn_taxis}
\begin{aligned}
\trho (\tbthvar|\thcvar)
&= \prod_{i=1}^N \trhoi (\thvarnd{i}|\thcvar), \\
\trhoi(y_i|\thcvar )=& \frac{1}{\sqrt{2\pi}} e^{-\frac{(y_i+\ktaxisnd 
\sin[\thcvar+\alpha_{i}-\gradang])^2}{2}}
\end{aligned}
\end{equation}

Averaging the drift of $\Theta^{(c)}$ in Eq.~\eqref{eg:nondimensional_rescaled_colony_model} with respect to this conditional stationary distribution, using the averaging formula $ \langle \sin Z \rangle = \sin \mu_Z \expe^{-\sigma_Z^2/2} $ for a normally distributed random variable $ Z \sim N(\mu_Z,\sigma_Z^2)$, we obtain:
\begin{equation}    \label{taxis_thetac_avg1}
\begin{aligned}
d\Thetacolavg(\hat{t})=-\epsilon \sigmaThe^{-1}\sum\limits_{i=1}^{N} \sin(-\sigma_\theta \ktaxisnd 
\sin[\Thetacolavg(\tilde{t})+\alpha_{j}-\gradang])e^{-\frac{1}{2}\sigma_\theta^2}d\tilde{t}
\end{aligned}
\end{equation}
where $\Thetacolavg$ is the averaged value of $\hat{\Theta}^{(c)}$ with respect to the fast variables.  
The thermal noise term has been omitted from this expression since fluctuation effects will be considered subsequently.
To simplify the expressions which follow, we assume $ \sigmaThe \ktaxisnd \ll 1$ so we can take a linear approximation of the outer sine:
\begin{equation}    \label{taxis_thetac_avg2}
\begin{aligned}
d\Thetacolavg(\hat{t})=\epsilon \ktaxisnd \sum\limits_{i=1}^{N}   
\sin[\Thetacolavg(\tilde{t})+\alpha_{j}-\gradang]e^{-\frac{1}{2}\sigma_\theta^2}d\tilde{t}
\end{aligned}
\end{equation}
We express the sum in polar form by trigonometric manipulation:
\begin{align}
\label{eq:trig_condense}
     \sum\limits_{j=1}^{N} 
\sin[\Thetacolavg (\tilde{t})+\alpha_{j}-\gradang)
&= \Imag \sum\limits_{j=1}^{N} 
\expe^{\mathi(\Thetacolavg (\tilde{t})+\alpha_{j}-\gradang)}
= \Imag \expe^{\mathi(\Thetacolavg (\tilde{t})-\gradang)}\sum\limits_{j=1}^{N} \expe^{\mathi \alpha_j} \\
&= -N^{-1/2}\Imag \expe^{\mathi(\Thetacolavg (\tilde{t})-\gradang)} \geotaxamp \expe^{\mathi\geotaxphase} = - N^{-1/2}
\geotaxamp \sin (\Thetacolavg (\tilde{t})-\gradang+\geotaxphase) 
\end{align}
to write the effective nondimensional equation for the averaged colony orientation as:
\begin{equation}
d\Thetacolavg(\hat{t})=-\frac{\epsilon \ktaxisnd \geotaxamp}{N^{1/2}} e^{-\frac{1}{2}\sigma_\theta^2}
\sin[\Thetacolavg(\tilde{t})-\gradang+\geotaxphase] \, \difd \hat{t}
\label{colony_orientation_avg}
\end{equation}
These averaged dynamics for the colony orientation would indicate that the colony orientation $ \Thetacol (\hat{t}) $ would approach a stable equilibrium value $ \gradang-\geotaxphase$ on a time scale $ \frac{N^{1/2}}{\epsilon \ktaxisnd \geotaxamp}$.  


To incorporate the effect of fluctuations, we  apply  the Central Limit Theorem for fast-slow stochastic systems~\citep{BouchetLDFS2016} which, via the calculation presented in Appendix~\ref{sec:appendix:homogenization_taxis},  
\begin{equation}
 \difd \Thetacol (\tilde{t}) =-\frac{\epsilon \ktaxisnd \geotaxamp}{N^{1/2}} e^{-\frac{1}{2}\sigma_\theta^2}
\sin[\Thetacol(\tilde{t})-\gradang+\geotaxphase] \, \difd \tilde{t} +\sqrt{2} \derat \epsilon \, dW^{\Theta,c}(\tilde{t}) + \sqrt{2N} \epsilon \, \difd W (\tilde{t}) \label{eq:thetacol_clt}
\end{equation}
for times $ \tilde{t} \sim O(\epsilon^{-1}) $.
Here we have the direct thermal noise on the colony orientation as well as the effective noise (represented by the unadorned Wiener process term $ \difd W(\tilde{t})$) which arises from the flagellar fluctuations.  Looking only at the formal appearance of the small parameter $ \epsilon$, we would see the drift term driving the colony orientation to a stable equilibrium on the time scale $ \epsilon^{-1}$, with the noise contributing fluctuations of order $ \sqrt{\epsilon^2 \epsilon^{-1}} \sim \epsilon^{1/2}$ on that same time scale.   The other factors, however, make the effects of the noise significant, as shown numerically in Figure~\ref{fig:thccomps} in Appendix~\ref{sec:appendix:homogenization_taxis}.  In particular, we see the appearance of the colony size $N$ amplifying the effective noise from active flagellar fluctuations relative to the restoring drift.  By linearization of the drift term to give an Ornstein-Uhlenbeck approximation, we can estimate the steady-state fluctuation magnitude of $ \Thetacol (\tilde{t})$ about its stable equilibrium of $ \gradang - \geotaxphase$ as:
\begin{equation*}
\Var (\Thetacol) \approx \frac{\epsilon^2(\derat^2+N)}{[\epsilon \ktaxisnd \geotaxamp N^{-1/2}]}
= \frac{\epsilon (\derat^2+N)N^{1/2}}{\ktaxisnd \geotaxamp}.
\end{equation*}
For the biophysical parameter estimates in Table~\ref{tab:params}, and the discussion at the preamble to this section, we have typically $ N \gtrsim \derat^2$ and $ \epsilon \sim (0.3-3) N^{-2}$.  We see thus that the fluctuation amplitude can actually work out to be effectively of order unity, particularly when the asymmetry measure $ \geotaxamp$ is small.  This does in principle call into question the accuracy of our asymptotic approximation, but our numerical comparisons in Subsection~\ref{subsec:simulations_taxis} indicate our theoretical results are capturing the leading order behavior reasonably well.  The fact the fluctuations in colony orientation are not highly localized near the stable equilibrium value also motivates the retention of the full nonlinear expression for the drift rather than the linearized expression which appears in the standard central limit theorem for fast-slow systems~\citep{BouchetLDFS2016}.

On a time scale $ \tilde{t} \sim O(\epsilon^{-1})$, the stochastic fluctuations together with the effective drift will drive the colony orientation to a statistically stationary distribution found from the its associated stationary Fokker-Planck equation 
\begin{align*}
0=\left[\frac{\epsilon\geotaxamp \ktaxisnd }{N^{1/2}} e^{-\frac{1}{2}\sigma_\Theta^2} 
\sin[\theta-\gradang+\geotaxphase]\pstatthc (\theta)\right]_{\theta}+\left[\epsilon^2(\derat^2+N)\pstatthc (\theta)\right]_{\theta\theta}
\end{align*}
with a periodic boundary condition from $-\pi$ to $\pi$.
Integrating twice, applying the periodic boundary conditions, and seeking a solution normalized to a unit integral yields the von-Mises distribution (see also~\citep{aerotax}): 
\begin{align}
\label{eq:pstattaxis}
\pstatthctax (\theta) &=
\frac{\exp\left(\frac{\ktaxisnd }{\epsilon(\derat^2+N) N^{1/2}}e^{-\sigma_\Theta^2/2}\geotaxamp \cos (\theta-\gradang+\geotaxphase)\right)}{2\pi I_0(\frac{\ktaxisnd}{\epsilon(\derat^2+N)N^{1/2}}e^{-\sigma_\Theta^2/2}\geotaxamp)}
\end{align}


We next turn to the translational dynamics~\eqref{colony_orientation_nondim}, which we first express in terms of the asymptotically small parameter $ \epsilon$:
\begin{equation*}
d\mathbf{\tilde{X}^{(c)}}(\tilde{t})=-\epsilon \sigmaThe^{-1}\sum\limits_{j'=1}^{N}  \begin{pmatrix}\cos(\alpha_{j'}+\sigmaThe \Thetacellnd_{j'} (\tilde{t}) +\Theta^{(c)} (\tilde{t}))\\
\sin(\alpha_{j'}+\sigmaThe \Thetacellnd_{j'} (\tilde{t})+\Theta^{(c)}(\tilde{t}))\end{pmatrix}\, d\tilde{t}+\sqrt{2} \derat \epsilon d\mathbf{W}^{X,c}(\tilde{t})
\end{equation*}
The flagellar angles $ \Thetacellnd_j$ undergo confined stochastic oscillations so their contribution to the long-time translational dynamics can be captured  by simply averaging over their fluctuations:
\begin{equation}
\begin{aligned}
d\Xcolavg(\tilde{t})=-\frac{\epsilon \expe^{-\sigmaThe^2/2}}{\sigmaThe}
\sum\limits_{j=1}^{N}  \begin{pmatrix}\cos(\alpha_{j}-\sigmaThe \ktaxisnd \sin (\Thetacol (\tilde{t}) + \flagang_j - \gradang)  +\Theta^{(c)} (\tilde{t}))\\
\sin(\alpha_{j}-\sigmaThe \ktaxisnd \sin (\Thetacol (\tilde{t}) + \flagang_j - \gradang)  +\Theta^{(c)} (\tilde{t}))
\end{pmatrix}\, d\tilde{t}
+\sqrt{2} \derat \epsilon d\mathbf{W}^{X,c}(\tilde{t})
\end{aligned}
\label{eq:x_avg}
\end{equation}
To obtain an explicit representation for the long-term drift, we will certainly be neglecting spatial variation in the log concentration gradient, so $ \gradnorm $ and $ \gradang$ are taken as constant.  As the colony orientation would then approach a unique stationary distribution~\eqref{eq:pstattaxis}, the long term drift of the colony would be expressed as the average of the drift term in Eq.~\eqref{eq:x_avg} with respect to the stationary distribution of the colony orientation.  Actually as cautioned in~\citet{newby2013breakdown}, one should  check whether the central limit theorem dynamics for $ \Thetacol (\tilde{t})$ presented in Eq.~\eqref{eq:thetacol_clt} is adequate, or if a large deviation calculation is required.  The central limit theorem dynamics suggest nominally small fluctuations of the colony orientation about the stable equilibrium at $ \gradang - \geotaxphase$, though in fact the fluctuation magnitude is not so numerically small for the parameter values  used as discussed above.  Large deviation phenomena would be concerned with the colony orientation making a (nominally) rare jump in phase by $ \pm 2 \pi $ between neighboring stable equilibria.  But accounting for such jumps, which would occur on fast time scales, would not materially change the effective translational dynamics from what would be computed by averaging over the typical orientation fluctuations.  Similarly, though Eq.~\eqref{eq:thetacol_clt} is based on a central limit theorem that guarantees the validity only for times $ \tilde{t} \sim O(\epsilon^{-1})$, the facts that $ \Thetacol (\tilde{t})$ reaches its stationary distribution on this time scale and large deviation phenomena have no significant impact imply we can readily work with the effective dynamics~\eqref{eq:thetacol_clt} to obtain the leading order fluctuating dynamics of $ \Thetacol (\tilde{t}) $ over arbitrarily long time scales.

This calculation of colony drift is expedited by expressing the resulting two-dimensional drift vector in complex form. 
We therefore consider:
\begin{equation} \label{Averaging_LOTUS_drift_taxis}
\begin{aligned}
&\int\limits_{-\pi}^{\pi}e^{i\alpha_{j}+i\theta-i\sigma_\theta \ktaxisnd\sin[\theta+\alpha_{j}-\gradang]}\pstatthctax (\theta) \, \difd \theta.
\end{aligned}
\end{equation}
This integral is intractable to evaluate analytically, but we can simplify the integrand by expanding it to first order in the force orientation variance $ \sigmaThe^2$, which is  small (see Table~\ref{tab:params}):
\begin{equation*}
e^{i\alpha_{j}+i\theta-i\sigma_\Theta \ktaxisnd\sin[\theta+\alpha_{j}-\gradang]}=
e^{i(\alpha_{j}+\theta)}
(1-i\sigma_\Theta \ktaxisnd\sin(\theta+\alpha_{j}-\gradang))+O(\sigma_\Theta^2 \ktaxisnd^2)
\end{equation*}
More precisely, we assume $ \sigmaThe \ktaxisnd = \ktaxis \gradnorm \ll 1$, meaning the taxis correction to the force orientation is small.  We do not, however, similarly expand the stationary distribution~\eqref{eq:pstattaxis} with respect to the same parameter group because of the large factor $ \delta^{-2}$ (see Eq.~\eqref{eq:def_delta}) and potentially large factor $ N^{1/2}$ which multiply it.
Substituting this asymptotic approximation into (\ref{Averaging_LOTUS_drift_taxis}), we compute for $ \geotaxamp \neq 0$:
\begin{equation} \label{Averaging_LOTUS_drift_taxis2}
\begin{aligned}
&\int\limits_{-\pi}^{\pi}e^{i\alpha_{j}+i\theta-i\sigma_\theta \ktaxisnd\sin[\theta+\alpha_{j}-\gradang]}\pstatthctax (\theta) \, \difd \theta = \frac{\sigmaThe \ktaxisnd \expe^{\mathi\gradang}}{2}\\
&+ \frac{1}{I_0(\frac{\ktaxisnd}{\epsilon(\derat^2+N)N^{1/2}}e^{-\sigma_\Theta^2/2}\geotaxamp)}\Big[ I_1(\frac{\ktaxisnd}{\epsilon(\derat^2+N)N^{1/2}}e^{-\sigma_\Theta^2/2}\geotaxamp)e^{i(\alpha_j+\gradang-\phi)}\\
 & - \frac{\sigmaThe \ktaxisnd \expe^{\mathi (2 \alpha_j +\gradang-2\geotaxphase)} }{2}I_2(\frac{\ktaxisnd}{\epsilon(\derat^2+N)N^{1/2}}e^{-\sigma_\Theta^2/2}\geotaxamp) \Big]+O(\sigma_\Theta^2 \ktaxisnd^2) \\
 &= \frac{\sigmaThe \ktaxisnd}{2} (\expe^{\mathi\gradang}-\expe^{\mathi (2 \alpha_j +\gradang-2\geotaxphase)}) \\
 &+ \frac{I_1(\frac{\ktaxisnd}{\epsilon(\derat^2+N)N^{1/2}}\expe^{-\sigma_\Theta^2/2}\geotaxamp)}{I_0(\frac{\ktaxisnd}{\epsilon(\derat^2+N)N^{1/2}}e^{-\sigma_\Theta^2/2}\geotaxamp)}\left[ \expe^{i(\alpha_j+\gradang-\phi)}
 + \frac{\sigmaThe\epsilon(\derat^2+N)N^{1/2}
 \expe^{\sigmaThe^2/2}}{\geotaxamp} \expe^{\mathi (2 \alpha_j +\gradang-2\geotaxphase)}\right]+O(\sigma_\Theta^2 \ktaxisnd^2)
 \end{aligned}
 \end{equation}

This yields the effective nondimensional drift: 
\begin{equation} \label{non-dim_taxis_drift}
\begin{aligned}
&\drifteffnd \equiv-\left(\epsilon \sigmaThe^{-1}\right)e^{-\sigma_\Theta^2/2}\sum\limits_{j=1}^{N}
  \left[ \frac{\sigmaThe \ktaxisnd}{2}\Big(\begin{pmatrix} \cos\theta_g\\ \sin\theta_g \end{pmatrix}- \begin{pmatrix} \cos(2\alpha_{j}+\gradang-2\phi)\\ \sin(2\alpha_{j}+\gradang-2\phi) \end{pmatrix}\Big)\right.
  \\
&+ \frac{I_1(\frac{\ktaxisnd}{\epsilon(\derat^2+N)N^{1/2}}e^{-\sigma_\Theta^2/2}\geotaxamp)}{I_0(\frac{\ktaxisnd}{\epsilon(\derat^2+N)N^{1/2}}e^{-\sigma_\Theta^2/2}\geotaxamp)}\left(\begin{pmatrix} \cos(\alpha_{j}+\gradang-\phi)\\ \sin(\alpha_{j}+\gradang-\phi) \end{pmatrix} +  \frac{\sigmaThe\epsilon(\derat^2+N)N^{1/2}
 \expe^{\sigmaThe^2/2}}{\geotaxamp}\begin{pmatrix} \cos(2\alpha_{j}+\gradang-2\phi)\\ \sin(2\alpha_{j}+\gradang-2\phi) \end{pmatrix}\right)
 \\
 & \qquad \qquad\left. +O(\sigma_\Theta^2 \ktaxisnd^2)\right]
\end{aligned}
\end{equation}
\begin{equation*} 
\begin{aligned}
 =&\left(\epsilon \sigmaThe^{-1}\right)e^{-\sigma_\Theta^2/2}
  \left[ -\frac{N \sigmaThe \ktaxisnd}{2}\begin{pmatrix} \cos(\theta_g)\\ \sin(\theta_g) \end{pmatrix}- \frac{ \geotaxamp_2 \sigmaThe \ktaxisnd}{2\sqrt{N}}\begin{pmatrix} \cos(\gradang-\geotaxphase_2)\\ \sin(\gradang-\geotaxphase_2) \end{pmatrix} \right.
  \\
&+ \frac{I_1(\frac{\ktaxisnd}{\epsilon(\derat^2+N)N^{1/2}}e^{-\sigma_\Theta^2/2}\geotaxamp)}{I_0(\frac{\ktaxisnd}{\epsilon(\derat^2+N)N^{1/2}}e^{-\sigma_\Theta^2/2}\geotaxamp)}
\left(\frac{\geotaxamp}{\sqrt{N}}\begin{pmatrix} \cos(\gradang)\\ \sin(\gradang) \end{pmatrix} + 
\frac{\sigmaThe\epsilon(\derat^2+N)
 \expe^{\sigmaThe^2/2}\geotaxamp_2}{\geotaxamp}
\begin{pmatrix} \cos(\gradang-\geotaxphase_2)\\ \sin(\gradang-\geotaxphase_2) \end{pmatrix}\right)
 \\&\left.+O(\sigma_\Theta^2 \ktaxisnd^2)\right]
\end{aligned}
\end{equation*}

Here $\geotaxamp_2$ and $\geotaxphase_2$ are defined in the preamble to section \ref{sec:taxisresults}.  When $ \geotaxamp =0$, then the stationary distribution~\eqref{eq:pstattaxis} for $ \Thetacol (t) $ is uniform so:
\begin{equation*} 
\begin{aligned}
&\int\limits_{-\pi}^{\pi}e^{i\alpha_{j}+i\theta-i\sigma_\theta \ktaxisnd\sin[\theta+\alpha_{j}-\gradang]}\pstatthctax (\theta) \, \difd \theta =
\frac{1}{2\pi} e^{i\gradang} 
&\int\limits_{-\pi}^{\pi} e^{i\theta^{\prime}-i\sigmaThe \ktaxisnd \sin (\theta^{\prime})} \, \difd \theta^{\prime}  = J_1 (\sigmaThe \ktaxisnd) e^{i\gradang}
\end{aligned}
\end{equation*}
and
\begin{equation*} 
\begin{aligned}
\drifteffnd &= -\left(\epsilon \sigmaThe^{-1}\right)e^{-\sigma_\Theta^2/2}\sum\limits_{j=1}^{N}
  J_1 (\sigmaThe \ktaxisnd)\begin{pmatrix} \cos\theta_g\\ \sin\theta_g \end{pmatrix}
  = -\left(\epsilon \sigmaThe^{-1}\right)Ne^{-\sigma_\Theta^2/2}
  J_1 (\sigmaThe \ktaxisnd)\begin{pmatrix} \cos\theta_g\\ \sin\theta_g \end{pmatrix}
\end{aligned}
\end{equation*} 

With the effective transport parameters calculated, we can do a formal self-consistency check of the fundamental asymptotic approximation that the rotational dynamics of the colony occur on a time scale large compared with that of an individual cell flagellum, which is $1 $ in our nondimensional time units.  From Eq.~\eqref{eq:thetacol_clt}, we see the nondimensional time scale of rotational drift is $ 
N^{1/2} \epsilon^{-1} \geotaxamp^{-1} \ktaxisnd^{-1}
=
N^{5/2} \sigmaThe^{-1} \zeta^{-1} \geotaxamp^{-1} \ktaxisnd^{-1}$, while that of rotational diffusion is $ 
(N+\derat^2)^{-1} \epsilon^{-2}= (\zeta \delta^2 + \sigmaThe^2 \zeta^2)^{-1} N^3$. 
Thus our asymptotic analysis is formally self-consistent provided:
\begin{equation}
    N \gg (\zeta\geotaxamp \ktaxisnd \sigmaThe)^{2/5}, (\zeta \delta^2+\sigmaThe^2 \zeta^2)^{1/3}.
    \label{eq:nondim_taxis_cond}
\end{equation}

\subsection{Kinesis}  \label{subsec:analysis_kinesis}
For kinesis, the flagellar force angles and consequently the torques on the colony have mean zero, so only contribute to a rotational diffusion of the colony. This rotational diffusion will have a time scale $ \sim \epsilon^{-2}$ so we rescale time $\hat{t}=\epsilon^{2} \tilde{t}$ to bring this colony rotation dynamics into focus.  On this time scale, we expect $ O(1) $ changes in colony orientation so we don't rescale this variable: $\hat{\Theta}^{(c)}=\Theta^{(c)}$.  The spatial coordinate has an $ O(\epsilon) $ drift term which should have nonzero mean, so we rescale it as:
$\hXcol=\epsilon\tilde{\mathbf{X}}^{(c)}$.
 The rescaling together with the notational changes from the beginning of Section~\ref{sec:analysis} will give the SDEs:
\begin{equation} \label{rescaled_nondim_cell_kinesis}
\begin{aligned}
d\hat{\Theta}_i(\hat{t})=&-\epsilon^{-2}{\hat{\Theta}_i}(\hat{t})d\hat{t}
+\sqrt{2}\epsilon^{-1}\Big[1+m_K\cos(\alpha_{j'}+{\Theta}^{(c)}(\hat{t})-\theta_g)\Big]^{\frac{1}{2}}dW_i^\Theta({\hat{t}})
\end{aligned}
\end{equation}
\begin{equation} \label{rescaled_nondim_colony_kinesis}
\begin{aligned}
\difd \hThetacol(\hat{t})=-\epsilon^{-1}\sum\limits_{j'=1}^{N} \sigmaThe^{-1} \sin{(\sigma_\Theta\hat{\Theta}_{j'}(\hat{t}))}d\hat{t}+\sqrt{2} \derat \, dW^{\Theta,c}(\hat{t})
\end{aligned}
\end{equation} 
\begin{equation}
\begin{aligned}
\difd \hXcol(\hat{t})=&\sum\limits_{j'=1}^{N} -\sigmaThe^{-1} \begin{pmatrix}\cos(\alpha_{j'}+\sigma_\Theta\hat{\Theta}_{j'}(\hat{t})+\hThetacol)(\hat{t})\\ \sin{(\alpha_{j'}+\sigma_\Theta\hat{\Theta}_{j'}(\hat{t})+\hThetacol(\hat{t}))}\end{pmatrix}d\hat{t}
+\sqrt{2} \derat \epsilon  \,  d\mathbf{W}^{X,c}(\hat{t})
\end{aligned}
\end{equation} 

We now proceed to homogenize the equation for $ \hThetacol $ using $ \epsilon \ll 1$ and $ \derat \sim O(1)$  to obtain (see Appendix~\ref{sec:appendix:homogenization}):
\begin{equation} \label{homogenized_colony_orientation}
\begin{aligned}
d{\hat{\Theta}}^{(c)}(\hat{t})\approx[2\beta^2+2N- \frac{2m_K \geotaxamp}{\sqrt{N}} \cos(\hThetacol(\hat{t})-\theta_g+\phi)]^{\frac{1}{2}}dW^{\Theta,c}(\hat{t}).
\end{aligned}
\end{equation}
where $ \geotaxamp $ and $ \geotaxphase $ are defined near the beginning of Sec.~\ref{sec:summary}.

The translational dynamics can be more simply averaged~\citep{homogenization} over the fast dynamics to obtain
\begin{equation}
\begin{aligned} \label{averaged_colony_position}
\difd \hXcol(\hat{t})=&-\sigmaThe^{-1} \sum\limits_{j'=1}^{N} e^{-\frac{1}{2}\sigma_\Theta ^2[1+m_K\cos(\alpha_{j'}+\hThetacol(\hat{t})-\theta_g)]}\begin{pmatrix}\cos(\alpha_{j'}+\hThetacol (\hat{t}))\\ \sin{(\alpha_{j'}+\hThetacol (\hat{t}))}\end{pmatrix}d\hat{t}
+\sqrt{2} \derat \epsilon  \,  d\mathbf{W}^{X,c}(\hat{t})
\end{aligned}
\end{equation}

Next, we compute the long-term drift of the colony, again temporarily passing to a representation of the drift vector in complex form as in Subsection~ \ref{subsec:analysis_taxis}.

Then we must calculate:
\begin{equation} \label{Averaging_LOTUS_drift_kinesis}
\begin{aligned}
\int_{-\pi}^{\pi} e^{\mathi\alpha_{j'}+\mathi\theta} e^{-\frac{1}{2}\sigma_\Theta ^2[1+m_K\cos(\alpha_{j'}+\theta-\theta_g)]} \pstatthc (\theta) \, \difd \theta
\end{aligned}
\end{equation}
where $\pstatthc$ denotes the stationary distribution of $ \Theta^{(c)} $ on the periodic domain $ (-\pi,\pi] $.  This stationary distribution is given by the steady-state Fokker-Planck equation associated to Eq.~\eqref{homogenized_colony_orientation}:
\begin{align*}
0=\left[(\beta^2+N -  \frac{\kkinnd \geotaxamp}{\sqrt{N}} \cos (\theta - \gradang + \geotaxphase))\pstatthc (\theta)\right]_{\theta\theta}
\end{align*}
with a periodic boundary condition from $-\pi$ to $\pi$.
Integrating twice, applying the periodic boundary conditions, and seeking a solution normalized to a unit integral yields:
\begin{align}
\pstatthc (\theta) 
= \frac{\sqrt{1-\frac{\kkinnd^2 \geotaxamp^2}{N(\beta^2+N)^2}}}{ 2\pi \left[1 -  \frac{\kkinnd \geotaxamp}{\sqrt{N}(\beta^2+N)} \cos (\theta - \gradang + \geotaxphase)\right]}
\end{align}
Evaluating analytically the average~\eqref{Averaging_LOTUS_drift_kinesis} via contour integration leads to a cumbersome expression, so to get a more transparent expression we now take a further assumption of $ \kkinnd \ll 1$ to get a first order perturbative effect of the kinesis modulation: 
\begin{equation}
\pstatthc (\theta) = 
 \frac{1}{2\pi}
+ \frac{\kkinnd \geotaxamp}{2\pi\sqrt{N} (\beta^2+N)} \cos (\theta - \gradang + \geotaxphase)
 +O\left(\frac{m_K^2}{N(\beta^2+N)^2}\right)\label{eq:pstat_kin}
\end{equation}

Next, we substitute this asymptotic approximation into Eq.~\eqref{Averaging_LOTUS_drift_kinesis} and
asymptotically expand the integrand in terms of $m_K$. We notice that the $ \ord(1) $ term naturally integrates to zero, as the colony would have no preferred direction absent a response to the environmental gradient.  
We thereby obtain:
\begin{equation} 
\begin{aligned}
&\int_{-\pi}^{\pi} e^{\mathi\alpha_{j'}+\mathi\theta} e^{-\frac{1}{2}\sigma_\Theta ^2[1+m_K\cos(\alpha_{j'}+\theta-\theta_g)]} \pstatthc (\theta) \, \difd \theta  \\
&\qquad  =e^{-\frac{1}{2}\sigma_\Theta^2}e^{i\theta_g} \kkinnd\left[\frac{\geotaxamp}{2(\beta^2+N)\sqrt{N}}e^{\mathi (\alpha_{\jp}-\geotaxphase)}-\frac{\sigma_\Theta^2}{4}+O\left(\kkinnd\left(\frac{1}{N(\beta^2+N)^2}+\frac{\sigmaThe^2}{\sqrt{N}(\beta^2+N)}\right)\right)
\right] \end{aligned}
\end{equation}
This therefore yields for the effective nondimensional drift: 
\begin{equation} \label{non-dim_kinesis_drift}
\begin{aligned}
\drifteffnd &\equiv \lim_{\tilde{t} \rightarrow \infty} \frac{\Xcolnd (\tilde{t})}{\tilde{t}} =
\epsilon \lim_{\hat{t}\rightarrow \infty} \frac{\hXcol (\hat{t})}{\hat{t}} \\
&=\frac{\zeta}{N^2} e^{-\frac{1}{2}\sigma_\Theta^2} m_K\left[\frac{1}{2N(\beta^2+N)} \geotaxamp^2
+\sigma_\Theta^2\frac{N}{4}+O\left(\kkinnd\left(\frac{1}{(\beta^2+N)^2}+\frac{\sqrt{N}\sigmaThe^2}{\beta^2+N}\right)\right)\right] \graddir .
\end{aligned}
\end{equation}

The chemotactic index follows from a similar average against the stationary distribution of the colony orientation:
\begin{align*}
\CI &= \int_{-\pi}^\pi \cos (\theta +\geotaxphase - \gradang) \pstatthc (\theta) \, \difd \theta = 
\frac{\sqrt{N}(\beta^2+N)}{\kkinnd \geotaxamp} \left(1 - \sqrt{1-\frac{\kkinnd^2 \geotaxamp^2}{N(\beta^2+N)^2}}\right) \\
&= \frac{\kkinnd \geotaxamp}{2\sqrt{N} (\beta^2+N)} + O \left(\frac{\kkinnd^3}{N^{3/2}(\beta^2+N)^3}\right)
\end{align*}

Now we check the formal self-consistency of our asymptotic assumption that the colony rotation time scale is large compared to the nondimensional time scale of the cellular flagella, which is order unity.  From Eq.~\eqref{homogenized_colony_orientation} and recalling the time rescaling $ \hat{t} = \epsilon^{2}\tilde{t}$, the nondimensional time scale of rotational diffusion is $ \epsilon^{-2}(\beta^2+N)^{-1} =(\zeta \delta^2 + \sigmaThe^2 \zeta^2)^{-1} N^3$
so the self-consistency requirement is:
\begin{equation}
    N \gg (\zeta \delta^2+\sigmaThe^2 \zeta^2)^{1/3},  \label{eq:nvalkin}
\end{equation}

\subsection{Asymmetric Statistic}
In both the taxis and kinesis model, we find the leading order effects of asymmetry on the colonial drift are proportional to $ \geotaxamp^2$.  If we take the displacement $ \{\flagdispi\}_{i=1}^N$ of the flagellar attachment points from the center of the exposed portion of each cell to be independently and uniformly distributed over a  range $(-u,u)$, with $ 0 \leq u \leq \celllen/2 $, we can compute the average of the factor $ \geotaxamp^2 $ in the drift depending on the flagellar arrangement:
\begin{align*}
\langle \geotaxamp^2 \rangle &= N
\sum\limits_{j=1}^N \sum\limits_{\jp=1}^N \expe^{\frac{2 \pi\mathi (j-\jp)}{N}}
\langle \expe^{\frac{2\pi\mathi}{N \celllen} (\flagdisp_j - \flagdisp_{\jp})} \rangle
=N\twoflagavg \sum\limits_{j=1}^N \sum\limits_{\jp=1}^N \expe^{\frac{2 \pi\mathi (j-\jp)}{N}}
+ N\sum\limits_{j'=1}^N (1- \twoflagavg) \\
&= \twoflagavg \left|\sum\limits_{j=1}^N \expe^{\frac{2 \pi\mathi j}{N}}\right|^2 
+ N^2(1- \twoflagavg) \\
&=0+ N^2(1- \twoflagavg),
\end{align*}
for $ N \geq 2$
where $ \langle \cdot \rangle $ denotes a statistical average over demographic randomness in the flagellar displacements $ \{\flagdispi\}_{i=1}^N$.
Here, for any index pair $ j \neq \jp $ corresponding to different flagella, we evaluate the average: 
\begin{equation}
    \twoflagavg \equiv 
    \langle \expe^{\frac{2\pi \mathi}{N \celllen} (\flagdisp_j - \flagdisp_{\jp})}\rangle = (2 u)^{-2} \int_{-u}^u 
\int_{-u}^u \expe^{\frac{2\pi\mathi}{N \celllen} (s - s^{\prime})} \, \difd s \, \difd s^{\prime}
= \frac{\sin^2(\frac{2\pi}{lN}u)}{(\frac{2\pi}{lN})^2u^2} \label{eq:twoflagavg}
\end{equation}
Thus:
\begin{equation}
    \langle \geotaxamp^2 \rangle  = N^2\left(1-\frac{\sin^2(\frac{2\pi}{lN}u)}{(\frac{2\pi}{lN})^2u^2}\right) \sim \frac{1}{3} \left(\frac{2\pi u}{l}\right)^2 + O\left(N^{-2}\left(\frac{ u}{l}\right)^4\right)
\end{equation}

To characterize demographic variations, we also compute:
\begin{align}
    \Var (\geotaxamp^2) &= N^2 \sum_{j=1}^N \sum_{\jp=1}^N \expe^{\frac{4 \pi\mathi (j-\jp)}{N}}
\Var \left[\expe^{\frac{2\pi\mathi}{N \celllen} (\flagdisp_j - \flagdisp_{\jp})} \right]
+N^2\sum_{\substack{j,\jp=1 \\ \jp \neq j}}^N 
\Cov \left(\expe^{\frac{2\pi\mathi}{N \celllen} (\flagdisp_j - \flagdisp_{\jp})} ,\expe^{\frac{2\pi\mathi}{N \celllen} (\flagdisp_{\jp} - \flagdisp_{j})} \right) 
\label{eq:vargeosum}
\\
& \qquad \qquad + N^2 \sum_{\substack{j,\jp,\jpp=1 \\ j\neq \jp \neq \jpp}}^N 
\expe^{\frac{2 \pi\mathi (2j-\jp-\jpp)}{N}}
\Cov\left(\expe^{\frac{2\pi\mathi}{N \celllen} (\flagdisp_j - \flagdisp_{\jp})} ,\expe^{\frac{2\pi\mathi}{N \celllen} (\flagdisp_j - \flagdisp_{\jpp})} \right) \nonumber \\
& \qquad \qquad + N^2 \sum_{\substack{j,\jp,\jpp=1 \\ j\neq \jp \neq \jpp}}^N 
\expe^{\frac{2 \pi\mathi (j+\jpp-2\jp)}{N}}
\Cov\left(\expe^{\frac{2\pi\mathi}{N \celllen} (\flagdisp_j - \flagdisp_{\jp})} ,\expe^{\frac{2\pi\mathi}{N \celllen} (\flagdisp_{\jpp} - \flagdisp_{\jp})} \right) \nonumber \\
& \qquad \qquad + N^2  \sum_{\substack{j,\jp,\jpp=1 \\ j\neq \jp \neq \jpp}}^N 
\expe^{\frac{2 \pi\mathi (\jpp-\jp)}{N}}
\Cov\left(\expe^{\frac{2\pi\mathi}{N \celllen} (\flagdisp_j - \flagdisp_{\jp})} ,\expe^{\frac{2\pi\mathi}{N \celllen} (\flagdisp_{\jpp} - \flagdisp_{j})} \right) \nonumber \\
&\qquad \qquad + N^2  \sum_{\substack{j,\jp,\jpp=1 \\ j\neq \jp \neq \jpp}}^N 
\expe^{\frac{2 \pi\mathi (j-\jpp)}{N}}
\Cov\left(\expe^{\frac{2\pi\mathi}{N \celllen} (\flagdisp_j - \flagdisp_{\jp})} ,\expe^{\frac{2\pi\mathi}{N \celllen} (\flagdisp_{\jp} - \flagdisp_{\jpp})} \right)\nonumber \\
&= N^2 \twoflagvar \left[\sum_{j=1}^N \sum_{\jp=1}^N \expe^{\frac{4 \pi\mathi (j-\jp)}{N}}
- N\right] +N^3 (N-1) \twoflagmod \nonumber\\
& + 2N^2 \Real \threeflagcov \left[ \sum_{j=1}^N \sum_{\jp=1}^N \sum_{\jpp=1}^N  \expe^{\frac{2 \pi\mathi (2j-\jp-\jpp)}{N}} -  \sum_{j=1}^{N} \sum_{\jp=1}^N \expe^{\frac{4 \pi\mathi (j-\jp)}{N}} - 2 \sum_{j=1}^{N} \sum_{\jp=1}^N \expe^{\frac{2 \pi\mathi (j-\jp)}{N}} + 2N\right] \nonumber\\
&+ 2N^2 \Real \threeflagcovconj\left[ \sum_{j=1}^N \sum_{\jp=1}^N \sum_{\jpp=1}^N  \expe^{\frac{2 \pi\mathi (\jpp-\jp)}{N}} - \sum_{j=1}^N \sum_{\jp=1}^N 1 - 2\sum_{j=1}^N \sum_{\jp=1}^{N}  \expe^{\frac{2 \pi\mathi (j-\jp)}{N}}  + 2N\right] \nonumber\\
&=  N^2 \twoflagvar \left[\left|\sum_{j=1}^N  \expe^{\frac{2 \pi\mathi j }{N}}\right|^2
- N\right] +(N^4-N^3) \twoflagmod\nonumber\\
& + 2N^2 \Real \threeflagcov \left[ \sum_{j=1}^N\expe^{\frac{4 \pi\mathi j}{N}} 
\left(\sum_{\jp=1}^N \expe^{-\frac{2 \pi\mathi\jp}{N}}\right)^2 - \left|\sum_{j=1}^N  \expe^{\frac{4 \pi\mathi j }{N}}\right|^2   - 2 \left|\sum_{j=1}^N  \expe^{\frac{2 \pi\mathi j }{N}}\right|^2 + 2N\right] \nonumber \\
& + 2N^2 \Real \threeflagcovconj\left[ \sum_{j=1}^N \left|\sum_{\jp=1}^N  \expe^{\frac{2 \pi\mathi \jp}{N}}\right|^2 -  N^2 - 2\left| \sum_{\jp=1}^{N} \expe^{\frac{2 \pi\mathi \jp)}{N}} \right|^2+ 2N\right] \nonumber \\
&=  
- N^3 \twoflagvar +(N^4-N^3) \twoflagmod
+ 2N^2 \Real \threeflagcov \left[  2N\right] 
+ 2N^2 \Real \threeflagcovconj\left[ -  N^2+ 2N\right] \nonumber \\
&= -N^3 \twoflagvar  +(N^4-N^3) \twoflagmod+ 4N^3 \Real \threeflagcov + (4N^3-2N^4) \Real \threeflagcovconj \nonumber
\end{align}
The statistical components are evaluated here as in Eq.~\eqref{eq:twoflagavg} as:
\begin{align*}
\twoflagvar &\equiv
    \Var \left[\expe^{\frac{2\pi\mathi}{N \celllen} (\flagdisp_j - \flagdisp_{\jp})} \right]
    = \langle \expe^{\frac{4\pi\mathi}{N \celllen} (\flagdisp_j - \flagdisp_{\jp})} \rangle 
    -  \langle \expe^{\frac{2\pi\mathi}{N \celllen} (\flagdisp_j - \flagdisp_{\jp})} \rangle^2 \\
    &= \frac{\sin^2(\frac{4\pi}{lN}u)}{(\frac{4\pi}{lN})^2u^2}-\frac{\sin^4(\frac{2\pi}{lN}u)}{(\frac{2\pi}{lN})^4u^4} \text{ for } j \neq \jp\\
    \twoflagmod &\equiv\Cov\left(\expe^{\frac{2\pi\mathi}{N \celllen} (\flagdisp_j - \flagdisp_{\jp})} ,\expe^{\frac{2\pi\mathi}{N \celllen} (\flagdisp_{\jp} - \flagdisp_{j})} \right) = 1-\frac{\sin^4(\frac{2\pi}{lN}u)}{(\frac{2\pi}{lN})^4u^4}\\
    \threeflagcov &\equiv \Cov\left(\expe^{\frac{2\pi\mathi}{N \celllen} (\flagdisp_j - \flagdisp_{\jp})} ,\expe^{\frac{2\pi\mathi}{N \celllen} (\flagdisp_j - \flagdisp_{\jpp})} \right)
    = \Var \left[\expe^{\frac{2\pi\mathi}{N \celllen} \flagdisp_j} \right] \frac{\sin^2(\frac{2\pi}{lN}u)}{(\frac{2\pi}{lN})^2u^2}  \text{ for } j \neq \jp\\
    &= \left[\frac{\sin(\frac{4\pi}{lN}u)}{(\frac{4\pi}{lN})u}-\frac{\sin^2(\frac{2\pi}{lN}u)}{(\frac{2\pi}{lN})^2u^2}\right]\frac{\sin^2(\frac{2\pi}{lN}u)}{(\frac{2\pi}{lN})^2u^2} \text{ for } j\neq \jp \neq \jpp, \\
    \threeflagcovconj &=\Cov\left(\expe^{\frac{2\pi\mathi}{N \celllen} (\flagdisp_j - \flagdisp_{\jp})} ,\expe^{\frac{2\pi\mathi}{N \celllen} (\flagdisp_{\jpp} - \flagdisp_{j})} \right)
    = \frac{\sin^2(\frac{2\pi}{lN}u)}{(\frac{2\pi}{lN})^2u^2} \left(1-\frac{\sin^2(\frac{2\pi}{lN}u)}{(\frac{2\pi}{lN})^2u^2}\right)  \text{ for } j\neq \jp \neq \jpp
\end{align*}
Substituting these expressions into Eq.~\eqref{eq:vargeosum}, we obtain 
\begin{align*}
     \Var (\geotaxamp^2) 
     &= N^4-N^3 -N^3\frac{\sin^2(\frac{4\pi}{lN}u)}{(\frac{4\pi}{lN})^2u^2}+
     (N^4 -6N^3)\frac{\sin^4(\frac{2\pi}{lN}u)}{(\frac{2\pi}{lN})^4u^4}
     +(4N^3 - 2N^4)\frac{\sin^2(\frac{2\pi}{lN}u)}{(\frac{2\pi}{lN})^2u^2} \\
     & \qquad \qquad +4N^3  \frac{\sin(\frac{4\pi}{lN}u)}{(\frac{4\pi}{lN})u} \frac{\sin^2(\frac{2\pi}{lN}u)}{(\frac{2\pi}{lN})^2u^2} \\
     & = N^4 \left[1- \frac{\sin^2(\frac{2\pi}{lN}u)}{(\frac{2\pi}{lN})^2u^2} \right]^2 \\
     & \qquad \qquad + N^3\left[-1 -\frac{\sin^2(\frac{4\pi}{lN}u)}{(\frac{4\pi}{lN})^2u^2}-6\frac{\sin^4(\frac{2\pi}{lN}u)}{(\frac{2\pi}{lN})^4u^4}
     +4\frac{\sin^2(\frac{2\pi}{lN}u)}{(\frac{2\pi}{lN})^2u^2} +4  \frac{\sin(\frac{4\pi}{lN}u)}{(\frac{4\pi}{lN})u} \frac{\sin^2(\frac{2\pi}{lN}u)}{(\frac{2\pi}{lN})^2u^2}\right]\\
     & \sim \frac{1}{36} \left(\frac{2\pi u}{l}\right)^4 + O \left(N^{-1} \left(\frac{u}{l}\right)^4\right)
\end{align*}
This implies that the standard deviation of $ \geotaxamp^2 $ across demographic variations would be, for sufficiently large colonies, approximately half the mean.   

. 


\section{Numerical Simulations} \label{sec:simulations}
In this section, we compare the theoretical formulas derived under various asymptotic approximations in previous sections to direct Monte Carlo simulations of the stochastic differential equation models described in Section~\ref{sec:model}.  Results from the taxis model simulations are presented in Subsection~\ref{subsec:simulations_taxis}, while those from the kinesis model are presented in Subsection~\ref{subsec:simulations_kinesis}.  Unless otherwise specified, all simulations are based on the physical parameters as given in Table~\ref{tab:params}, with flagellar force magnitude 5 pN. 
In presenting the theoretical and numerical comparisons, we use the nondimensionalization of variables and associated nondimensional groups from Sec.~\ref{sec:nondim}.  
The attractant profile $ c(x_1 ,x_2) $ for the simulations is taken as proportional to the following non-dimensional affine function:
\begin{equation}  \label{sim. chemical profile}
  \begin{aligned}
        c(x_1,x_2) \propto \cos(\theta_g)\frac{x_1}{\gradlen}+\sin(\theta_g)\frac{x_2}{\gradlen}+1
  \end{aligned} 
\end{equation}
where the gradient length scale $ \gradlen = 1 \si{\metre}$.
The constant of proportionality is irrelevant in determining 
the magnitude of the logarithmic gradient:
\begin{equation*}
    \gradnorm = \frac{1}{\cos(\theta_g)x_1+\sin(\theta_g)x_2+\gradlen}
\end{equation*}
  We restrict our simulation times so that the  distance moved is not much more than a few $ \si{\mm}$, so really the concentration gradient is effectively constant, and we will report results in terms of the nondimensional taxis and kinesis strengths based on the nominal logarithmic concentration gradient value $ g \approx 1 \si{\metre}^{-1}.$   The only estimates for taxis and kinesis response strengths for colonial protists of which we are aware are some values fit to colonies of \emph{S. rosetta} in~\citet{aerotax}.  We will resort to using these values fit for the colonies to motivate, as very rough estimates, the response coefficients for cells in our model.  

We use the standard Euler-Marayama method with time step $ \Delta t = 0.1 \gamma_\Theta^{-1}$. 
In all simulations, we initialize the colony orientation $ \Thetacol (0) $ uniformly and independently across simulations, the center of mass of the colony at $ \Xcol (0) = \bm{0}$ and the flagellar angles at $ \Thetacelli (0)=0$.   The nondimensional rotational diffusion time scale from Eq.~\eqref{colony_orientation_nondim} is $ \sim 0.5 N^3$, and our simulations are generally run for nondimensional time $ \tilde{t}=50000$ which is at least 100 times as long for the range of colony sizes $ N\leq 10$ considered.  We also verified from statistical analysis of simulations over this time scale for the more challenging kinesis model that the colonies did typically make several rotations, and that differences in initial conditions did not significantly affect the computed drift.

The placement of the flagellar attachment points $ \alpha_j$ on the colony is simulated in two ways, following a demographic stochasticity model~\eqref{eq:unifmodel} in which each flagellar attachment point is displaced by a random distance $ \flagdispi $ uniformly distributed over an interval $ (-u,u) $ with $ 0 < u < l/2$ where $l$ is the length of the arc describing the exposed cell surface.  The first method simulates a population average by choosing the flagellar displacements independently in each simulation, thereby effectively simulating different colonies (of a specified size $N$).  The second method simulates an individual colony sample by generating just one set of the flagellar displacements $ \{\flagdispi\}_{i=1}^N$, which are held fixed for all simulations, effectively simulating different trajectories of a given colony. Under both approaches, we hold these flagellar attachment variables fixed when the colony behavior is compared, at fixed colony sizes, over different physical parameters.



\subsection{Taxis}\label{subsec:simulations_taxis}
\begin{figure}[H]    
{\begin{multicols}{2}
    \includegraphics[height=80mm,width=\linewidth]{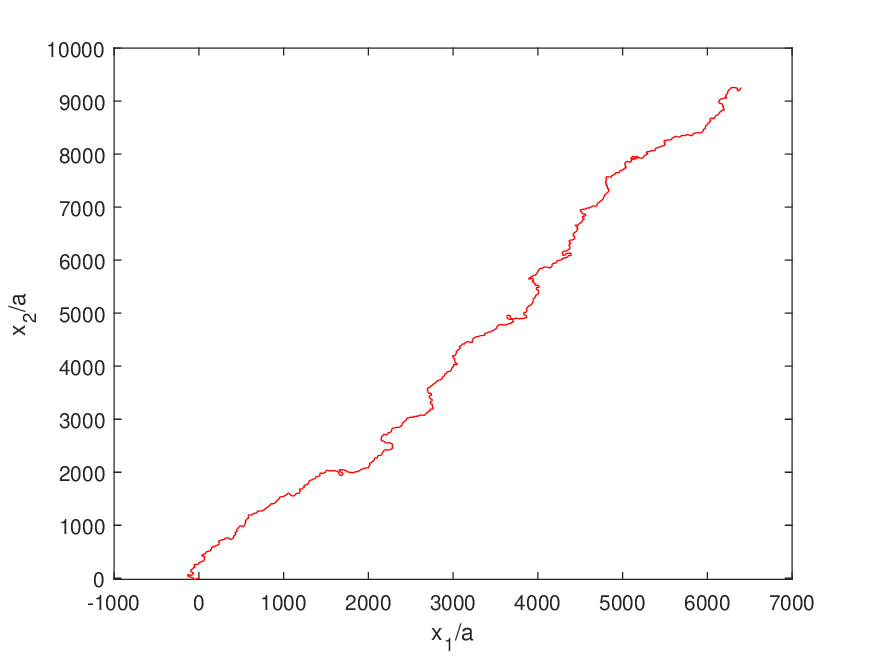}\par 
\includegraphics[height=80mm,width=\linewidth]{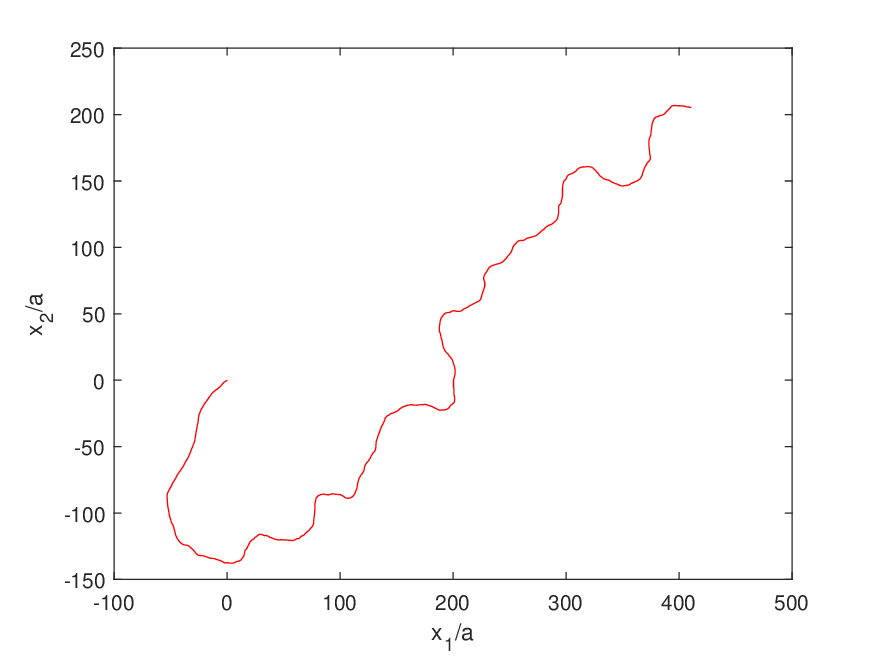}
\end{multicols}
\caption{Sample trajectories starting from $(0,0) $ of a colony of  $N=3$ (left) and $N=10$ (right) cells with each cell flagellum randomly displaced by $\flagdispi \sim U (-l/2,l/2)$ over nondimensional time $\tilde{t}=10^5$ with nondimensional taxis response factor $ m_T = 0.67 $ 
to the attractant gradient directed along  $\gradang=\frac{\pi}{4}$.}
\label{fig:taxis_trajectory}}
\end{figure}

First we plot 
in Figure~\ref{fig:taxis_trajectory} Monte Carlo simulations of two colonies exhibiting taxis going up a chemo-attractant gradient with an affine profile using relative (logarithmic) gradient sensing.  The fit reported in~\citet{aerotax} suggests  $ \ktaxis \gradnorm \gamma_{\Theta} \approx 0.3 \si{\sec}^{-1}$ in our model, so together with the biophysical parameter values reported in Table~\ref{tab:params}, we correspondingly take $ \ktaxisnd = \ktaxis \gradnorm \sigmaThe^{-1} =0.67$ in most simulations.  We generally report the taxis behavior in terms of the effective drift projected (with sign) along the attractant gradient.  We note that the theoretical prediction~\eqref{non-dim_taxis_drift} does not align exactly with the attractant gradient, but the cross-gradient drift component for the parameters explored has at most a couple percent of the drift along the attractant gradient for the vast majority (95\%) of the individual colonies sampled.

\label{fig:taxisTraj}

 \begin{figure}[H]    
 \centering{
    \begin{multicols}{2}   
    \includegraphics[scale=0.6]{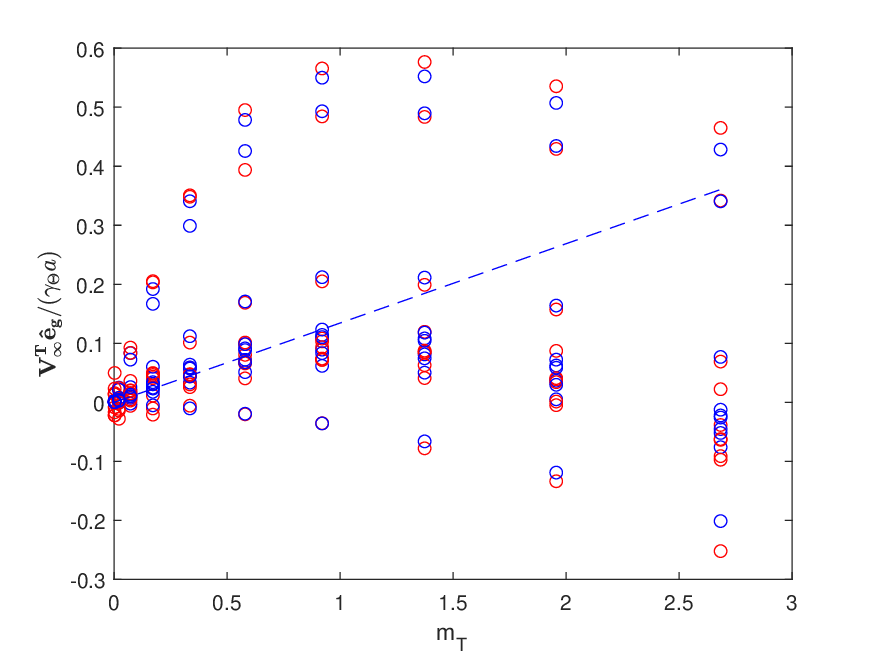}\par 
    \includegraphics[scale=0.6]{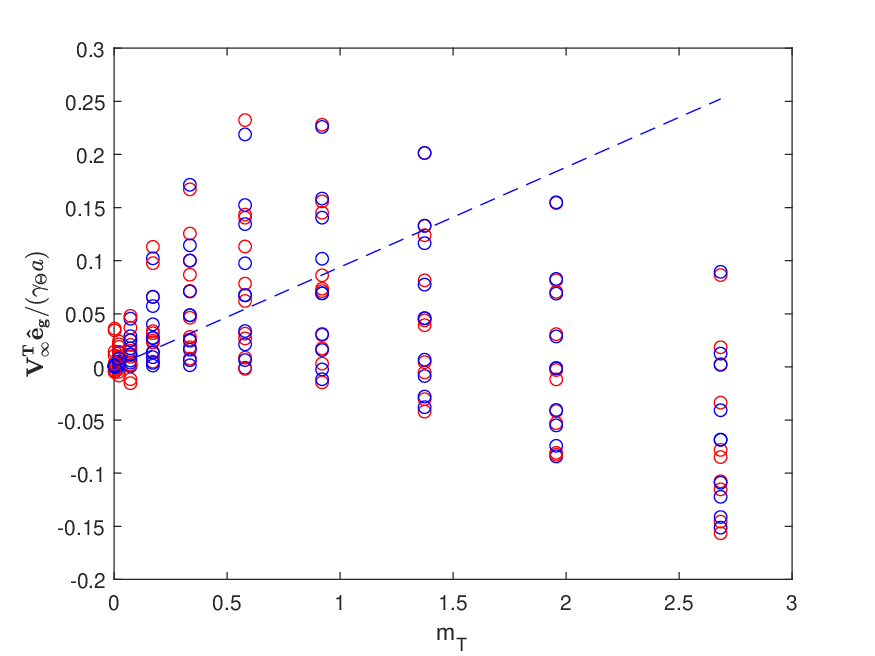} 
\end{multicols}}
\caption{ 
Projection of nondimensional effective drift $\drifteffnd = \drifteff/(\gamma_\Theta a) $ along attractant gradient $\graddir=(1,0)$ for colonies of $N=7 $ cells (left) and $N=10$ (right) cells exhibiting taxis with various values for the nondimensional taxis response factor $m_T$. The simulations consider ten cases of the flagella arrangements. The flagellar displacement variables are generated independently for each individual cell in independent colonies with  $\flagdispi \sim U (-l/2,l/2)$. The results of 10 Monte Carlo simulations for each indicated value of $m_T$ over nondimensional time $\tilde{t}=50000$ are plotted in red. The analytical predictions (\ref{non-dim_taxis_drift}) for each sample colony is plotted as blue circles, with the broken blue line being the demographic average~\eqref{eq:meantaxisdrift}. Parameter values are as listed in Table~\ref{tab:params}, with flagellar force magnitude $F= 5 \,\si{\pico\newton}$.
}
 \label{fig:taxis_responsefactor}
\end{figure}

Next we compare the parametric dependence of the theoretical predictions of colony taxis drift against numerical simulations.
We see in Figure~\ref{fig:taxis_responsefactor}  a linear dependence of the average effective drift up the attractant gradient at small  taxis sensitivity factors $m_T$, but then a deterioration of effective taxis and a trend toward motion down the gradient as the taxis sensitivity factor increases through order unity values.  This nonlinear dependence can be understood from the theoretical relation~\eqref{non-dim_taxis_drift}, in which the term giving rise to productive motion of the colony up the gradient is multiplied by the ratio of modified Bessel functions, while the term corresponding to the perverse drift down the gradient is linear in the taxis sensitivity factor $ \ktaxisnd$. The ratio of modified Bessel functions is approximately linear for small values of the argument, but approaches unity as its argument (proportional to $ \ktaxisnd$) moves through order unity values. We see individual colonies (with particularly diverse values of their asymmetry statistics $ \geotaxamp$) have a wide distribution of drift speeds about the demographic mean, with good quantitative agreement over all values of $ \ktaxisnd $ between the distribution of drifts computed analytically from Eq.~\eqref{non-dim_taxis_drift} and the direct numerical simulations.  In particular, we see that some colonies at all values of $ \ktaxisnd >0$, particularly larger values, will drift down the gradient rather than up it.  The theoretical demographic average speed up the attractant gradient~\eqref{eq:meantaxisdrift}, plotted as a dashed line, was derived under a first order expansion with respect to the taxis response factor $ \ktaxisnd$ and consequently only works well for $ \ktaxisnd \lesssim 1$.  
The conditions~\eqref{eq:nondim_taxis_cond} for the validity of our asymptotic theory would read, for the parameter choices in Figure~\ref{fig:taxis_responsefactor}:  $ N \gg (\geotaxamp \ktaxisnd)^{2/5}, 2^{1/3}$, which are plausibly satisfied for the values of $N$ and $ \ktaxisnd$ depicted.


Next, we verify the relationship found in (\ref{eff_taxis_drift}) between colony population size and effective drift along  the attractant gradient with Monte Carlo simulations. Here we see that the asymptotic results agree reasonably well with the simulations for colonies of all sizes, and that the taxis becomes rather ineffective at large colony sizes.
 \begin{figure}[H]   
 \centering
  \begin{multicols}{2}   
\includegraphics[scale=0.6]
{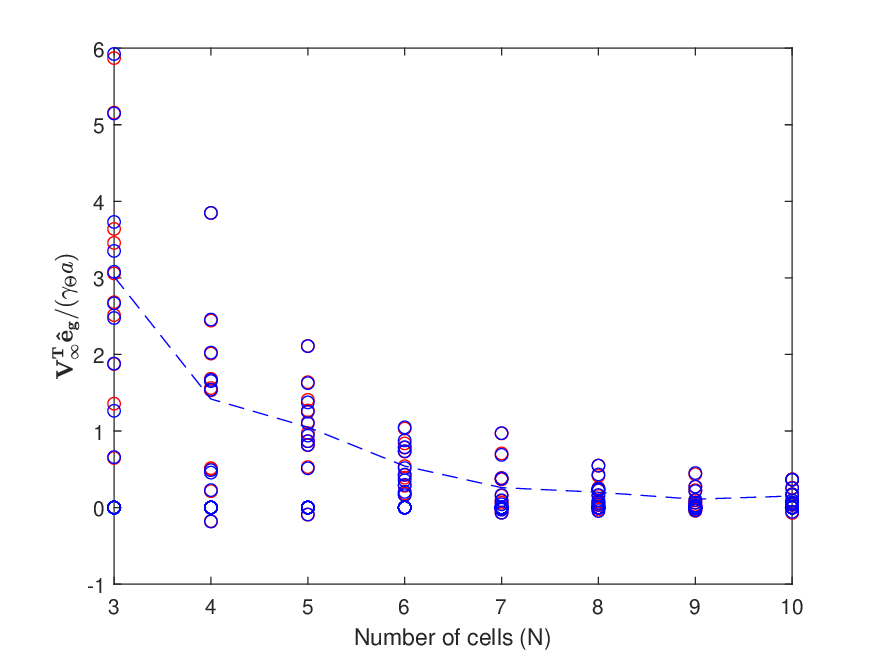} \par     
\includegraphics[scale=0.6]{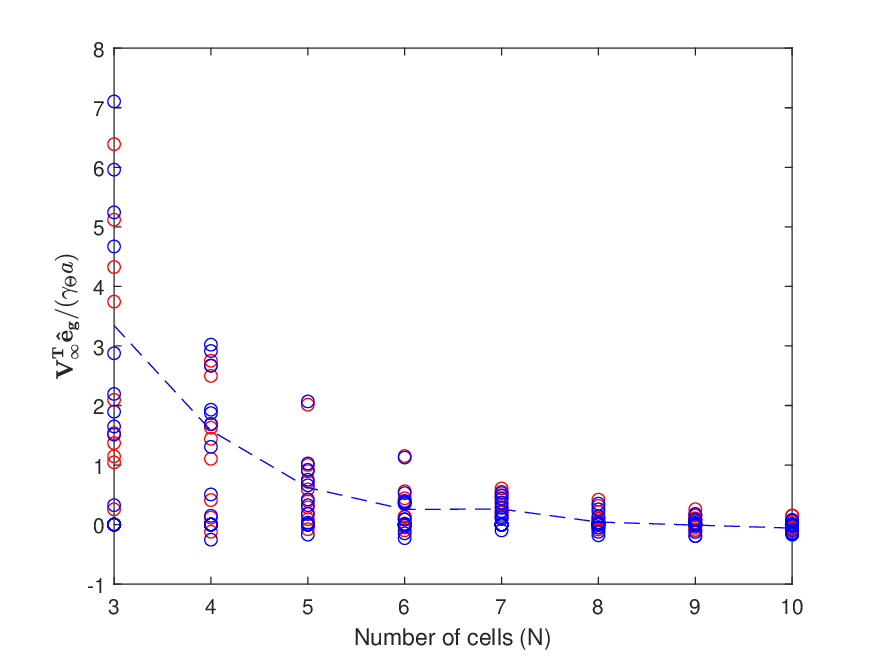}
\end{multicols}
\caption{Projection of the nondimensional effective drift $\drifteffnd = \drifteff/(\gamma_\Theta a) $ along attractant gradient $\graddir=(1,0)$ for colonies with nondimensional taxis response factor $m_T=1$ (left) and $m_T=2$ (right) exhibiting taxis with various population size $N$. The placement variables are generated independently for each individual cells in each colony with the placement variables randomly displaced by $\flagdispi \sim U (-l/2,l/2)$.  The results of 10 Monte Carlo simulations based on 10 different colonies for each indicated value of $N$ over nondimensional time $\tilde{t}=50000$ are plotted in red circles and the corresponding analytical results~\eqref{eff_taxis_drift} are plotted in blue circles. The analytical theory's demographic average (\ref{non-dim_taxis_drift})  is plotted as broken blue curves. Parameter values are as listed in Table~\ref{tab:params}, with flagellar force magnitude $F= 5 \,\si{\pico\newton}$.
}
\label{fig:taxis_popnsize}
\end{figure}


\subsection{Kinesis}  \label{subsec:simulations_kinesis}





\begin{figure}[H]    
{\centering{
\includegraphics[height=80mm,width=120mm]{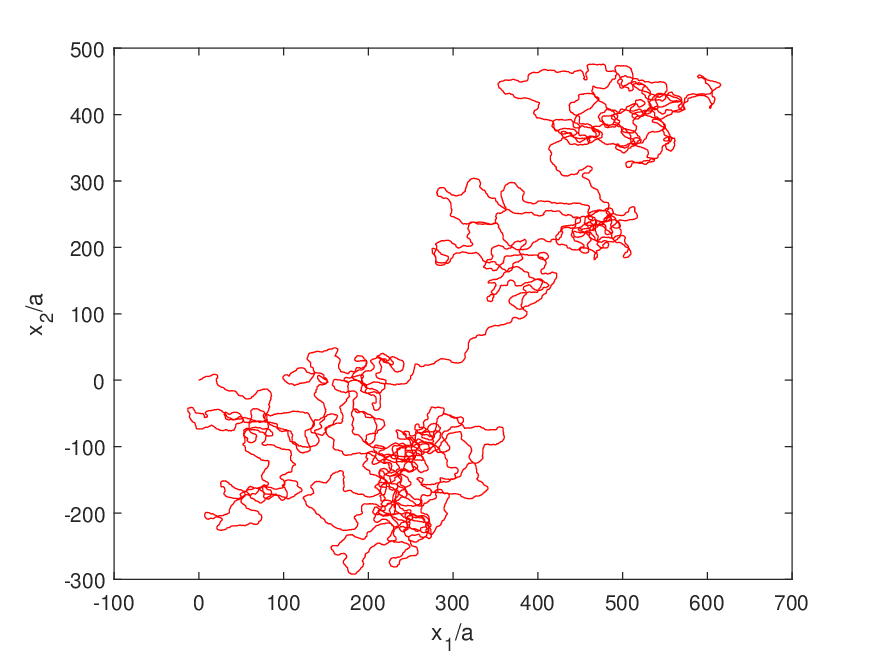}}\caption{Sample trajectory starting from $(0,0)$ of a colony of $N=7$ with each cell flagellum randomly displaced by $\flagdispi \sim U (-l/2,l/2)$ over nondimensional time $\tilde{t}=10^5$ with nondimensional kinesis response factor $ \kkinnd = 0.55 $ to the attractant gradient directed along  $\gradang=\frac{\pi}{4}$.    The flagellar force is taken as $ F=\SI{5}{\pico\newton}$.
    The other parameters are as specified in Table~\ref{tab:params}.}\label{fig:kinesis_trajectory}}
\end{figure}

\begin{figure}[H]   
\begin{multicols}{2}
    \includegraphics[scale=0.6]{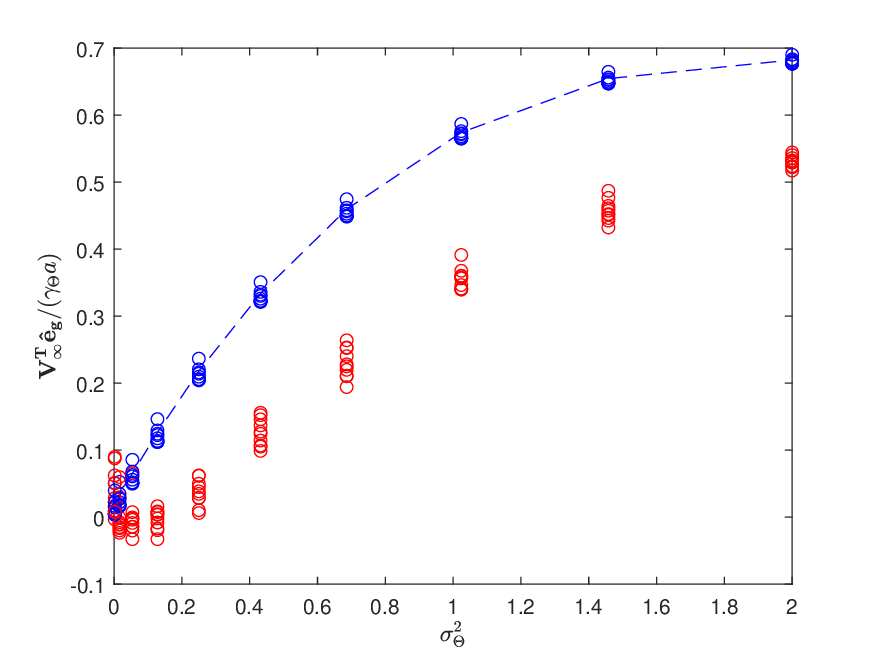} 
    \includegraphics[scale=0.6]{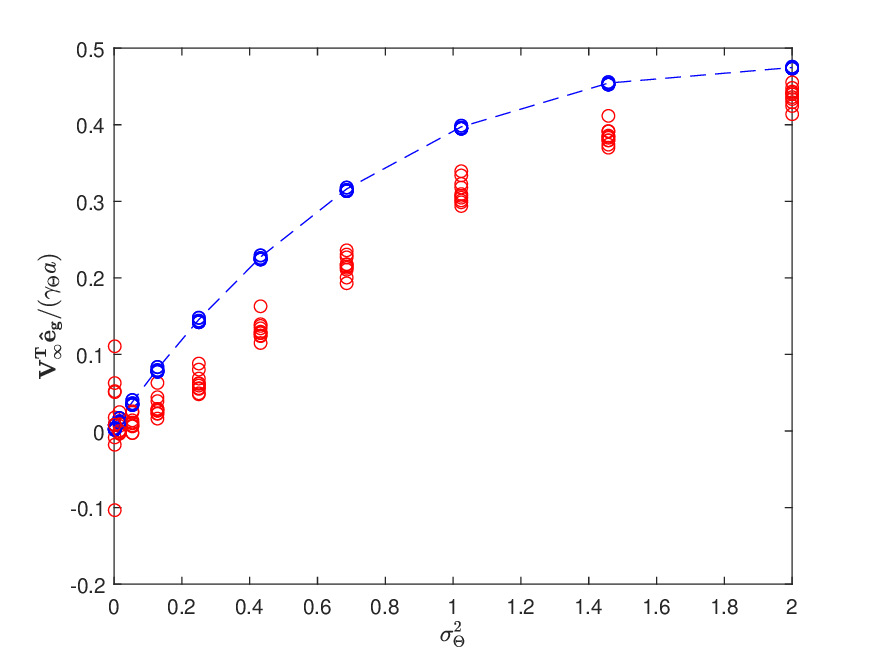} 
\end{multicols}
\caption{
Projection of nondimensional effective drift $\drifteffnd = \drifteff/(\gamma_\Theta a) $ along attractant gradient $ \graddir$
for colonies of $N=7 $ cells (left panel) and $N=10$ cells (right panel) exhibiting kinesis with various flagellar force variances $\sigma_\Theta^2 $.  The nondimensional kinesis response factor is chosen as $ m_K= 0.55$. 
Simulation results over nondimensional time $\tilde{t}=50000$  for  $10$ colonies with each cell flagellum randomly displaced by $\flagdispi \sim U (-l/2,l/2)$ are plotted in red. The analytical predictions (\ref{non-dim_kinesis_drift}) for each sample colony is plotted as blue circles, with the broken blue line being the demographic average~\eqref{eff_kinesis_drift}. 
 Parameter values are as listed in Table~\ref{tab:params}, with flagellar force magnitude $F= 5 \,\si{\pico\newton}$.  
}
\label{fig:kinesis_variancea}
\end{figure}

First, we show in Figure~\ref{fig:kinesis_trajectory} 
the simulated trajectory of a colony reacting by kinesis to an environmental gradient directed along $ \gradang=\pi/4$.  We see a kinesis drift up the gradient, though with noisy excursions.
Next, we numerically verify in Figure~\ref{fig:kinesis_variancea}  the analytical relationship~\eqref{non-dim_kinesis_drift} between the rate of progress up the attractant gradient and the flagellar force angle fluctuation magnitude $ \sigmaThe$.  Here we explore a range of $ \sigmaThe $ that are in fact much larger than our biophysical estimate in Table~\ref{tab:params}, to test the theory more broadly.  As  $\beta^2 \sim 0.0004 N/\sigmaThe^2 $ for the parameter values chosen here, the geometry-dependent term $ \geotaxamp^2/(N\derat^2 + N^2) \approx \geotaxamp^2/N^2$ is generally small compared to the term $ \sigma_{\Theta}^2 N/4$ in Eq.~\eqref{non-dim_kinesis_drift} over most of the range of values of $ \sigmaThe $ presented.  Thus, the effective drift should be primarily determined  by the size of the colony, not its detailed geometric arrangement (summarized in $ \geotaxamp$), which explains the small demographic variability for most of the values of $ \sigmaThe^2 $ considered in Figure~\ref{fig:kinesis_variancea}.

We see generally that the effectiveness of the kinesis of the colony rises linearly over sufficiently small $ \sigmaThe^2$ but eventually turns over to decay at larger values.  
From Table~\ref{tab:params},  $\sigma_\Theta^2 \sim 0.002$ for protozoa like Choanoflagellates, which is far below the optimal value for colonial kinesis. Greater capacity for variability in the flagellar angle with the cell body could significantly enhance the kinesis of the colony.  We see general agreement between the asymptotic theory and the simulations, except for a quantitative but not qualitative overprediction at colony sizes $ N= 7$.  This can be understood from the self-consistency condition~\eqref{eq:nvalkin} for our asymptotic analysis, which in particular requires $ N \gg \sigmaThe^{2/3} \zeta^{2/3}$.  For the parameter values used in Figure~\ref{fig:kinesis_variancea}, $ \zeta \approx 40$ so $ N \gg 10 \sigmaThe^{2/3}$ would be required for self-consistency of the theory.  This is easily satisfied for the biophysical value of $ \sigmaThe $ from Table~\ref{tab:params} but not so well over the wider range of $ \sigmaThe$ considered in Figure~\ref{fig:kinesis_variancea}.  We do note significant improvement of agreement with theory at $ N=10$  cells even though the self-consistency condition is still not well-satisfied.

 \begin{figure}[H]    
 \centering 
\begin{multicols}{2}
  \includegraphics[scale=0.35]{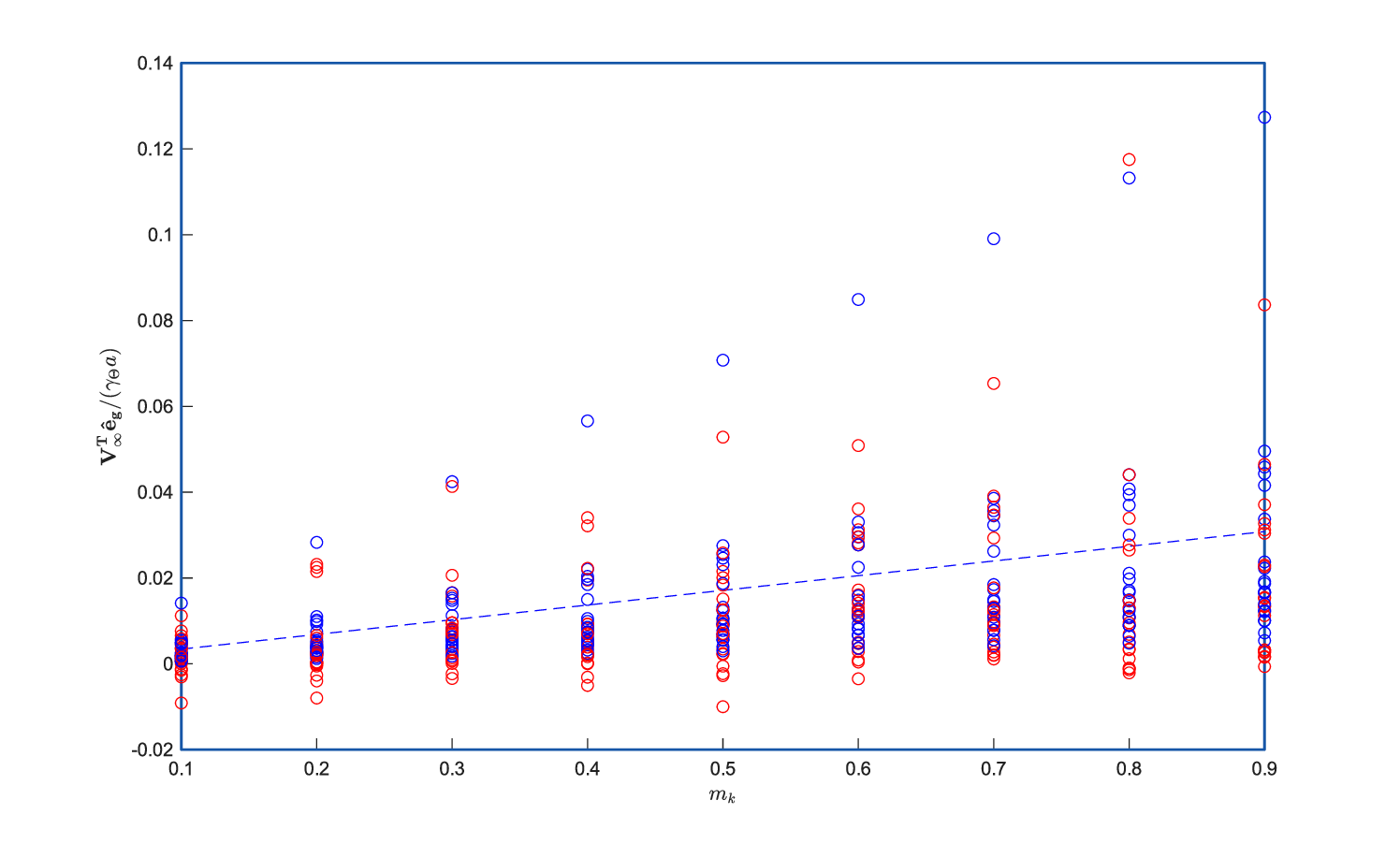}
  \includegraphics[scale=0.35]{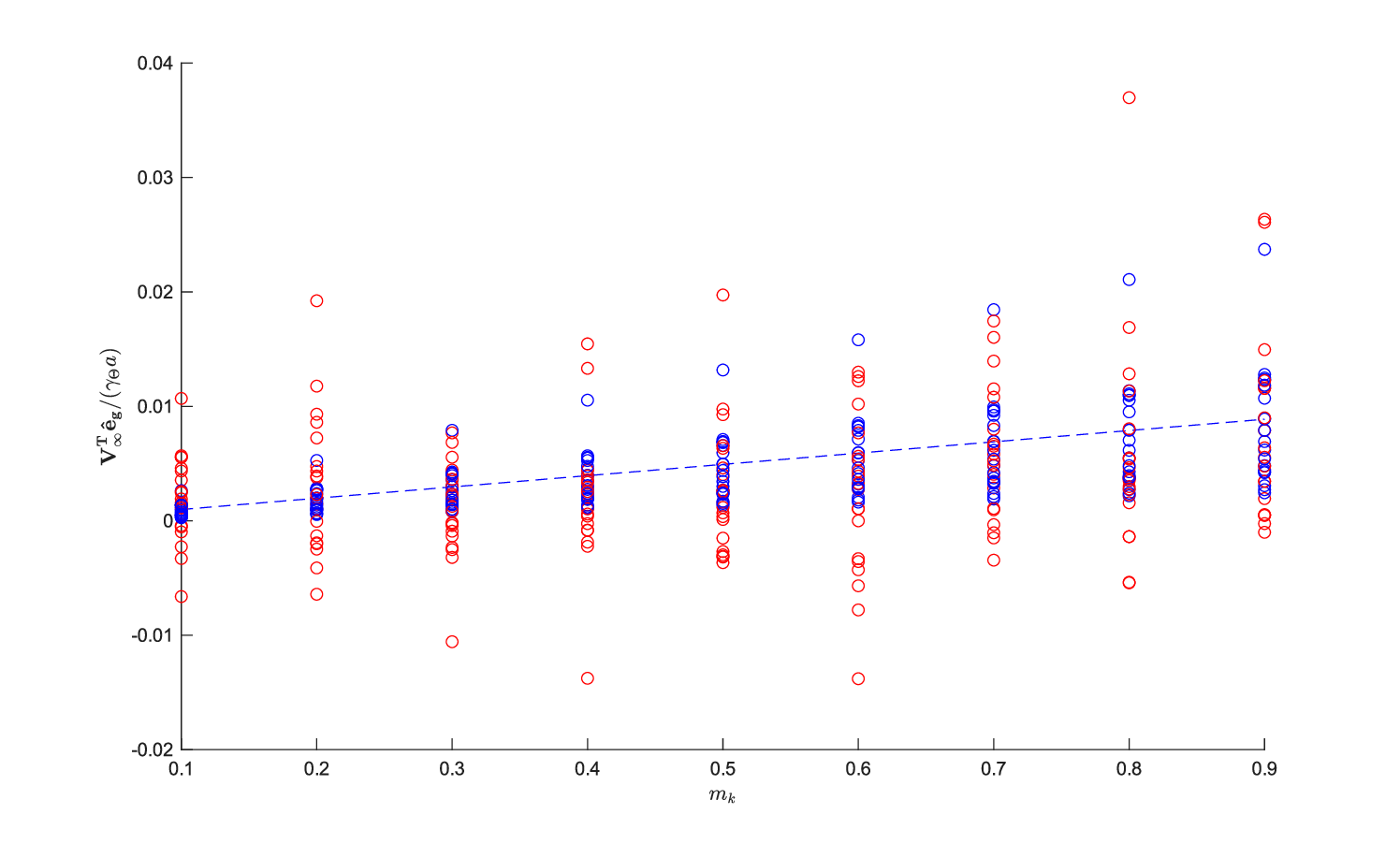}
\end{multicols}
\caption{
Projection of 
nondimensional effective drift $\drifteffnd = \drifteff/(\gamma_\Theta a) $ along attractant gradient $\graddir$
for colonies of $N=7 $ cells (left panel) and $N=10$ cells (right panel) exhibiting kinesis with various nondimensional kinesis response factors $m_K$. The placement variables are generated independently for each individual cell in independent colonies with the placement variables randomly displaced by $\flagdispi \sim U (-l/2,l/2)$. The results of 20 Monte Carlo simulations for each indicated value of $m_K$ over nondimensional time $\tilde{t}=500000$ are plotted in red circles.    The analytical predictions (\ref{non-dim_kinesis_drift}) for each sample colony is plotted as blue circles, with the broken blue line being the demographic average~\eqref{avgdrift_kinesis}. 
 Parameter values are as listed in Table~\ref{tab:params}, with flagellar force magnitude $F= 5 \,\si{\pico\newton}$.}
 \label{fig:kinesis_responsefactor}
\end{figure}

\begin{figure}[H]  
 \centering
    \includegraphics[scale=0.6]{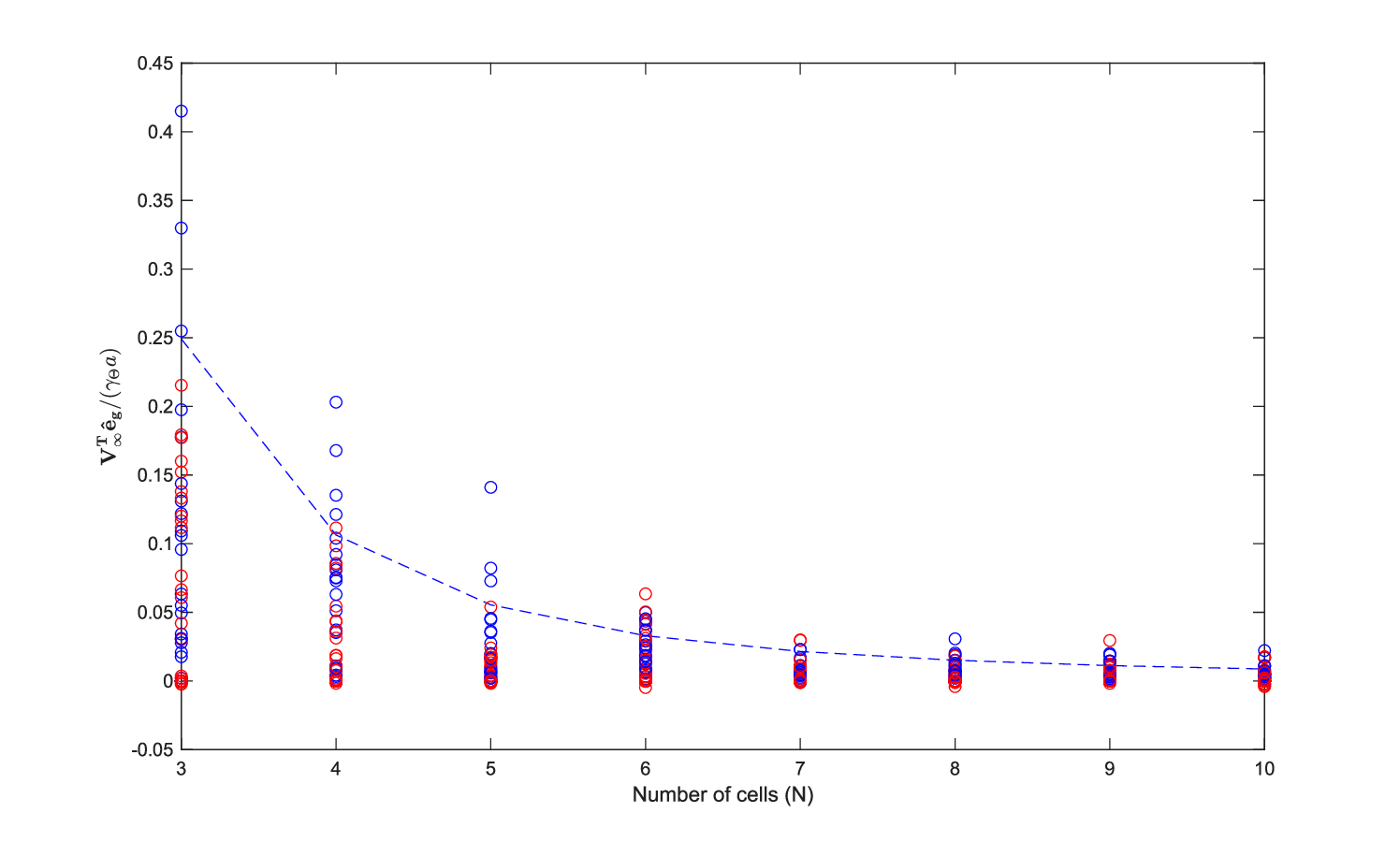} 
\caption{
Projection of nondimensional effective drift $\drifteffnd = \drifteff/(\gamma_\Theta a) $
of colonies of $m_K=0.55$ cells 
exhibiting kinesis with various population sizes $N$. The placement variables are generated independently for each individual cells in independent colonies with the placement variables randomly displaced by $\flagdispi \sim U (-l/2,l/2)$. The results of 20 Monte Carlo simulations over nondimensional time $\tilde{t}=500000$
are plotted in red circles. The analytical predictions (\ref{non-dim_kinesis_drift}) for each sample colony is plotted as blue circles, with the broken blue line being the demographic average~\eqref{avgdrift_kinesis}.  Parameter values are as listed in Table~\ref{tab:params}, with flagellar force magnitude $F= 5 \,\si{\pico\newton}$.
}
\label{fig:kinesis_popnsize}
\end{figure}

In Figure~\ref{fig:kinesis_responsefactor},
we numerically verify the linear proportionality between the non-dimensional kinesis response factor ($m_K$) and the effective speed along the environmental gradient.  While this linearity was derived under the assumption that $\kkinnd \ll 1$, we see it holds reasonably well as a general trend over the whole range of admissible $ 0\leq \kkinnd \leq 1$.  
Next, we verify in Figure~\ref{fig:kinesis_popnsize} the relationship found in~\eqref{non-dim_kinesis_drift} between colony population size and effective drift. Here we see that the asymptotic results agree with the simulations 
and that the non-dimensional kinesis performance ($V_\infty/ (\gamma_\Theta a)$ along the attractant gradient) has a simple inverse relationship with colony size. As the colony radius $ a \propto N$, this is consistent with the dimensional drift $ V_{\infty}$ being approximately independent of colony size.
The theoretical results were derived under the restriction on colony size $N \gg (\zeta \delta^2 + \sigmaThe^2 \zeta^2)^{1/3}$ from Eq.~\eqref{eq:nvalkin}, which implies $N \gg 1.4$ for $F=5 \,\si{\pico\newton}$.  For $ 3 \leq N \leq 5$ we see the simulations have typically smaller drift than the theoretical predictions, but for $ N\geq 6$ the simulations and theoretical predictions are in reasonable agreement.

We remark that the kinesis simulations in Figures~\ref{fig:kinesis_responsefactor} and~\ref{fig:kinesis_popnsize} were run for 10 times longer than the taxis simulations were run in Figures~\ref{fig:taxis_responsefactor} and~\ref{fig:taxis_popnsize}.  In Appendix~\ref{sec:appendix:homogenization} we report the results of kinesis simulations run for the same nondimensional time $ \tilde{t}=50000$ as the taxis simulations; these show good agreement of the means but poor agreement between individual colonies.  This discrepancy can be traced to the empirical distribution of colony orientation not having relaxed adequately to the theoretical stationary distribution.  Better convergence was evident when simulations were run for $ \tilde{t}=500000$, though we see the agreement between the simulated and theoretical values for each colony is less good than for the taxis simulations in Figures~\ref{fig:taxis_responsefactor} and~\ref{fig:taxis_popnsize}.  The slower convergence over time for the kinesis simulations can be understood from the theoretical developments in Subsections~\ref{subsec:analysis_taxis} and~\ref{subsec:analysis_kinesis}, which show that the orientational dynamics take place on time scale $ \epsilon^{-1}$ for taxis and $ \epsilon^{-2}$ for kinesis.

\section{Discussion}  \label{sec:conclusion}
The question of the  performance of colonies of protozoan cells in life functions relative to the individual cells is naturally connected toward understanding evolution.  On one level, a cost-benefit analysis could help explain why some protozoan cells have evolved colony forming behavior either through aggregation or incomplete division~\citep{KoehlSelective,SolariAllometric2013,SolariHydrodynamics2006}.  More profoundly, organisms such as choanoflagellates and life functions such as stimulus response provide a means for exploring how and why multicellular organisms may have evolved out of unicellular organisms in apparently multiple independent events~\citep{King-urmetazoa}. Such analysis also aids in understanding the navigational possibilities of synthetic assemblies of microswimmers. By working out the multiscale asymptotic analysis and simulations of the stochastic models for colonial taxis and kinesis in Section~\ref{sec:model} we aimed to address the question of the consequence of colony formation for stimulus driven movement among relative gradient sensing eukaryotic cells. In particular, we were interested in seeing the relationship between colony size and the ability to properly react to directional stimulus. We  find that for slowly varying attractant profiles in an effectively two-dimensional medium, the effectiveness of our taxis model decreases inversely with colony size while the effectiveness of our kinesis model for large colonies approaches a positive value independent of colony size and geometry.  The above findings that kinesis is not thwarted by colonial aggregration are consistent with an experimental study comparing the prevalence of taxis and kinesis in colonial Choanoflagellates showing a clear pervasiveness of kinesis over taxis \citep{aerotax}.


For both our taxis and kinesis models, the effective rotational dynamics of the colony did take a structurally similar form to the individual cell models.  Under taxis, the colony tended to steer so its nominal swimming direction is up the stimulus gradient.  Under kinesis, the colony's effective rotational diffusion increased as the nominal swimming direction of the colony deviated from being up the stimulus gradient. The nominal swimming direction here is the direction the colony would travel when all flagella are in their normal positions, and is only defined when the flagella have some asymmetric displacement.  Moreover, the strength of the effective taxis and kinesis on the colony orientation is proportional to an asymmetry measure (Subsection~\ref{sec:asymmetry}) and the inverse square root of the colony size.  Despite the relatively straightforward coarse-graining of taxis and kinesis responses at the cell level to the colony orientation, the effect on the translational dynamics is more subtle due to the translational forces applied by the flagella whose orientation is influenced by the orientation of its associated cell, not the colony as a whole.  

For our taxis model (Subsection~\ref{subsec:analysis_taxis}), we found that the steering mechanisms beneficial for cells to orient themselves up a stimulus gradient can produce counterproductive drift down the stimulus gradient for the colony when the asymmetry measure is low, the colony size is large, and/or the individual taxis response is large (see Figure~\ref{fig:taxis_responsefactor}).   For the kinesis model (Subsection~\ref{subsec:analysis_kinesis}), the effective colony drift is always up the gradient even for symmetric colonies, with asymmetry improving the drift though with less effect for large colony sizes.  The shape-independent component of the colony drift arises from the flagella of adversely oriented cells fluctuating more due to the kinesis response and thus averaging our their directed force on the colony more than those oriented more favorably along the stimulus gradient. Furthermore, the drift along environmental gradients has a typically monotone positive relationship with the kinesis response factor. 
In our study then, colonies displaying kinesis are able to move effectively up environmental gradients, indicating that colony formation does not thwart their effective movement.


In a quite distinct model for multicellular taxis, \citet{Colizzi-Pottsmodel} emphasized the importance of the persistence of the direction of motion relative to the instantaneous speed in determining the overall success of the colonies in navigating up the environmental gradient.  Our modeling framework directly represents the persistence of motion through the dynamics of the colony orientation $ \Thetacol$ in Eq.~\eqref{Colony_orientationb}, and the effective colonial taxis and kinesis drift rates account for the effects of the colonial rotation.    \citet{vuijk2021carrying} considered the effects of chaining simple active Brownian particles together or to passive cargo to induce effective taxis behavior, meaning in their context the particles spending more times in regions where their swimming speed is higher (due to perhaps excitation by a stimulus).  The framework of the present work could be readily adapted to explore effective taxis of rigid colonial assemblies of such active Brownian particles.



The ``emergent chemotaxis'' (EC) model studied in~\citet{MuglerChemotaxis}, while focused  on sensing effectiveness of a cellular cluster, has some formal mathematical similarities to our microswimming colony model.  While the cells in the EC model don't have flagella, they are assumed to be polarized in an outwardly normal direction, and the colony velocity is modeled as proportional to the vectorial sum of the cell polarizations.  This much is mathematically identical to the swimming colony velocity being determined by the vectorial sum of the flagellar forces, which are approximately normal, with stochastic fluctuations, to the colony surface.  The flagellar forces are directed inward rather than outward, but this does not make a big difference.  In both models we have
a cancellation effect of the influence of cells on opposite sides of the cluster.  Also, in both models, the cellular responses have noisy dynamical fluctuations -- due to concentration fluctuations in the EC model, and due to  flagellar  fluctuations in our swimming colony model.
The larger difference is that, in the EC model, the cancellation is largely mitigated because the cells polarize in proportion to the measured concentration (not gradient) so the polarization of the cells will be systematically stronger along the edges facing up the environmental gradient.  The analog in our swimming colony model would be if the propulsive forces $ F_i (t) $ from each cell varied substantially depending on how the cell faces the environmental gradient.  We are not aware of evidence, at least in choanoflagellates, for such observed dynamical variability in the flagellar force, and following~\citet{aerotax}, we modeled taxis and kinesis response in terms of a steering response of the flagellar orientation.  This makes the breaking of the force cancellation more subtle.

To focus on our main technical consideration of coarse-graining cellular flagellar dynamics to the colony scale, we have taken minimal models for the taxis and kinesis response. More detailed mathematical models of taxis and kinesis are of course available (for example~\citep{Othmer}), and a question for future work would be to what extent aspects of the more detailed flagellar dynamics would influence the colony-scale dynamics under our prevailing assumption of separation of flagellar and colony dynamical time scales.  We have also simplified our swimming colony model from~\citet{Ashenafi-mobility} by neglecting demographic stochasticity in the orientation and magnitude of the flagellar forces.  These effects could be readily included at the cost of a more elaborate calculation, whose most important effect would be incorporating mean rotation of the colony due to flagellar torque imbalances~\citep{Ashenafi-mobility}.  Taxis under rotation has been found in previous theoretical work to be efficient in locating favorable environments~\citep{Larson2023Protists}.
Another compelling direction for extension would be the inclusion of correlations between the flagella of different cells, in an effort to model phototaxis of larger volvocine green algae colonies, which exhibit some metachronal phase synchronization of the flagellar beating  but no apparent coordination of taxis response across cells
~\citep{GoldsteinVolvocine,BrumleyVolvox}. 

Extending the asymptotic analysis presented here to a three-dimensional polyhedral version of the model would have to contend with considerably more technical complexity with regard to the stochastic rotational dynamics.  We would expect the general aspects of our conclusions should carry over.  Because the colony radius $ \colrad $ would scale more like $ \sqrt{N} $ in three dimensions rather than $ N$ in our two-dimensional model, the criteria on minimal colony size described in Section~\ref{sec:analysis} for the validity of our analysis, based on the rotational drag being large enough to induce sufficiently slow colony rotation, would become somewhat more stringent in three dimensions. \citet{MuglerChemotaxis} do find a qualitative difference in the relative efficiency of their sensing models in two or three dimensions, due largely to how the spatial distribution of cells responds to negative spatial correlations in the concentration of the environmental field.  No such dynamical correlation effect would seem to be present in our taxis and kinesis models that would suggest qualitatively different behavior in three dimensions. 

\section{Acknowledgments}
The authors would like to thank Julius Kirkegaard for crucially helpful discussions in the formulation stage of this work.

\appendix

\section{Effective Noise Amplitude of Colony Dynamics in Taxis Model} \label{sec:appendix:homogenization_taxis} 

The general central limit theorem~\citep{BouchetLDFS2016} for describing stochastic fluctuations in fast-slow systems about the averaged behavior of the slow variable would state that we can represent $ \Thetacol (\tilde{t}) $
for $\tilde{t} \sim O(\epsilon^{-1}) $ time scales as $ \Thetacol (\tilde{t}) = \Thetacolavg (\tilde{t}) + \sqrt{\epsilon} \Thetacolfluc (\tilde{t}) $ with 
\begin{equation}
\difd \Thetacolfluc (\tilde{t}) =
B(\Thetacolavg(\tilde{t})\Thetacolfluc (\hat{t}) \, \difd \tilde{t}+ \eta (\Thetacolavg (\tilde{t})) \, \difd W (\tilde{t}) +\sqrt{2 \epsilon} \derat \, dW^{\Theta,c}(\tilde{t}) \label{eq:thetafluc_clt}
\end{equation} 
for a suitable gain function $ B(\thetacol)$ and effective noise coefficient $ \eta (\thetacol)$. 
General formulas for these coefficients are derived and presented in~\citet{BouchetLDFS2016}, but it's more transparent for our system to obtain them by direct manipulation of the fast-slow system~\eqref{eg:nondimensional_rescaled_taxis_model} and~\eqref{eg:nondimensional_rescaled_colony_model}.  First, we can express the flagellar angle orientations as integrals over the history of the colony orientation:
\begin{equation}
\Thetacellndi (\tilde{t}) = \expe^{-\tilde{t}} \Thetacellndi (0) - \int_{0}^{\tilde{t}} \expe^{-(\tilde{t}-\tilde{s})} \ktaxisnd 
\sin[\Thetacol (\tilde{s})+\alpha_{i}-\gradang]) \, \difd \tilde{s}
+ \sqrt{2} W_i^{\Theta} (\tilde{t}) 
\label{eq:stochintcells}
 \end{equation}
Recalling that the dynamics of the quantities associated with the colony rotation, $\Thetacolavg (\tilde{t})$ and $ \Thetacolfluc (\tilde{t})$, evolve on the slower time scale $ \epsilon^{-1}$, the flagellar orientations will  reach on an $ O(1) $ time scale a quasi-stationary distribution governed by these slow variables:
\begin{equation}
\Theta_i (\tilde{t}) \sim N\left(-\ktaxisnd\sin(\Theta^{(c)}(\tilde{t})+\alpha_i-\gradang),1\right) \label{eq:flagang_fastdist}
\end{equation}
Averaging the drift term in Eq.~\eqref{eg:nondimensional_rescaled_colony_model} using these quasi-stationary distributions, applying the same condensation of the trigonometric sum as in Eq.~\eqref{eq:trig_condense}, and substituting the representation $ \Theta^{(c)} (\tilde{t})= \Thetacolavg (\tilde{t}) + \sqrt{\epsilon} \Thetacolfluc (\tilde{t})$, and expanding to first order gives the gain function as just first the first order Taylor coefficient of the averaged drift in Eq.~\eqref{taxis_thetac_avg2} with respect to the colony orientation $ \Thetacolavg $,
which moreover agrees with the calculation from the rigorous and more cumbersome expression in~\citet{BouchetLDFS2016}.  

\begin{figure}[H]
    \centering
    \includegraphics{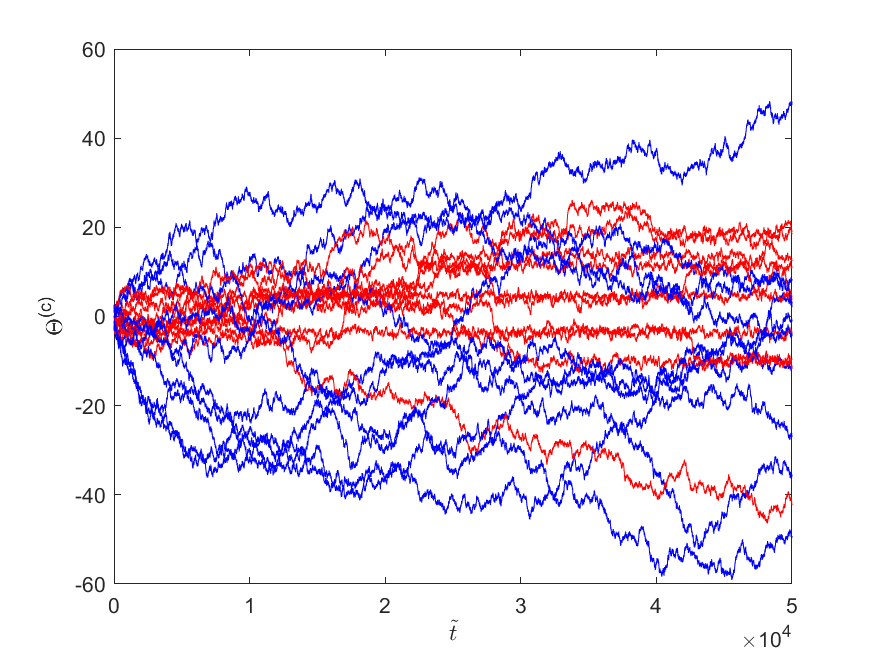}
    \caption{Comparison of the colony orientation $ \Thetacol$ as a function of nondimensional time $ \tilde{t}$  from direct simulations of the model~\eqref{eg:nondimensional_rescaled_colony_model} (red) and from the reduced effective dynamics~\eqref{eq:thetacol_clt} (blue).  These simulations were conducted using our standard parameter choices from Table~\ref{tab:params} for a colony of 10 cells with nondimensional taxis factor $m_T=0.3$.} \label{fig:thccomps}
\end{figure}

The effective noise amplitude $ \eta (\thetacol) $ induced on the colony orientation from the stochastic fluctuations of the flagellar angles is obtained in terms of the stationary autocorrelation function $ \Cordrift (\tilde{\tau};\thetacol) $ of the drift term on the slow colony orientation variable in Eq.~\eqref{eg:nondimensional_rescaled_colony_model} when this slow variable is held constant at $ \Thetacol (\tilde{t}) = \thetacol$:
\begin{equation*}
\epsilon \eta^2(\thetacol) = 2 \int_{0}^{\infty} \Cordrift (\tilde{\tau};\thetacol)\,\difd \tilde{\tau}.
\end{equation*}
The flagellar angles $ \Thetacellndi (\tilde{t}) $ then act as independent Gaussian random variables with marginal probability distributions given by Eq.~\eqref{eq:flagang_fastdist} with $ \Thetacol (\tilde{t}) $ replaced by $ \thetacol$, and the stationary autocorrelation function of these angles is readily found from Eq.~\eqref{eq:stochintcells} to be:
\begin{equation*}
\Cov (\Thetacellndi (\tilde{t}),\Thetacellndi (\tilde{t}+\tilde{\tau}) = \expe^{-\tilde{\tau}}.
\end{equation*}
Then via standard averaging of trigonometric functions of Gaussian random variables, we obtain:
\begin{align*}
\Cordrift (\tilde{\tau};\thetacol) &= 
\epsilon^2 \sigmaThe^{-2} \sum_{i=1}^N 
\frac{1}{2}
\left[-\cos (2 \ktaxisnd \sigmaThe \sin(\thetacol + \alpha_i-\gradang))\expe^{-\sigmaThe^2(1+\expe^{-\tilde{\tau}})}
+\expe^{-\sigmaThe^2(1-\expe^{-\tilde{\tau})}}\right. \\
& \left.\qquad \qquad - 2\sin^2 (\sigmaThe \ktaxisnd \sin(\thetacol + \alpha_i-\gradang))\expe^{-\sigmaThe^2}\right] \\
& \approx \frac{\epsilon^2}{2} \expe^{-\sigmaThe^2}\sum_{i=1}^N (1+\cos (2 \ktaxisnd \sigmaThe \sin(\thetacol + \alpha_i-\gradang)))\expe^{-\tilde{\tau}} \approx N \epsilon^2 \expe^{-\tilde{\tau}}
\end{align*}
so $ \eta (\thetacol) = \sqrt{2N\epsilon}  $. We summarize the calculations in this appendix compactly via the single nonlinear stochastic differential equation~\eqref{eq:thetacol_clt} rather than via decomposition into the solution of the averaged equation~\eqref{colony_orientation_avg} and a linearized fluctuation dynamics~\eqref{eq:thetafluc_clt}.  The descriptions are equivalent to leading order in the fluctuation magnitude.
In Figure~\ref{fig:thccomps} we show that the reduced description~\eqref{eq:thetacol_clt} of the sample trajectories of the colony orientation $ \Thetacol (\tilde{t}) $ visually resemble those from direct simulation of our full model.  Just using the averaged dynamics~\eqref{colony_orientation_avg} would clearly be inadequate as the stochastic fluctuations are prominent, though the asymptotic analysis treats them as formally small.

\section{Coarse-Graining of Colony Dynamics in Kinesis Model} \label{sec:appendix:homogenization}
\begin{figure}[H]
    \centering
    \includegraphics[scale=0.6]{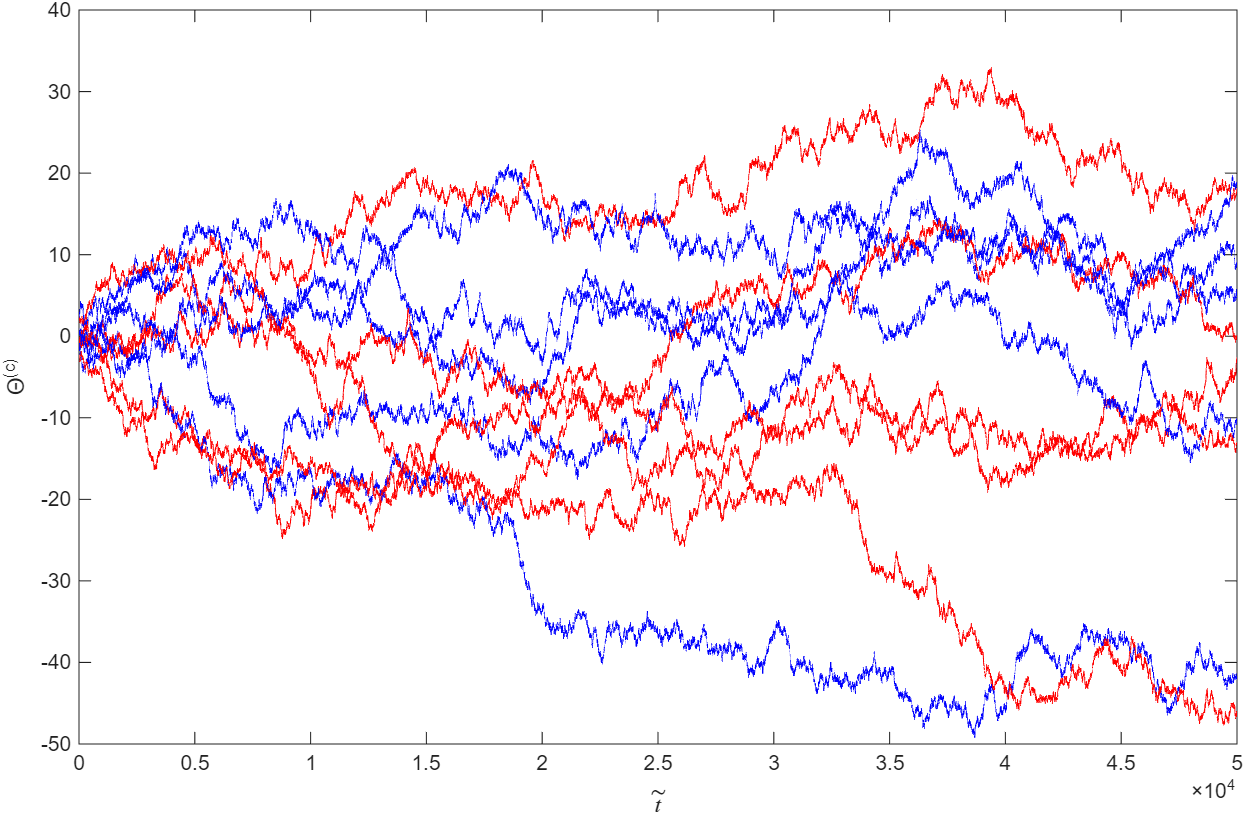}
    \caption{Comparison of the colony orientation $ \Thetacol$ as a function of nondimensional time $ \tilde{t}$  from direct simulations of the model~\eqref{rescaled_nondim_colony_kinesis} (red) and from the reduced effective dynamics~\eqref{eq:hom_kin_angle} (blue).  These simulations were conducted using our standard parameter choices from Table~\ref{tab:params} for a colony of 10 cells with nondimensional taxis factor $m_K=0.5$. 
    } \label{fig:thccompskin}
\end{figure}

As the flagellar angles have zero mean and small variance $ \sigmaThe^2$ in the kinesis model (Table~\ref{tab:params}), we use the approximation $\sin{(\sigma_\Theta{\Theta}_i)}\sim \sigma_\Theta{\Theta}_i$. 
Following the homogenization procedure from~\citep{homogenization}, we must solve first for  $\rho (\hbthvar|\hthcvar)$, the joint stationary distribution of $ \{\hat{\Theta}_i\}_{i=1}^N $, conditioned on a fixed value  $\hat{\Theta}^{(c)} = \hthcvar$. Since {$\hat{\Theta}_i$} are conditionally independent, $\rho (\hbthvar;\hthcvar)=\prod\limits_{i=1}^{N}\rho_i (\hthvar{i};\hthcvar)$, with $\rho_i$ given by the  stationary solution to the following Fokker Planck equation: 
 \begin{equation} \label{Flagella_orientation_FPE_kinesis}
\begin{aligned}
\epsilon^2(\rho_i)_{\hat{t}}=(y_i\rho_i)_{y_i}+[(1+m_K\cos(\alpha_{j'}+\hthcvar-\theta_g))\rho_i]_{y_iy_i}
\end{aligned}
\end{equation}
The solution is readily obtained to this one-dimensional problem:
\begin{equation} \label{Flagella_orientation_FPE_kinesis_soln}
 \begin{aligned}
(\rho_i) (y_i|\hthcvar)=&(2\pi [1+m_K\cos(\alpha_{j'}+\hthcvar-\theta_g)])^{-\frac{1}{2}}
e^{-y_i^2/2[1+m_K\cos(\alpha_{j'}+\hthcvar-\theta_g)]}
\end{aligned}
\end{equation}
Next we solve the following cell problem associated to equation~\eqref{rescaled_nondim_cell_kinesis} and~\eqref{rescaled_nondim_colony_kinesis}:
\begin{align*} 
\sum\limits_{j=1}^{N}\hthvar{j}\Phi_{\hthvar{j}}-\sum\limits_{j=1}^{N}[1+m_K\cos(\alpha_{j}+\hthcvar-\theta_g)]^{\frac{1}{2}}[1+m_K\cos(\alpha_j+\hthcvar-\theta_g)]^{\frac{1}{2}}\Phi_{\hthvar{j}\hthvar{j}}&=-\sum\limits_{j=1}^N \hthvar{j}, \\\hspace{2 cm}
\int\limits_{\mathcal{R}^N}\Phi(\hthcvar,\hbthvar)\rho (\hbthvar|\hthcvar)\, \difd \hbthvar&=0
\end{align*}
We get the simple linear solution $ \Phi (\hthcvar,\hbthvar) = -\sum\limits_{j=1}^N y_j$.\\
\\
For $\epsilon = \sigma_\Theta\frac{\zeta}{N^2}\ll 1$ and times $\hat{t}$ up to $\mathcal{O}(1)$, homogenization theory~\citep{homogenization} shows that the solution $\hat{\Theta}^{(c)} (\hat{t})$ to Eq.~\eqref{rescaled_nondim_colony_kinesis}, is well approximated by the solution of the homogenized equation:
\begin{align*}
d{\hat{\Theta}}^{(c)}(\hat{t})\approx&[\int\limits_{\mathbb{R}^N} \Big(2\beta^2+2\sum\limits_{j=1}^{N}(y_j)^2\Big)\rho(\mathbf{y}|\hThetacol (\hat{t}))d\mathbf{y}]^{\frac{1}{2}}dW^{\Theta,c} (\hat{t})
\end{align*} 
(where the stochastic noise is actually a different but equivalent Brownian motion).  Evaluating the integral using the above expression for the stationary distribution yields
the simplified form of the SDE  below. 
\begin{equation} \label{eq:hom_kin_angle}
\begin{aligned}
d{\hat{\Theta}}^{(c)}(\hat{t})\approx[2\beta^2+2 \sum\limits_{j=1}^{N}[1+m_K\cos(\alpha_{j}+\hThetacol(\hat{t})-\theta_g)]]^{\frac{1}{2}}dW^{\Theta,c}(\hat{t})
\end{aligned}
\end{equation}
In Figure~\ref{fig:thccompskin} we show that the reduced description~\eqref{eq:hom_kin_angle} of the sample trajectories of the colony orientation $ \Thetacol (\tilde{t}) $ visually resemble those from direct simulation of our full model~\eqref{rescaled_nondim_colony_kinesis}. 
Expressing the sum in polar form we have:
\begin{equation} 
\begin{aligned}
&\sum\limits_{j=1}^{N}[1+m_K\cos(\alpha_{j}+\hThetacol(\hat{t})-\theta_g)] = N+ m_K\sum\limits_{j=1}^{N}\Real (e^{i[\alpha_{j}+\hThetacol(\hat{t})-\theta_g]})\\&
=N+m_K\Real (e^{i[\hThetacol(\hat{t})-\theta_g]}\sum\limits_{j=1}^{N}e^{i\alpha_{j}})=N- \frac{m_K}{ \sqrt{N}} \geotaxamp \Real (e^{i[\hThetacol(\hat{t})-\theta_g+\phi]})\\&
=N- \frac{m_K}{ \sqrt{N}} \geotaxamp \cos(\hThetacol(\hat{t})-\theta_g+\phi)
\end{aligned}
\end{equation}
Using this simplified sum we get the expression in (\ref{homogenized_colony_orientation}).

The translational dynamics can be approximated more simply by averaging~\citep{homogenization}  against the fluctuations of the fast flagellar angles:
\begin{equation} \label{Colony_position_averaged_kinesis}
\begin{aligned}
\difd \hXcol(\hat{t})=&\left[\int\limits_{\mathbb{R}^N}\sum\limits_{j=1}^N -\sigmaThe^{-1} \begin{pmatrix}\cos(\alpha_{j}+\sigma_\Theta\hthvar{j}+\hThetacol (\hat{t}))\\ \sin{(\alpha_{j}+\sigma_\Theta\hthvar{j}+\hThetacol (\hat{t})}\end{pmatrix}
\rho (\hbthvar;\hThetacol (\hat{t})) \, \difd \hbthvar\right] \, \difd\hat{t}
+\sqrt{2} \derat \epsilon  \,  d\mathbf{W}^{X,c}(\hat{t})
\end{aligned}
\end{equation} 
The simplified SDE in Eq.~\eqref{homogenized_colony_orientation} is obtained by computing the integrals against the Gaussian stationary distribution~\eqref{Flagella_orientation_FPE_kinesis_soln}, using the averaging formula $ \langle \mathrm{e}^{\mathi Z} \rangle = \mathrm{e}^{\mathi \mu_Z - \frac{1}{2} \sigma_Z^2}$ for a Gaussian random variable $ Z $ with mean $ \mu_Z $ and standard deviation $ \sigma_Z$.

The theorems for stochastic averaging and homogenization~\citep{homogenization} generally impose limits on validity to order unity time for the slow variables (here $ \hat{t} \sim O(1)$).  
The only complexity in the coarse-grained equations Eqs.~\eqref{homogenized_colony_orientation} and~\eqref{averaged_colony_position} that could potentially obstruct their validity for arbitrarily large times is when the colony has moved far enough to make the spatial dependence of the logarithmic concentration gradient significant (so $\kkinnd $ can no longer be treated as constant, as discussed at the end of Subsection~\ref{subsec:model_col}).  

\begin{figure}[H]    
 \centering 
\begin{multicols}{2}
  \includegraphics[scale=0.55]{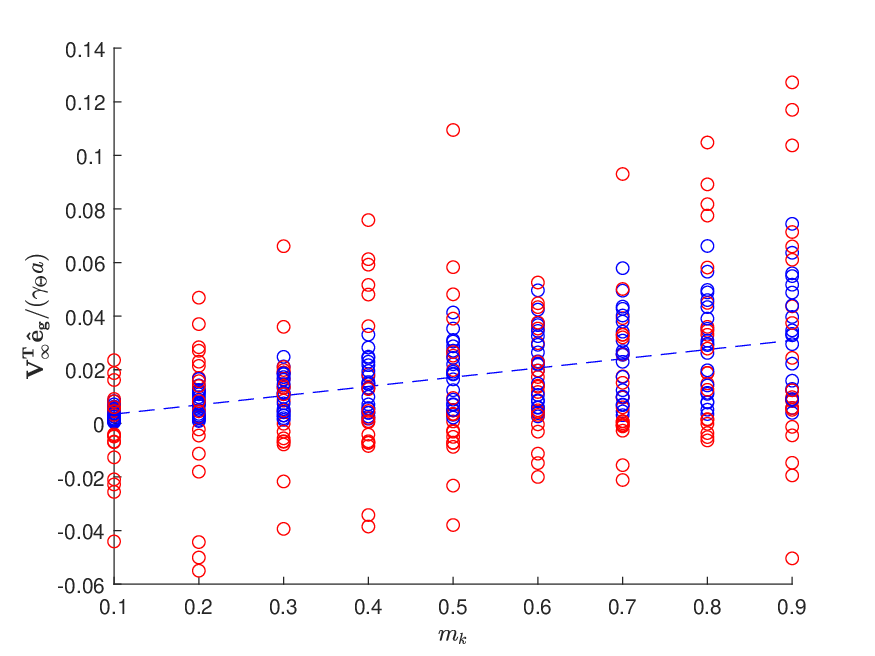}
  \includegraphics[scale=0.55]{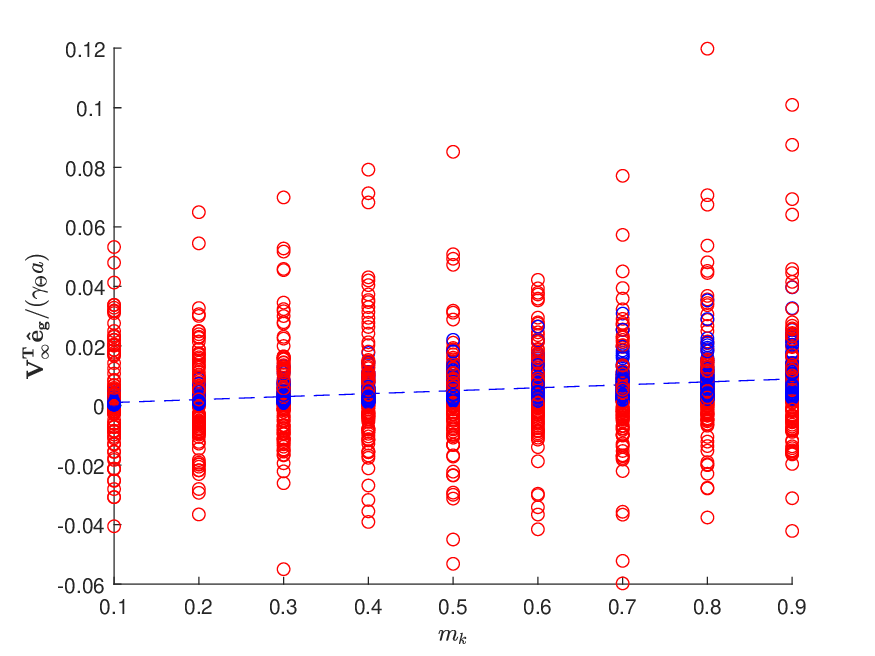}
\end{multicols}
\caption{Projection of 
nondimensional effective drift $\drifteffnd = \drifteff/(\gamma_\Theta a) $ along attractant gradient $\graddir$
for colonies of $N=7 $ cells (left panel) and $N=10$ cells (right panel) exhibiting kinesis with various nondimensional kinesis response factors $m_K$. The placement variables are generated independently for each individual cell in independent colonies with the placement variables randomly displaced by $\flagdispi \sim U (-l/2,l/2)$. The results of 20  Monte Carlo simulations for each indicated value of $m_K$ over nondimensional time $\tilde{t}=50000$ are plotted in red circles.    The analytical predictions (\ref{non-dim_kinesis_drift}) for each sample colony is plotted as blue circles, with the broken blue line being the demographic average~\eqref{avgdrift_kinesis}. 
 Parameter values are as listed in Table~\ref{tab:params}, with flagellar force magnitude $F= 5 \,\si{\pico\newton}$.}
 \label{fig:kinesis_responsefactor_short}
\end{figure}

\begin{figure}[H]  
 \centering
    \includegraphics[scale=0.8]{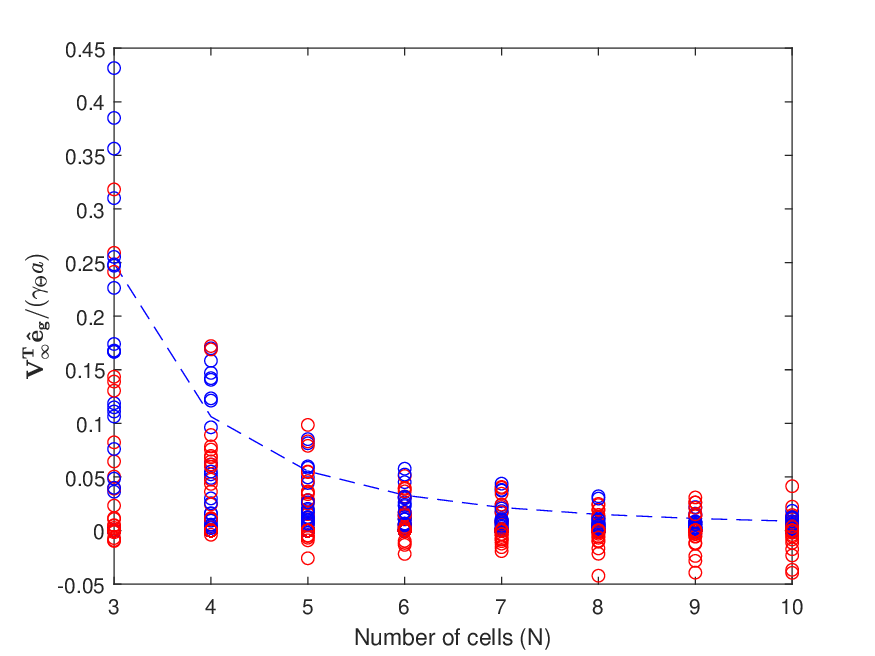} 
\caption{Projection of nondimensional effective drift $\drifteffnd = \drifteff/(\gamma_\Theta a) $
of colonies of $m_K=0.55$ cells 
exhibiting kinesis with various population sizes $N$. The placement variables are generated independently for each individual cells in independent colonies with the placement variables randomly displaced by $\flagdispi \sim U (-l/2,l/2)$. The results of 20 Monte Carlo simulations over nondimensional time $\tilde{t}=50000$
are plotted in red circles. The analytical predictions (\ref{non-dim_kinesis_drift}) for each sample colony is plotted as blue circles, with the broken blue line being the demographic average~\eqref{avgdrift_kinesis}.  Parameter values are as listed in Table~\ref{tab:params}, with flagellar force magnitude $F= 5 \,\si{\pico\newton}$.
}
\label{fig:kinesis_popnsize_short}
\end{figure}

In Figures~\ref{fig:kinesis_responsefactor} and~\ref{fig:kinesis_popnsize} of the main text, we presented results of the kinesis simulations for ten times longer than the taxis simulations.  We show in Figures~\ref{fig:kinesis_responsefactor_short} and~\ref{fig:kinesis_popnsize_short} the results from running the kinesis simulations over the same time ($\tilde{t}=50000$) as the taxis simulations.  We see the simulated drifts are much more variable than those seen in the corresponding Figures~\ref{fig:kinesis_responsefactor} and~\ref{fig:kinesis_popnsize} run for ten times longer, indicating the slower convergence for $\Thetacol$ to reach its stationary distribution in the kinesis model.  This suggests the discrepancies in Figures~\ref{fig:kinesis_responsefactor} and~\ref{fig:kinesis_popnsize} between the theoretical and simulation results are likely owing at least in part to finite-time sampling error in the simulations.


\bibliography{ref}
\bibliographystyle{unsrtnat}

\end{document}